\documentclass[10pt, journal, letterpaper]{IEEEtran}

\usepackage{picins}

\usepackage[utf8]{inputenc}
\usepackage{epsfig}
\usepackage{amsmath}
\usepackage{graphicx}
\usepackage{amssymb}
\usepackage{fixmath}
\usepackage[belowskip=-5pt, font=small]{caption}
\usepackage{algorithmic}
\usepackage{algorithm}
\usepackage{multirow}
\usepackage[font=small]{subcaption}

\PassOptionsToPackage{hyphens}{url}
\usepackage{hyperref}
\usepackage[numbers,sort&compress]{natbib}
\usepackage{xspace}

\usepackage{xcolor}
\usepackage{makecell}

\usepackage[acronyms,nonumberlist,nopostdot,nomain,nogroupskip,acronymlists={hidden}]{glossaries}
\newglossary[algh]{hidden}{acrh}{acnh}{Hidden Acronyms}
\newacronym{3gpp}{3GPP}{3rd Generation Partnership Project}
\newacronym{4g}{4G}{4th generation}
\newacronym{5g}{5G}{5th generation}
\newacronym{5gc}{5GC}{5G Core}
\newacronym{adc}{ADC}{Analog to Digital Converter}
\newacronym{aerpaw}{AERPAW}{Aerial Experimentation and Research Platform for Advanced Wireless}
\newacronym{ai}{AI}{Artificial Intelligence}
\newacronym{aimd}{AIMD}{Additive Increase Multiplicative Decrease}
\newacronym{am}{AM}{Acknowledged Mode}
\newacronym{amc}{AMC}{Adaptive Modulation and Coding}
\newacronym{amf}{AMF}{Access and Mobility Management Function}
\newacronym{aops}{AOPS}{Adaptive Order Prediction Scheduling}
\newacronym{api}{API}{Application Programming Interface}
\newacronym{apn}{APN}{Access Point Name}
\newacronym{aqm}{AQM}{Active Queue Management}
\newacronym{ausf}{AUSF}{Authentication Server Function}
\newacronym{avc}{AVC}{Advanced Video Coding}
\newacronym{awgn}{AGWN}{Additive White Gaussian Noise}
\newacronym{balia}{BALIA}{Balanced Link Adaptation Algorithm}
\newacronym{bbu}{BBU}{Base Band Unit}
\newacronym{bdp}{BDP}{Bandwidth-Delay Product}
\newacronym{bf}{BF}{Beamforming}
\newacronym{bler}{BLER}{Block Error Rate}
\newacronym{brr}{BRR}{Bayesian Ridge Regressor}
\newacronym{bsr}{BSR}{Buffer Status Report}
\newacronym{bss}{BSS}{Business Support System}
\newacronym{ca}{CA}{Carrier Aggregation}
\newacronym{caas}{CaaS}{Connectivity-as-a-Service}
\newacronym{cb}{CB}{Code Block}
\newacronym{cc}{CC}{Congestion Control}
\newacronym{ccid}{CCID}{Congestion Control ID}
\newacronym{cco}{CC}{Carrier Component}
\newacronym{cdd}{CDD}{Cyclic Delay Diversity}
\newacronym{cdf}{CDF}{Cumulative Distribution Function}
\newacronym{cdn}{CDN}{Content Distribution Network}
\newacronym{cn}{CN}{Core Network}
\newacronym{codel}{CoDel}{Controlled Delay Management}
\newacronym{comac}{COMAC}{Converged Multi-Access and Core}
\newacronym{cord}{CORD}{Central Office Re-architected as a Datacenter}
\newacronym{cornet}{CORNET}{COgnitive Radio NETwork}
\newacronym{cosmos}{COSMOS}{Cloud Enhanced Open Software Defined Mobile Wireless Testbed for City-Scale Deployment}
\newacronym{cots}{COTS}{Commercial Off-the-Shelf}
\newacronym{cp}{CP}{Control Plane}
\newacronym{cpu}{CPU}{Central Processing Unit}
\newacronym{cqi}{CQI}{Channel Quality Information}
\newacronym{cr}{CR}{Cognitive Radio}
\newacronym{cran}{CRAN}{Cloud \gls{ran}}
\newacronym{crs}{CRS}{Cell Reference Signal}
\newacronym{csi}{CSI}{Channel State Information}
\newacronym{csirs}{CSI-RS}{Channel State Information - Reference Signal}
\newacronym{cu}{CU}{Central Unit}
\newacronym{d2tcp}{D$^2$TCP}{Deadline-aware Data center TCP}
\newacronym{d3}{D$^3$}{Deadline-Driven Delivery}
\newacronym{dac}{DAC}{Digital to Analog Converter}
\newacronym{dag}{DAG}{Directed Acyclic Graph}
\newacronym{das}{DAS}{Distributed Antenna System}
\newacronym{dash}{DASH}{Dynamic Adaptive Streaming over HTTP}
\newacronym{dc}{DC}{Dual Connectivity}
\newacronym{dccp}{DCCP}{Datagram Congestion Control Protocol}
\newacronym{dce}{DCE}{Direct Code Execution}
\newacronym{dci}{DCI}{Downlink Control Information}
\newacronym{dctcp}{DCTCP}{Data Center TCP}
\newacronym{dl}{DL}{Downlink}
\newacronym{dmr}{DMR}{Deadline Miss Ratio}
\newacronym{dmrs}{DMRS}{DeModulation Reference Signal}
\newacronym{drlcc}{DRL-CC}{Deep Reinforcement Learning Congestion Control}
\newacronym{drs}{DRS}{Discovery Reference Signal}
\newacronym{du}{DU}{Distributed Unit}
\newacronym{e2e}{E2E}{end-to-end}
\newacronym{ecaas}{ECaaS}{Edge-Cloud-as-a-Service}
\newacronym{ecn}{ECN}{Explicit Congestion Notification}
\newacronym{edf}{EDF}{Earliest Deadline First}
\newacronym{embb}{eMBB}{Enhanced Mobile Broadband}
\newacronym{empower}{EMPOWER}{EMpowering transatlantic PlatfOrms for advanced WirEless Research}
\newacronym{enb}{eNB}{evolved Node Base}
\newacronym{endc}{EN-DC}{E-UTRAN-\gls{nr} \gls{dc}}
\newacronym{epc}{EPC}{Evolved Packet Core}
\newacronym{eps}{EPS}{Evolved Packet System}
\newacronym{es}{ES}{Edge Server}
\newacronym{etsi}{ETSI}{European Telecommunications Standards Institute}
\newacronym[firstplural=Estimated Times of Arrival (ETAs)]{eta}{ETA}{Estimated Time of Arrival}
\newacronym{eutran}{E-UTRAN}{Evolved Universal Terrestrial Access Network}
\newacronym{faas}{FaaS}{Function-as-a-Service}
\newacronym{fapi}{FAPI}{Functional Application Platform Interface}
\newacronym{fdd}{FDD}{Frequency Division Duplexing}
\newacronym{fdm}{FDM}{Frequency Division Multiplexing}
\newacronym{fdma}{FDMA}{Frequency Division Multiple Access}
\newacronym{fed4fire}{FED4FIRE+}{Federation 4 Future Internet Research and Experimentation Plus}
\newacronym{fit}{FIT}{Future \acrlong{iot}}
\newacronym{fpga}{FPGA}{Field Programmable Gate Array}
\newacronym{fr2}{FR2}{Frequency Range 2}
\newacronym{fs}{FS}{Fast Switching}
\newacronym{fscc}{FSCC}{Flow Sharing Congestion Control}
\newacronym{ftp}{FTP}{File Transfer Protocol}
\newacronym{fw}{FW}{Flow Window}
\newacronym{ge}{GE}{Gaussian Elimination}
\newacronym{gnb}{gNB}{Next Generation Node Base}
\newacronym{gop}{GOP}{Group of Pictures}
\newacronym{gpr}{GPR}{Gaussian Process Regressor}
\newacronym{gpu}{GPU}{Graphics Processing Unit}
\newacronym{gtp}{GTP}{GPRS Tunneling Protocol}
\newacronym{gtpc}{GTP-C}{GPRS Tunneling Protocol Control Plane}
\newacronym{gtpu}{GTP-U}{GPRS Tunneling Protocol User Plane}
\newacronym{gtpv2c}{GTPv2-C}{\gls{gtp} v2 - Control}
\newacronym{gw}{GW}{Gateway}
\newacronym{harq}{HARQ}{Hybrid Automatic Repeat reQuest}
\newacronym{hetnet}{HetNet}{Heterogeneous Network}
\newacronym{hh}{HH}{Hard Handover}
\newacronym{hol}{HOL}{Head-of-Line}
\newacronym{hqf}{HQF}{Highest-quality-first}
\newacronym{hss}{HSS}{Home Subscription Server}
\newacronym{http}{HTTP}{HyperText Transfer Protocol}
\newacronym{ia}{IA}{Initial Access}
\newacronym{iab}{IAB}{Integrated Access and Backhaul}
\newacronym{ic}{IC}{Incident Command}
\newacronym{ietf}{IETF}{Internet Engineering Task Force}
\newacronym{imsi}{IMSI}{International Mobile Subscriber Identity}
\newacronym{imt}{IMT}{International Mobile Telecommunication}
\newacronym{iot}{IoT}{Internet of Things}
\newacronym{ip}{IP}{Internet Protocol}
\newacronym{itu}{ITU}{International Telecommunication Union}
\newacronym{kpi}{KPI}{Key Performance Indicator}
\newacronym{kvm}{KVM}{Kernel-based Virtual Machine}
\newacronym{los}{LOS}{Line-of-Sight}
\newacronym{lsm}{LSM}{Link-to-System Mapping}
\newacronym{lstm}{LSTM}{Long Short Term Memory}
\newacronym{lte}{LTE}{Long Term Evolution}
\newacronym{lxc}{LXC}{Linux Containers}
\newacronym{m2m}{M2M}{Machine to Machine}
\newacronym{mac}{MAC}{Medium Access Control}
\newacronym{manet}{MANET}{Mobile Ad Hoc Network}
\newacronym{mano}{MANO}{Management and Orchestration}
\newacronym{mc}{MC}{Multi-Connectivity}
\newacronym{mcc}{MCC}{Mobile Cloud Computing}
\newacronym{mchem}{MCHEM}{Massive Channel Emulator}
\newacronym{mcs}{MCS}{Modulation and Coding Scheme}
\newacronym{mec}{MEC}{Multi-access Edge Computing}
\newacronym{mec2}{MEC}{Mobile Edge Cloud}
\newacronym{mfc}{MFC}{Mobile Fog Computing}
\newacronym{mi}{MI}{Mutual Information}
\newacronym{mib}{MIB}{Master Information Block}
\newacronym{miesm}{MIESM}{Mutual Information Based Effective SINR}
\newacronym{mimo}{MIMO}{Multiple Input, Multiple Output}
\newacronym{ml}{ML}{Machine Learning}
\newacronym{mlr}{MLR}{Maximum-local-rate}
\newacronym[plural=\gls{mme}s,firstplural=Mobility Management Entities (MMEs)]{mme}{MME}{Mobility Management Entity}
\newacronym{mmtc}{mMTC}{Massive Machine-Type Communications}
\newacronym{mmwave}{mmWave}{millimeter wave}
\newacronym{mpdccp}{MP-DCCP}{Multipath Datagram Congestion Control Protocol}
\newacronym{mptcp}{MPTCP}{Multipath TCP}
\newacronym{mr}{MR}{Maximum Rate}
\newacronym{mrdc}{MR-DC}{Multi \gls{rat} \gls{dc}}
\newacronym{mse}{MSE}{Mean Square Error}
\newacronym{mss}{MSS}{Maximum Segment Size}
\newacronym{mt}{MT}{Mobile Termination}
\newacronym{mtd}{MTD}{Machine-Type Device}
\newacronym{mtu}{MTU}{Maximum Transmission Unit}
\newacronym{mumimo}{MU-MIMO}{Multi-user \gls{mimo}}
\newacronym{mvno}{MVNO}{Mobile Virtual Network Operator}
\newacronym{nalu}{NALU}{Network Abstraction Layer Unit}
\newacronym{nas}{NAS}{Non-Access Stratum}
\newacronym{nbiot}{NB-IoT}{Narrow Band IoT}
\newacronym{nfv}{NFV}{Network Function Virtualization}
\newacronym{nfvi}{NFVI}{Network Function Virtualization Infrastructure}
\newacronym{nic}{NIC}{Network Interface Card}
\newacronym{nlos}{NLOS}{Non-Line-of-Sight}
\newacronym{now}{NOW}{Non Overlapping Window}
\newacronym{nsm}{NSM}{Network Service Mesh}
\newacronym[type=hidden]{nr}{NR}{New Radio}
\newacronym{nrf}{NRF}{Network Repository Function}
\newacronym{nsa}{NSA}{Non Stand Alone}
\newacronym{nse}{NSE}{Network Slicing Engine}
\newacronym{nssf}{NSSF}{Network Slice Selection Function}
\newacronym{o2i}{O2I}{Outdoor to Indoor}
\newacronym{oai}{OAI}{OpenAirInterface}
\newacronym{oaicn}{OAI-CN}{\gls{oai} \acrlong{cn}}
\newacronym{oairan}{OAI-RAN}{\acrlong{oai} \acrlong{ran}}
\newacronym{oam}{OAM}{Operations, Administration and Maintenance}
\newacronym{ofdm}{OFDM}{Orthogonal Frequency Division Multiplexing}
\newacronym{olia}{OLIA}{Opportunistic Linked Increase Algorithm}
\newacronym{omec}{OMEC}{Open Mobile Evolved Core}
\newacronym{onap}{ONAP}{Open Network Automation Platform}
\newacronym{onf}{ONF}{Open Networking Foundation}
\newacronym{onos}{ONOS}{Open Networking Operating System}
\newacronym{oom}{OOM}{\gls{onap} Operations Manager}
\newacronym{opnfv}{OPNFV}{Open Platform for \gls{nfv}}
\newacronym[type=hidden]{oran}{O-RAN}{Open \gls{ran}}
\newacronym{orbit}{ORBIT}{Open-Access Research Testbed for Next-Generation Wireless Networks}
\newacronym{os}{OS}{Operating System}
\newacronym{oss}{OSS}{Operations Support System}
\newacronym{pa}{PA}{Position-aware}
\newacronym{pase}{PASE}{Prioritization, Arbitration, and Self-adjusting Endpoints}
\newacronym{pawr}{PAWR}{Platforms for Advanced Wireless Research}
\newacronym{pbch}{PBCH}{Physical Broadcast Channel}
\newacronym{pcef}{PCEF}{Policy and Charging Enforcement Function}
\newacronym{pcfich}{PCFICH}{Physical Control Format Indicator Channel}
\newacronym{pcrf}{PCRF}{Policy and Charging Rules Function}
\newacronym{pdcch}{PDCCH}{Physical Downlink Control Channel}
\newacronym{pdcp}{PDCP}{Packet Data Convergence Protocol}
\newacronym{pdsch}{PDSCH}{Physical Downlink Shared Channel}
\newacronym{pdu}{PDU}{Packet Data Unit}
\newacronym{pf}{PF}{Proportional Fair}
\newacronym{pgw}{PGW}{Packet Gateway}
\newacronym{phich}{PHICH}{Physical Hybrid ARQ Indicator Channel}
\newacronym{phy}{PHY}{Physical}
\newacronym{pmch}{PMCH}{Physical Multicast Channel}
\newacronym{pmi}{PMI}{Precoding Matrix Indicators}
\newacronym{powder}{POWDER}{Platform for Open Wireless Data-driven Experimental Research}
\newacronym{ppp}{PPP}{Poisson Point Process}
\newacronym{prach}{PRACH}{Physical Random Access Channel}
\newacronym{prb}{PRB}{Physical Resource Block}
\newacronym{psnr}{PSNR}{Peak Signal to Noise Ratio}
\newacronym{pss}{PSS}{Primary Synchronization Signal}
\newacronym{pucch}{PUCCH}{Physical Uplink Control Channel}
\newacronym{pusch}{PUSCH}{Physical Uplink Shared Channel}
\newacronym{qam}{QAM}{Quadrature Amplitude Modulation}
\newacronym{qci}{QCI}{\gls{qos} Class Identifier}
\newacronym{qoe}{QoE}{Quality of Experience}
\newacronym{qos}{QoS}{Quality of Service}
\newacronym{quic}{QUIC}{Quick UDP Internet Connections}
\newacronym{rach}{RACH}{Random Access Channel}
\newacronym{ran}{RAN}{Radio Access Network}
\newacronym[firstplural=Radio Access Technologies (RATs)]{rat}{RAT}{Radio Access Technology}
\newacronym{rcn}{RCN}{Research Coordination Network}
\newacronym{rec}{REC}{Radio Edge Cloud}
\newacronym{red}{RED}{Random Early Detection}
\newacronym{renew}{RENEW}{Reconfigurable Eco-system for Next-generation End-to-end Wireless}
\newacronym{rf}{RF}{Radio Frequency}
\newacronym{rfc}{RFC}{Request for Comments}
\newacronym{rfr}{RFR}{Random Forest Regressor}
\newacronym{ric}{RIC}{\gls{ran} Intelligent Controller}
\newacronym{rlc}{RLC}{Radio Link Control}
\newacronym{rlf}{RLF}{Radio Link Failure}
\newacronym{rlnc}{RLNC}{Random Linear Network Coding}
\newacronym{rmse}{RMSE}{Root Mean Squared Error}
\newacronym{rnis}{RNIS}{Radio Network Information Service}
\newacronym{rr}{RR}{Round Robin}
\newacronym{rrc}{RRC}{Radio Resource Control}
\newacronym{rrm}{RRM}{Radio Resource Management}
\newacronym{rru}{RRU}{Remote Radio Unit}
\newacronym{rs}{RS}{Remote Server}
\newacronym{rsrp}{RSRP}{Reference Signal Received Power}
\newacronym{rsrq}{RSRQ}{Reference Signal Received Quality}
\newacronym{rss}{RSS}{Received Signal Strength}
\newacronym{rssi}{RSSI}{Received Signal Strength Indicator}
\newacronym{rtt}{RTT}{Round Trip Time}
\newacronym{ru}{RU}{Radio Unit}
\newacronym{rw}{RW}{Receive Window}
\newacronym{rx}{RX}{Receiver}
\newacronym{s1ap}{S1AP}{S1 Application Protocol}
\newacronym{sa}{SA}{standalone}
\newacronym{sack}{SACK}{Selective Acknowledgment}
\newacronym{sap}{SAP}{Service Access Point}
\newacronym{sc2}{SC2}{Spectrum Collaboration Challenge}
\newacronym{scef}{SCEF}{Service Capability Exposure Function}
\newacronym{sch}{SCH}{Secondary Cell Handover}
\newacronym{scoot}{SCOOT}{Split Cycle Offset Optimization Technique}
\newacronym{sctp}{SCTP}{Stream Control Transmission Protocol}
\newacronym{sdap}{SDAP}{Service Data Adaptation Protocol}
\newacronym{sdk}{SDK}{Software Development Kit}
\newacronym{sdm}{SDM}{Space Division Multiplexing}
\newacronym{sdma}{SDMA}{Spatial Division Multiple Access}
\newacronym{sdn}{SDN}{Software-defined Networking}
\newacronym{sdr}{SDR}{Software-defined Radio}
\newacronym{seba}{SEBA}{SDN-Enabled Broadband Access}
\newacronym{sgsn}{SGSN}{Serving GPRS Support Node}
\newacronym{sgw}{SGW}{Service Gateway}
\newacronym{si}{SI}{Study Item}
\newacronym{sib}{SIB}{Secondary Information Block}
\newacronym{sinr}{SINR}{Signal to Interference plus Noise Ratio}
\newacronym{sip}{SIP}{Session Initiation Protocol}
\newacronym{siso}{SISO}{Single Input, Single Output}
\newacronym{sla}{SLA}{Service Level Agreement}
\newacronym{sm}{SM}{Saturation Mode}
\newacronym{smf}{SMF}{Session Management Function}
\newacronym{sms}{SMS}{Short Message Service}
\newacronym{smsgmsc}{SMS-GMSC}{\gls{sms}-Gateway}
\newacronym{snr}{SNR}{Signal-to-Noise-Ratio}
\newacronym{son}{SON}{Self-Organizing Network}
\newacronym{sptcp}{SPTCP}{Single Path TCP}
\newacronym{srb}{SRB}{Service Radio Bearer}
\newacronym{srs}{SRS}{Sounding Reference Signal}
\newacronym{ss}{SS}{Synchronization Signal}
\newacronym{sss}{SSS}{Secondary Synchronization Signal}
\newacronym{st}{ST}{Spanning Tree}
\newacronym{svc}{SVC}{Scalable Video Coding}
\newacronym{tb}{TB}{Transport Block}
\newacronym{tcp}{TCP}{Transmission Control Protocol}
\newacronym{tdd}{TDD}{Time Division Duplexing}
\newacronym{tdm}{TDM}{Time Division Multiplexing}
\newacronym{tdma}{TDMA}{Time Division Multiple Access}
\newacronym{tfl}{TfL}{Transport for London}
\newacronym{tfrc}{TFRC}{TCP-Friendly Rate Control}
\newacronym{tft}{TFT}{Traffic Flow Template}
\newacronym{tip}{TIP}{Telecom Infra Project}
\newacronym{tm}{TM}{Transparent Mode}
\newacronym{tr}{TR}{Technical Report}
\newacronym{trp}{TRP}{Transmitter Receiver Pair}
\newacronym{ts}{TS}{Technical Specification}
\newacronym{tti}{TTI}{Transmission Time Interval}
\newacronym{ttt}{TTT}{Time-to-Trigger}
\newacronym{tx}{TX}{Transmitter}
\newacronym{uas}{UAS}{Unmanned Aerial System}
\newacronym{uav}{UAV}{Unmanned Aerial Vehicle}
\newacronym{udm}{UDM}{Unified Data Management}
\newacronym{udp}{UDP}{User Datagram Protocol}
\newacronym{udr}{UDR}{Unified Data Repository}
\newacronym{ue}{UE}{User Equipment}
\newacronym{uhd}{UHD}{\gls{usrp} Hardware Driver}
\newacronym{ul}{UL}{Uplink}
\newacronym{um}{UM}{Unacknowledged Mode}
\newacronym{uml}{UML}{Unified Modeling Language}
\newacronym{upa}{UPA}{Uniform Planar Array}
\newacronym{upf}{UPF}{User Plane Function}
\newacronym{urllc}{URLLC}{Ultra Reliable and Low Latency Communication}
\newacronym{usa}{U.S.}{United States}
\newacronym{usim}{USIM}{Universal Subscriber Identity Module}
\newacronym{usrp}{USRP}{Universal Software Radio Peripheral}
\newacronym{utc}{UTC}{Urban Traffic Control}
\newacronym{vim}{VIM}{Virtualization Infrastructure Manager}
\newacronym{vm}{VM}{Virtual Machine}
\newacronym{vnf}{VNF}{Virtual Network Function}
\newacronym{volte}{VoLTE}{Voice over \gls{lte}}
\newacronym{voltha}{VOLTHA}{Virtual OLT HArdware Abstraction}
\newacronym{vr}{VR}{Virtual Reality}
\newacronym{vran}{vRAN}{Virtualized \gls{ran}}
\newacronym{vss}{VSS}{Video Streaming Server}
\newacronym{wbf}{WBF}{Wired Bias Function}
\newacronym{wf}{WF}{Wired-first}
\newacronym{wlan}{WLAN}{Wireless Local Area Network}
\newacronym{osm}{OSM}{Open Source \gls{nfv} Management and Orchestration}
\newacronym{pnf}{PNF}{Physical Network Function}

\makeglossaries
\glsdisablehyper

\newcommand{\ie}{i.e.,\xspace}
\newcommand{\eg}{e.g.,\xspace}
\newcommand{\enb}{\gls{enb}\xspace}
\newcommand{\enbs}{\glspl{enb}\xspace}

\newcommand{\fig}[1]{Figure~\ref{#1}}
\newcommand{\gnb}{\gls{gnb}\xspace}

\newcommand{\powderrenew}{\acrshort{powder}-\acrshort{renew}\xspace}

\renewcommand{\sec}[1]{Section~\ref{#1}}

\newcommand{\srsue}{srsUE\xspace}

\newcommand{\ue}{\gls{ue}\xspace}
\newcommand{\ues}{\glspl{ue}\xspace}

\begin{document}


\title{Open, Programmable, and Virtualized 5G Networks:\\State-of-the-Art and the Road Ahead}

\author{Leonardo Bonati,
Michele Polese,
Salvatore D'Oro,
Stefano Basagni,
Tommaso Melodia\\
\vspace{4pt}
Institute for the Wireless Internet of Things, Northeastern University, Boston, MA 02115, USA\\
\vspace{2pt}
Email: \{l.bonati, m.polese, s.doro, s.basagni, t.melodia\}@northeastern.edu
\thanks{This work was supported in part by the US National Science Foundation under Grant CNS-1618727 and in part by the US Office of Naval Research under Grants N00014-19-1-2409 and N00014-20-1-2132.}
}

\glsunset{nr} 
\glsunset{oran}

\maketitle

\begin{abstract}
     Fifth generation (5G) cellular networks will serve a wide variety of heterogeneous use cases, including mobile broadband users, ultra-low latency services and massively dense connectivity scenarios.
     The resulting diverse communication requirements will demand networking with unprecedented flexibility, not currently provided by the monolithic black-box approach of 4G cellular networks.
     The research community and an increasing number of standardization bodies and industry coalitions have recognized softwarization, virtualization, and disaggregation of networking functionalities as the key enablers of the needed shift to flexibility.
     Particularly, software-defined cellular networks are heralded as the prime technology to satisfy the new application-driven traffic requirements and to support the highly time-varying topology and interference dynamics, because of their \textit{openness} through well-defined interfaces, and \textit{programmability}, for swift and responsive network optimization.
     Leading the technological innovation in this direction, several 5G software-based projects and alliances have embraced the open source approach, making new libraries and frameworks available to the wireless community.
     This race to open source softwarization, however, has led to a deluge of solutions whose interoperability and interactions are often unclear.
     This article provides the first cohesive and exhaustive compendium of recent open source software and frameworks for 5G cellular networks, with a full stack and end-to-end perspective.
     We detail their capabilities and functionalities focusing on how their constituting elements fit the 5G ecosystem, and unravel the interactions among the surveyed solutions.
     Finally, we review hardware and testbeds on which these frameworks can run, and provide a critical perspective on the limitations of the state-of-the-art, as well as feasible  directions toward fully open source, programmable 5G networks. 
\end{abstract}


\begin{picture}(0,0)(20,-510)
\put(0,0){
\put(0,20){\footnotesize This article has been published on Computer Networks. Please cite it as L. Bonati, M. Polese, S. D'Oro, S. Basagni, and T. Melodia, ``Open, Programmable,}
\put(0,10){\footnotesize and Virtualized 5G Networks: State-of-the-Art and the Road Ahead,'' Computer Networks, vol. 182, December 2020, doi: \href{https://doi.org/10.1016/j.comnet.2020.107516}{10.1016/j.comnet.2020.107516}.}
\put(0,0){\tiny \copyright 2020. This manuscript version is made available under the CC-BY-NC-ND 4.0 license \url{http://creativecommons.org/licenses/by-nc-nd/4.0/}}
\put(0,-5){\scriptsize}}
\end{picture}

\section{Introduction}
\label{sec:intro}
 

The potential of \gls{5g} communications is being unleashed into the fabric of cellular networks, enabling unprecedented technological advancements in the networking hardware and software ecosystems~\cite{parkvall2017nr}.
Applications such as virtual reality, telesurgery, high-resolution video streaming, and private cellular networking---just to name a few---will be freed from the shadows of the spectrum crunch and resource-scarcity that have haunted \gls{4g} networks for years.
By unbridling the sheer power of these applications, \gls{5g} will usher unparalleled business opportunities for infrastructure and service providers, and foster unrivaled cellular networking-based innovation~\cite{boccardi2014five}.

The journey to achieve the~\gls{5g} vision, however, is still beset by many research and development challenges.
Traditional cellular networks are characterized by an inflexible and monolithic infrastructure, incapable of meeting the heterogeneity and variability of~\gls{5g} scenarios and the strict requirements of its applications~\cite{martin2009way}. 
Now more than ever, the limitations of the ``black-box'' approaches of current cellular deployments, where hardware and software are \textit{plug-and-play} with little or no reconfiguration capabilities, are manifest. 
The lack of full control of the vast amount of available resources and network parameters makes it hard to adapt network operations to real-time traffic conditions and requirements, resulting in ineffective resource management, sub-optimal performance, and inability to implement \gls{caas} technologies such as private cellular networking~\cite{oranwp}.
The inflexibility of current approaches is even more harmful in 5G scenarios, where densification and the need for directional communications call for fine-grained network control \cite{doroInfocom2019Slicing,doro2020slicing,zambianco2020interference}, resources are scarce and spectrum availability and energy consumption are strictly regulated~\cite{bhushan2014network}. 

%
Both industry and academia now agree that the practical realization of 5G systems needs a radical overhaul of all plug-and-play approaches in favor of new, agile and open paradigms for network deployment, control and management.
In this context, revolutionary and innovative networking solutions based upon \textit{programmability}, \textit{openness}, \textit{resource sharing} and \textit{edgefication} are welcome to the cellular arena~\cite{haavisto2019opensource,berde2014onos}. 
New networking principles such as \gls{sdn}~\cite{mckeown2008openflow}, network virtualization~\cite{CHOWDHURY2010862}, and \gls{mec}~\cite{mao2017survey} have demonstrated that dynamic network control and agile management (\eg frequency planning, user scheduling, mobility management, among others) is possible. 
Similarly, the emergence of network slicing and cloud \gls{ran} technologies have made it clear that infrastructure sharing not only maximizes resource utilization, but also opens new market opportunities (\eg differentiated services, infrastructure leasing, \gls{caas}), \textit{thus representing a desirable solution for network operators and infrastructure providers alike}~\cite{zhang2017survey,BARAKABITZE2020106984}.

Following the growing interest in softwarization and virtualization technologies, the~5G ecosystem has witnessed the exponential growth of dedicated solutions for~5G applications~\cite{sun2015integrating}. 
These solutions include software and hardware tailored to specific tasks~\cite{kaltenberger2020openairinterface} and full-fledged multitasking frameworks spanning the whole infrastructure~\cite{onap_architecture}.
Despite their diversity in structure and purpose, the majority of these solutions has two important aspects in common: They are \textit{open source} and fully \textit{programmable}. 
\textit{These two aspects together are bringing unprecedented flexibility to~5G systems, making them accessible to a much broader community of researchers and developers.}

Just a few years ago, the majority of researchers had no access to actual cellular networks. 
When they did, access was limited to individual network components or functionalities.  
Today, the software-defined paradigm as made popular by the GNU Radio libraries~\cite{gnu-radio}, has been easily adopted by software bundles such as \gls{oai}~\cite{kaltenberger2020openairinterface} and srsLTE~\cite{gomez2016srslte} for swift instantiation of fully-functional cellular networks on commercial \gls{sdr} devices. 
Software frameworks such as O-RAN~\cite{oran_website, onap_website}, which run on ``white-box'' servers, allow reconfiguration and optimization of network and transceiver functionalities.
These new software and hardware components have radically changed the way the research community and the telecom industry plan, deploy, and interact with cellular systems.
Prototyping, testing, and deploying new algorithms and protocols for cellular networks enjoys now unprecedented ease and time to market.
The advantage of this revolutionary approach is twofold: (i)~\textit{Openness} allows researchers to evaluate and analyze their solutions on a real-world setup~\cite{bertizzolo2020arena}, and enables telecom operators to directly interact and control networking equipment~\cite{oranwp}.
Also, (ii)~\textit{programmability} fosters the design of novel and advanced algorithms that optimize network performance by efficiently and dynamically allocating network resources and controlling software and hardware functionalities, even in real time, if appropriate. 
For instance, telecom operators such as Rakuten are leveraging microservices to separate the user and control planes in their network deployments, thus endowing them with unprecedented flexibility~\cite{rekuten2020deployment}.
Programs like \gls{pawr} by the U.S.\ National Science Foundation~\cite{pawr}, are bringing programmable wireless testing infrastructure at scale to broad communities of researchers---thus creating a fertile ground for software-based open innovation.

The race to the open source and programmable Holy Grail has generated a plethora of heterogeneous software and hardware components and frameworks, whose functionality, scope, and interoperability with other solutions are often obscure and hard to assess.
%
\textit{This article organizes the multiplicity of solutions into the appropriate building blocks of the open source and programmable~5G ecosystem.
We detail how each components fits into a~5G network, highlight the interactions among solutions, and unfold their capabilities and functionalities, highlighting strengths and limitations.
%
Our survey provides the first cohesive and exhaustive recount and taxonomy of open, programmable, and virtualized solutions for 5G networks.}
As most frameworks and devices serve specific purposes in the~5G architectures, we also provide usage directives and how-to guidelines to combine different components into full-fledged open source~5G systems. 

With respect to previous survey efforts~\cite{5gamericas2019, chen2015software} we provide extensive details and commentary on the architecture of softwarized 5G networks, their building blocks, the software frameworks developed so far, and their interactions.

\begin{figure}[ht]
    \centering
    \includegraphics[width=\columnwidth]{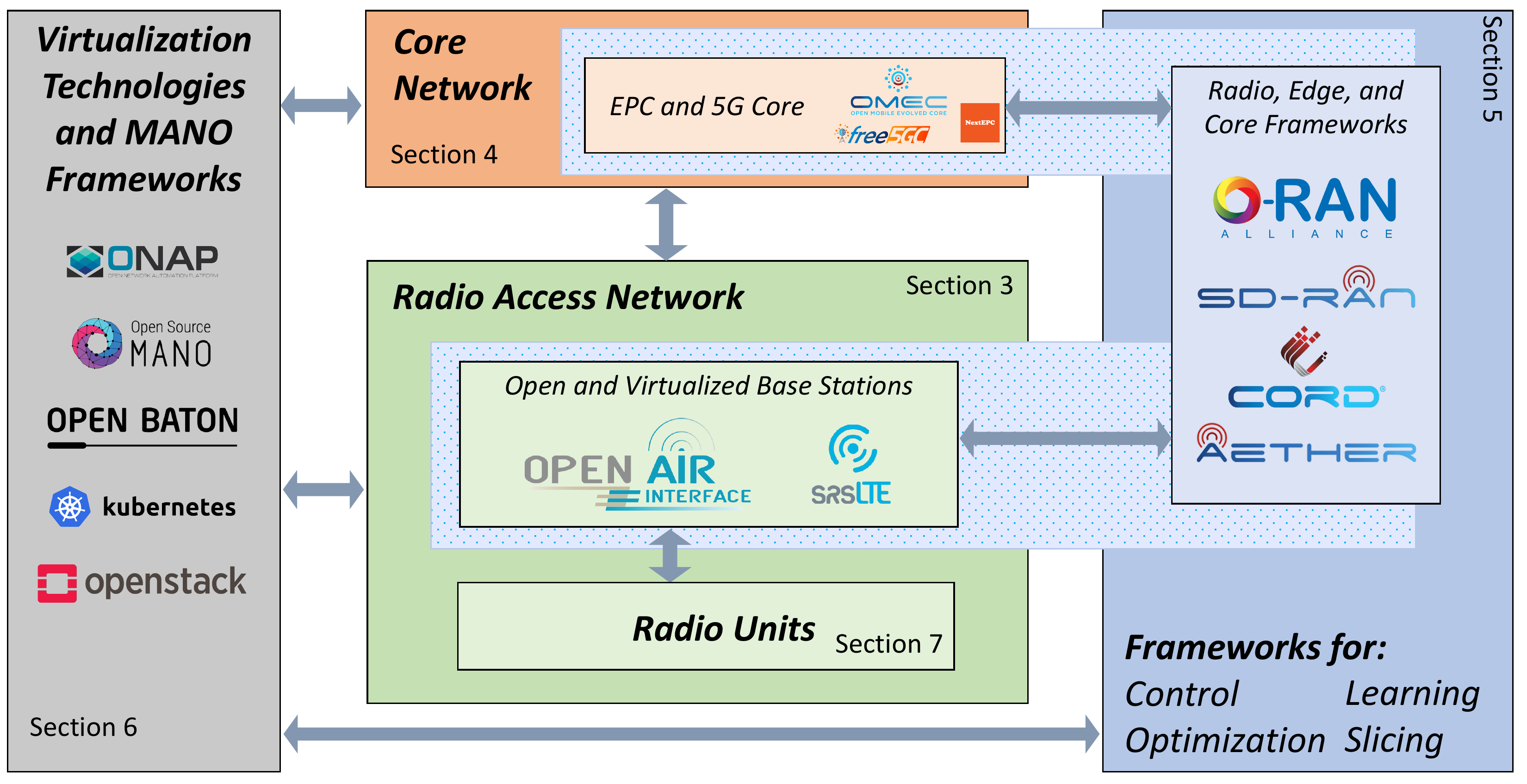}
    \caption{The main building blocks of open source, programmable and virtualized 5G networks with their components and technologies.}
    \label{fig:paperstructure}
\end{figure}

Figure~\ref{fig:paperstructure} provides a visual guide to how the topics surveyed in our work relate to one another, as well as to the structure of the remainder of this article.
%

\sec{sec:architecture} provides a bird's-eye view of the architecture of~5G systems, describing its components and technologies.
Sections~\ref{sec:cloud_ran} and~\ref{sec:core_network} introduce and describe open source solutions for the \gls{ran} and \gls{cn} portions of the infrastructure, respectively. 
General open source frameworks inclusive of both \gls{ran} and \gls{cn} functionalities are discussed in \sec{sec:frameworks}. 
Virtualization and management frameworks are provided in details in \sec{sec:management}. 
\sec{sec:hardware} describes software-defined hardware platforms for open source radio units, highlighting their features and their suitability for~5G applications. 
\sec{sec:testbeds} presents a variety of experimental testbeds allowing instantiation of softwarized 5G networks and testing of new solutions.
Finally, in \sec{sec:limitations} we conclude this article by identifying limitations of the current 5G open source ecosystem and discuss the road ahead, with its unanswered research questions.
A list of acronyms used throughout the article is provided in Appendix~\ref{sec:appendix_acronyms}.

\section{Architectural Enablers of 5G Cellular Networks} \label{sec:architecture}

Mobile networks are transitioning from monolithic architectures, based on dedicated ``\textit{black-box}'' hardware with proprietary firmware and software, to disaggregated deployments based on open source software that runs on generic \gls{sdr} or ``agnostic'' computing devices~\cite{condoluci2018softwarization,afolabi2018networks, thembelihle2017softwarization}. 
This trend is not new to cellular networking, as it has been part of the general discussion around~\gls{4g} cellular networks.
However, while software-based design represents a relatively recent evolution in the context of~\gls{4g} networks, \gls{5g} specifications have foreseen the flexible deployment of agile, softwarized services already in their early stages, with their application to key infrastructure components such as the core, the \gls{ran} and the edge cloud~\cite{38300}. 
This ``flexibility-by-design'' puts~\gls{5g} networks in the privileged position to meet the requirements of heterogeneous traffic classes, mobility and advanced applications through design that is unified, open and dynamically changeable. 

In this section we provide an overview of \gls{4g} and \gls{5g} cellular network architectures, as well as their main components and building blocks (\fig{fig:paperstructure}). 
We start by describing radio access and core network elements and general deployment paradigms. 
We then discuss the architectural and technology enablers such as \acrfull{sdn}, \gls{nfv}, network slicing, \gls{mec}, and intelligent networks. 
Our aim is to provide a reference architecture to map the different open source software libraries and frameworks surveyed in this article to specific network functionalities.

\subsection{Architecture of 4G and 5G Cellular Networks} \label{sec:4g_5g_architecture}

\begin{figure*}[h]
    \centering
    \includegraphics[width=0.8\textwidth]{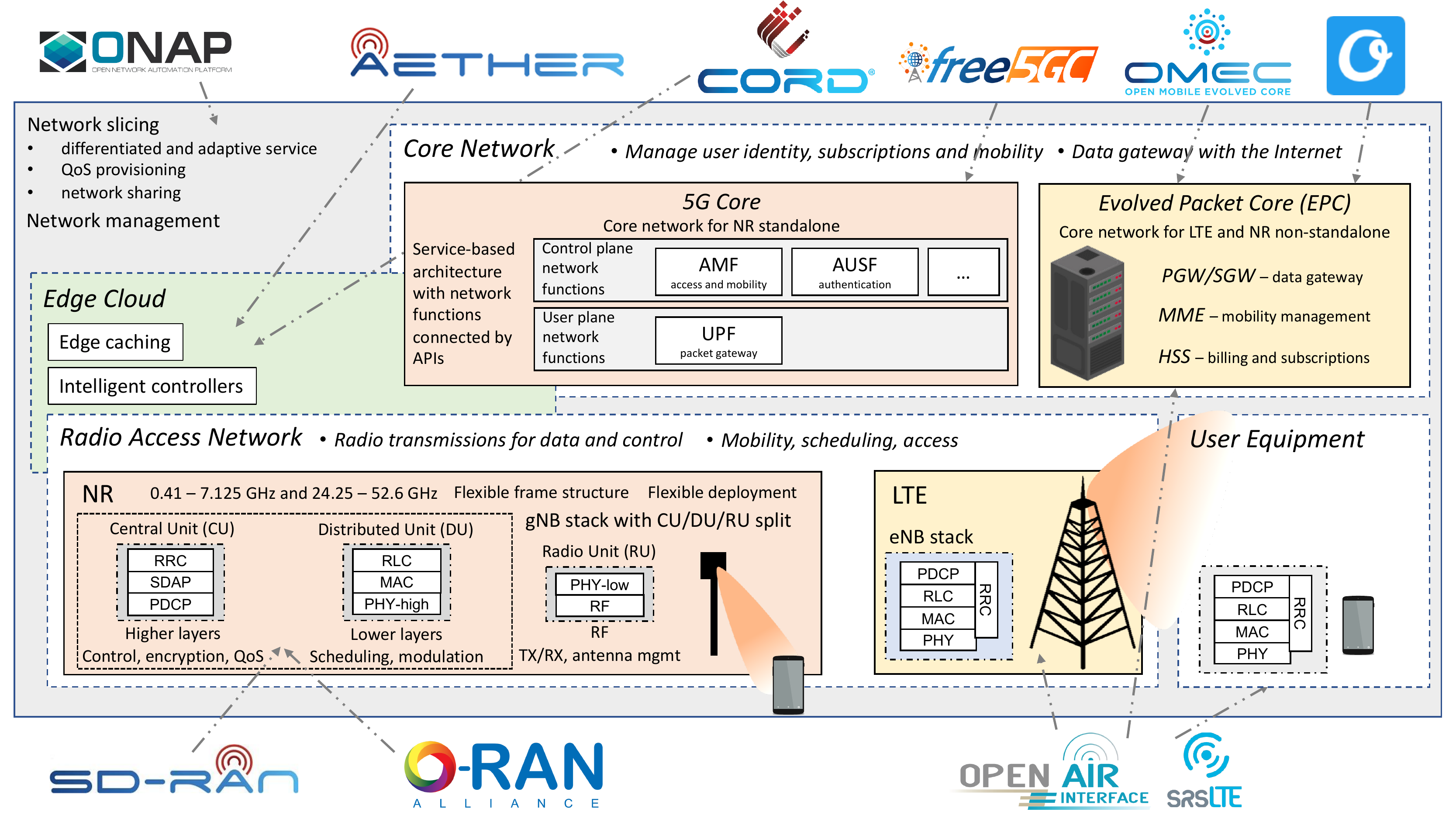}
    \caption{Cellular network architecture.}
    \label{fig:architecture}
\end{figure*}

\fig{fig:architecture} provides a high-level overview of the 4G and 5G cellular architectures, along with some of the open source software frameworks envisioned as their components.

Cellular networks consist of a \acrfull{ran} and a \acrfull{cn}. 
Even though this separation remains unaltered in~4G and~5G deployments, the actual implementation and configuration of these core components differ greatly. 
Particularly, they comply with the \gls{3gpp} \gls{lte} and \gls{nr}\footnote{Although initially introduced as ``New Radio'' in~\cite{38913}, NR has lost its original meaning in the latest \gls{3gpp} specifications~\cite{38300} where it now refers to the 5G RAN.} specifications for the \gls{ran}, and the \gls{epc} and \gls{5gc} for the \gls{cn}, respectively.\footnote{Notice that, while \gls{lte} has been originally associated with 4G networks, its evolution (e.g., LTE-A) will be part of the air interface of 5G networks, together with \gls{nr}~\cite{3gpp2017itu}.}
\fig{fig:architecture} highlights the differences, in terms of flexibility, between the deployments of 4G (in the yellow boxes) and 5G (in the orange boxes) networks. 
For the \gls{cn}, the 4G \gls{epc} has multiple components that have been traditionally executed on dedicated hardware, and only recently have transitioned to software-based deployments. 
The \gls{5gc}, instead, has been designed according to a service-based approach from the get-go. 
The \gls{epc} servers are split into multiple virtual network functions providing specific functionalities.
They are connected to each other through open and standardized interfaces. 
A similar separation principle has been considered for the \gls{5g} \gls{ran}, now designed to provide a functional split among heterogeneous parts of the base stations (\eg control, computing and radio units), with different layers of the protocol stack instantiated on different elements located in different parts of the network.

\paragraph*{\textbf{\acrshort{lte} and \acrshort{epc}}}

The \gls{lte} \gls{ran} is composed of \glspl{enb}, \ie the \gls{lte} base stations, which provide wireless connectivity to the mobile \glspl{ue}. 
The \glspl{enb} are generally deployed as a single piece of equipment on dedicated hardware components, and are networked together and to the core network. 
\gls{lte} operates on a frame structure with 10 subframes of $1\:\mathrm{ms}$ per frame, and 12 to 14 OFDM symbols for each subframe. 
The maximum carrier bandwidth is $20\:\mathrm{MHz}$.
Up to~$5$ carriers can be aggregated for a total of~$100\:\mathrm{MHz}$~\cite{3gpp.36.300}. 

The \gls{lte} protocol stack for the user plane (also known as \gls{eutran}, bottom right corner of the \gls{ran} box in \fig{fig:architecture}) consists of:

\begin{itemize}
    \item The \gls{pdcp} layer, which implements security functionalities (\eg ciphering of packets), performs header compression, and takes care of the end-to-end packet delivery between the \gls{enb} and the \gls{ue}~\cite{3gpp.36.323}.
    
    \item The \gls{rlc} layer, which provides data link layer services (\eg error correction, packet fragmentation and reconstruction).
    It supports three different configurations: 
    The \gls{tm}, to simply relay packets between the \gls{mac} and \gls{pdcp} layers; 
    the \gls{um}, for buffering, segmentation, concatenation and reordering, and the \gls{am}, for retransmitting packets via a ACK/NACK feedback loop~\cite{3gpp.36.322}.
    
    \item The \gls{mac} layer, which performs scheduling, interacts with \gls{rlc} to signal transmissions, forwards the transport blocks to the physical layer, and performs retransmissions via \gls{harq}~\cite{3gpp.36.321}.
    
    \item The \gls{phy} layer, which takes care of channel coding, modulates the signal, and performs transmissions in an OFDM-based frame structure~\cite{3gpp.36.201}.
    
\end{itemize}

These layers also perform control plane functionalities, which concern measurement collection and channel quality estimation. Additionally, the \gls{rrc} layer manages the life cycle of the \gls{enb} to \gls{ue} connection, and it is a point of contact with the core network for control functionalities. 

The main components of the \gls{epc} (in the top right corner of \fig{fig:architecture}) are: (i)~The \gls{pgw} and \gls{sgw}, which are packet gateways to and from the Internet; (ii)~the \gls{mme}, which handles handovers and the \gls{ue} connection life cycle from the core network point of view, and (iii)~the \gls{hss}, which manages subscriptions and billing~\cite{3gpp.23.002}.

\paragraph*{\textbf{NR}}

The \gls{3gpp} \gls{nr} \gls{ran} represents quite the evolution of the 4G \gls{lte}, especially in terms of protocol stack, functionalities and capabilities.
First, it supports a wider range of carrier frequencies, which include part of the \gls{mmwave} spectrum~\cite{rangan2017potentials}. 
Second, the frame structure, while still OFDM-based, is more flexible, with a variable number of symbols per subframe, the option to use much larger bandwidths than \gls{lte} (up to $400\:\mathrm{MHz}$ per carrier), and the integration of signals and procedures to manage directional transmissions at \glspl{mmwave}~\cite{giordani2018tutorial}. 
Third, the 5G \gls{ran} can be connected either to the 4G \gls{epc} (\textit{non-standalone configuration}) or to the new \gls{5gc} (\textit{standalone configuration}). 
Finally, the \gls{nr} base stations (\glspl{gnb}) allows distributed deployment, with different parts of the protocol stack in different hardware components.

The \gls{nr} protocol stack (bottom left corner of \fig{fig:architecture}) features a new layer on top of the \gls{pdcp}, \ie the \gls{sdap} layer~\cite{3gpp.37.324}, which manages the \gls{qos} of end-to-end flows, and maps them to local resources in the \gls{gnb}-\gls{ue} link.
The design of the remaining layers has been updated to support the aforementioned \gls{nr} features~\cite{3gpp.38.201,3gpp.38.321,3gpp.38.322,3gpp.38.323,3gpp.38.331}.

\paragraph*{\textbf{CU/DU Split and the Virtualized RAN  Architecture}}

The main innovation introduced by \gls{nr} comes from the possibility of splitting the higher layers of the \gls{3gpp} stack (\gls{pdcp}, \gls{sdap}, and \gls{rrc}) and the lower layers (\gls{rlc}, \gls{mac}, and \gls{phy}) into two different logical units, called \textit{\gls{cu}} and the \textit{\gls{du}}, which can be deployed at separate locations. 
Moreover, the lower part of the physical layer can be separated from the \gls{du} in a standalone \textit{\gls{ru}}. 
The \gls{cu}, \gls{du} and \gls{ru} are connected through well-defined interfaces operating at different data rates and latency (with tighter constraints between the \gls{du} and \gls{ru}).

This architecture, proposed by \gls{3gpp} in~\cite{3gpp.38.816}, enables the \gls{vran} paradigm. Specifically, the antenna elements (in the \gls{ru}) are separated from the baseband and signal processing units (in the \gls{du} and \gls{cu}), which are hosted on generic, even multi-vendor, hardware. 
If the interfaces between the different \gls{ran} components are open, the 5G deployment follows the Open \gls{ran} model, which defines open and standardized interfaces among the elements of the disaggregated \gls{ran}~\cite{5gwp}.
A notable example of Open \gls{ran} is currently being promoted by the \gls{oran} Alliance~\cite{oranwp}. 
This consortium has defined a set of interfaces between \gls{cu}, \gls{du}, \gls{ru}, and a \gls{ric} that can be deployed at the edge of the network (see also \sec{sec:oran}).

\paragraph*{\textbf{The 5G Core}}

Openness and flexibility have guides the design of the \gls{5gc}, now realized according to a service-based approach~\cite{3gpp.23.501}. 
Control and user plane core functionalities have been split into multiple network functions~\cite{3gpp.23.742}. 
The \gls{3gpp} has also defined interfaces and \glspl{api} among the network functions, which can be instantiated on the fly, enabling elastic network deployments and network slicing (Section \ref{sec:slicing}). 
The \gls{upf} is a user plane gateway to the public Internet that acts as mobility anchor and \gls{qos} classifier for the incoming flows. 
On the control plane side, most of the \gls{mme} functions (\eg mobility management) are assigned to the \gls{amf}.
The \gls{smf} allocates IP addresses to the \glspl{ue}, and orchestrates user plane services, including the selection of which \gls{upf} a \gls{ue} should use. 
For a detailed overview of all~5G core functions the reader is referred to~\cite{3gpp.23.501,kim2017sa}.



\subsection{Enabling technologies for softwarized 5G cellular networks}

5G networks will embody heterogeneous network components and technologies to provide unprecedented performance levels and a unique experience to subscribers. 
Managing the integration of such a menagerie of technologies, controlling such variegate infrastructure and orchestrating network services and functionalities is clearly no trivial feat. 
%
To solve this management and control problem, 5G networks have borrowed widespread and well-established processes and architectures from the cloud-computing ecosystem, where softwarization and virtualization are merged together to abstract services and functionalities from the hardware where they are executed. In the following, we introduce two of these technologies and how they integrate with future 5G systems.

\paragraph*{\textbf{Softwarization and \acrlong{sdn}}}

\begin{figure}[ht]
    \centering
    \includegraphics[width=0.8\columnwidth]{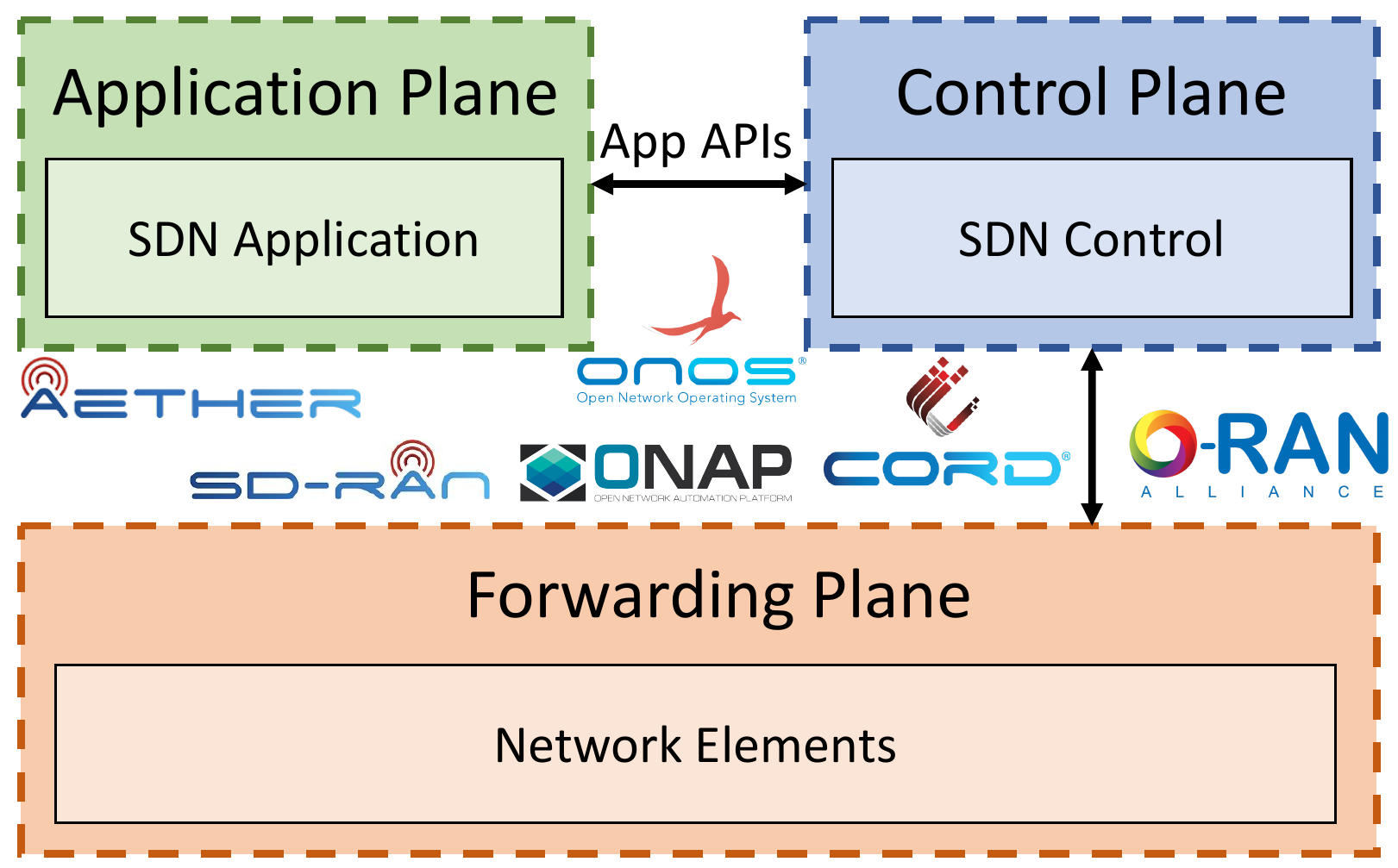}
    \caption{High-level overview of the \gls{sdn} architecture.}
    \label{fig:sdn_high_level}
\end{figure}

In order to integrate hardware components produced by multiple vendors with different functionalities and configuration parameters, 5G systems rely on \emph{softwarization}. 
This technology concept grew in popularity in the second decade of the 21st century thanks to the \acrfull{sdn} architectural paradigm and the widespread adoption of the now well-established OpenFlow protocol. 
As shown in \fig{fig:sdn_high_level}, \gls{sdn} leverages softwarization to decouple network control from the forwarding (or data) plane, thus separating routing and control procedures from specialized hardware-based forwarding operations.
By decoupling the functions of these two planes, network control dynamics can be directly programmed in software with an abstract view of the physical infrastructure.
Then, a centralized network controller runs the network intelligence, retains a global view of the network, and makes decisions on policies regarding automated network optimization and management, among others. 

The fundamental principle of \gls{sdn}, namely, the separation of data and control, has been adopted by 5G networks to detach \gls{ran} and edge hardware components from their networking and service capabilities. 
In fact, 5G systems takes softwarization to a broader and comprehensive application range, where it is leveraged to effectively put into practice \gls{ran} disaggregation, where \glspl{ru} operate as basic transceivers and all control and processing operations are performed in software through open interfaces and \glspl{api}.
This is witnessed by the plethora of open source \gls{sdn} solutions for mobile networks, also shown in \fig{fig:sdn_high_level}, which include \gls{onos}~\cite{berde2014onos}, \gls{cord}~\cite{peterson2016central}, O-RAN~\cite{oranwp}, \gls{onap}~\cite{onap_architecture}, Aether~\cite{aetherwp}, and SD-RAN~\cite{sdranwp}.

\paragraph*{\textbf{\acrlong{nfv}}}

\begin{figure}[ht]
    \centering
    \includegraphics[width=0.9\columnwidth]{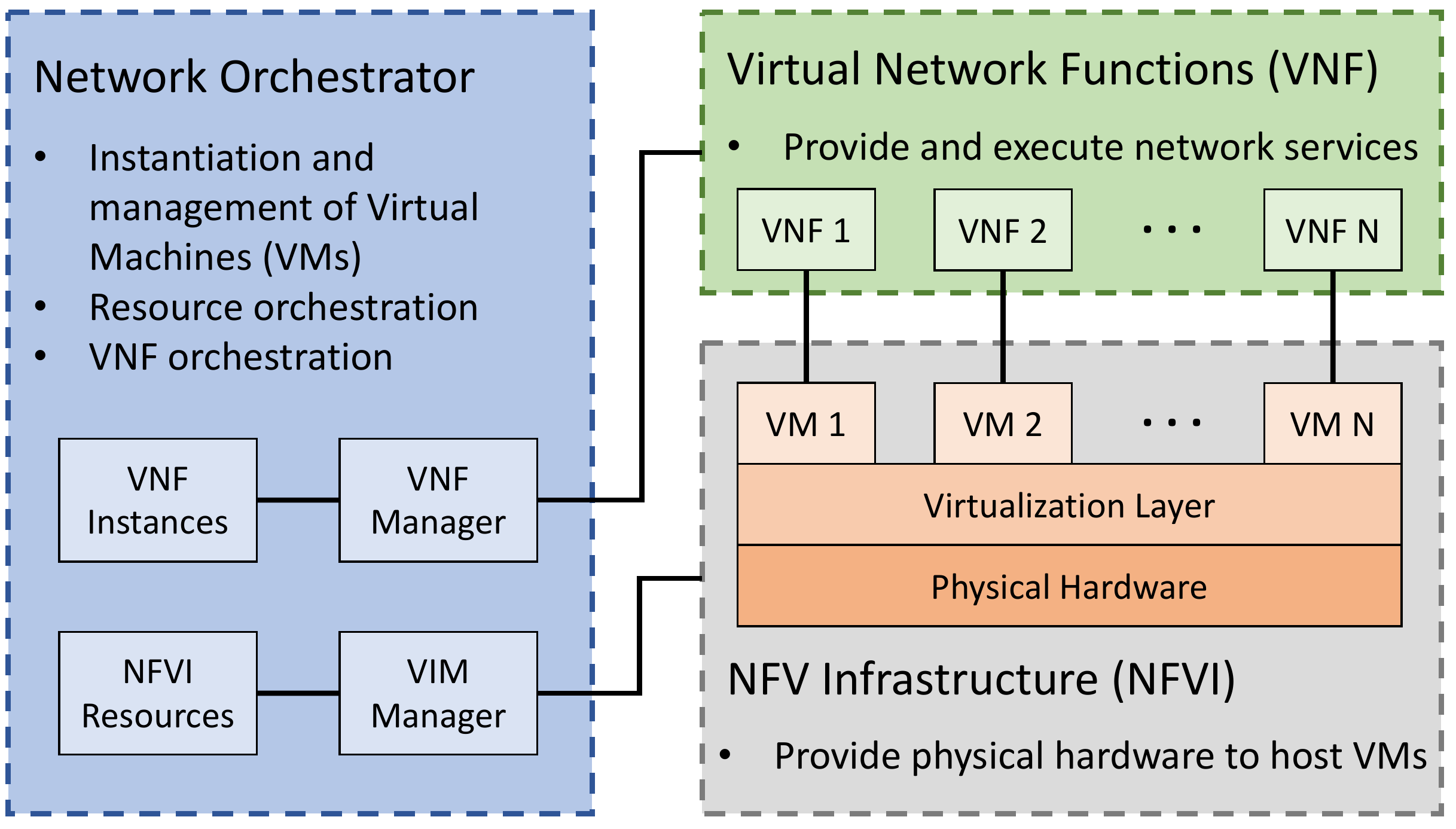}
    \caption{High-level overview of the \gls{nfv} architecture.}
    \label{fig:nfv_high_level}
\end{figure}

\gls{nfv} brings scalable and flexible management and orchestration to softwarized networks. 
This is achieved by virtualizing network services and functionalities and decoupling them from the hardware where they are executed. 
Each functionality is implemented in software via \glspl{vnf}, which are executed on \glspl{vm} instantiated on general-purpose hardware. 
One of the main advantages of \gls{nfv} is that each \gls{vnf} provides atomic functionalities. Therefore, multiple \glspl{vnf} can be combined together to create more complex and customized network services.
\fig{fig:nfv_high_level} depicts the main components of the \gls{nfv} architecture.

These are: (i) The network orchestrator, which instantiates and manages the \glspl{vm} on the physical infrastructure and the services they run; (ii) the \glspl{vnf}, which are executed on the \glspl{vm} and implement the network services, and (iii) the \gls{nfvi}, which consists of the general purpose physical hardware hosting the \glspl{vm} deployed by the network orchestrator.

An example of open source network virtualization project is \gls{opnfv}, which facilitates the adoption and development of a common \gls{nfvi}~\cite{opnfv}.
\gls{opnfv} also provides testing tools, compliance and verification programs to accelerate the transition of enterprise and service provider networks to the \gls{nfv} paradigm.

\subsection{RAN and Core Network Slicing} \label{sec:slicing}




The whole~5G network design is rooted in softwarization, virtualization and sharing principles. 
This strategic design choice paved the way toward a new generation of more efficient, dynamic and profitable networks. 
Such a revolution has also been made possible by the concepts of \textit{network slicing}.

Network slicing is a multi-tenancy virtualization technique where network functionalities are abstracted from hardware and software components, and are provided to the so-called \textit{tenants} as \textit{slices}~\cite{doro2020slicing}.
The physical infrastructure (e.g., base stations, optical cables, processing units, routers, etc.) is shared across multiple tenants, each of which may receive one or more slices. 
Each slice is allocated a specific amount of physical resources and operates as an independent virtual network instantiated on top of the physical one. 
Although tenants have full control over their slices and the resources therein, they cannot interact with other slices, a concept known as \textit{slice isolation} or \textit{orthogonality}~\cite{doroInfocom2019Slicing}. 

\begin{figure}[ht]
	\centering
	\includegraphics[width=\columnwidth]{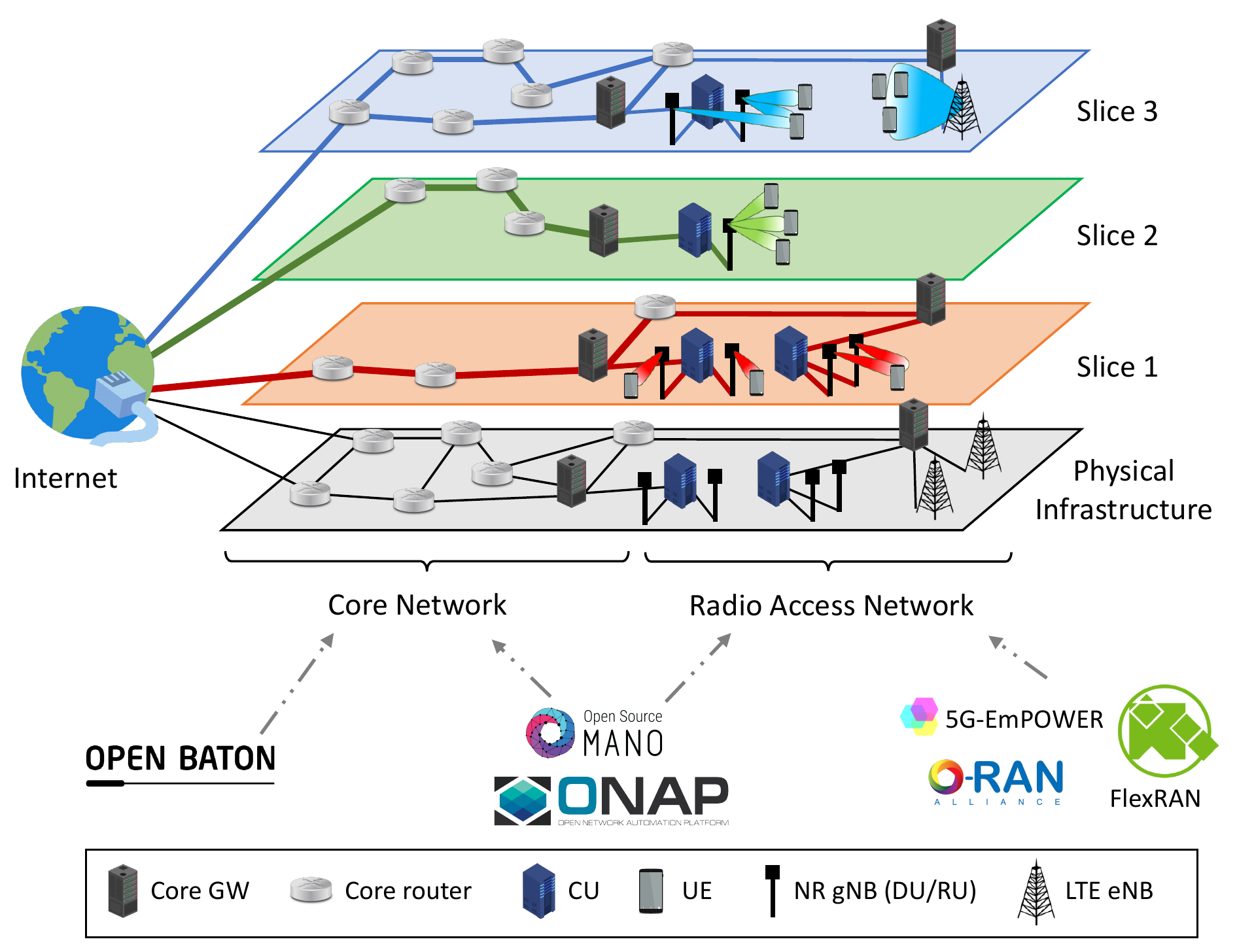}
	\caption{An example of RAN and \gls{cn} slicing.}
	\label{fig:slicing}
\end{figure}

Each slice provides specific functionalities covering the RAN and the core portions of the network. For example, tenants can be granted RAN slices instantiated on selected base stations providing \gls{caas} (\eg for private cellular networking) to mobile users~\cite{doro2019tnet}. 
They can also instantiate network slices dedicated to specific services, users and applications. 
Such a flexible approach makes it possible to instantiate slices dedicated to resource-hungry applications, such as virtual and augmented reality, while simultaneously controlling another slice carrying low-priority traffic generated by browsing activities. 
An example of practical interest is shown in \fig{fig:slicing}, depicting how slicing technologies enable infrastructure sharing and support the instantiation of multiple slices embedding different infrastructure components. 

The figure also lists relevant and well-established open source software projects for effective instantiation, control and configuration of network slices in different portions of the infrastructure (see also Sections~\ref{sec:frameworks} and~\ref{sec:management}).

The benefits of network slicing include: (i) Each slice can be reserved to handle specific traffic classes with diverse security requirements and is allocated with a different amount of resources, thus enabling \textit{service differentiation} at the infrastructure level; (ii) slicing is controlled by software components, which enable real-time and on-demand instantiation, reconfiguration, and revocation of network slices to adapt to time-varying traffic demand and/or fulfill \glspl{sla}, and (iii) underutilized resources can be leased to \glspl{mvno} in the form of network slices, thus maximizing resource utilization and generating new profit opportunities for infrastructure providers~\cite{rost2017network}. 

In this context, it is worth mentioning how \glspl{oss} and \glspl{bss} will play a vital role in the 5G ecosystem and will determine the success of network slicing. Network slices must be orchestrated, instantiated and revoked dynamically, must satisfy \gls{sla} agreements and should be robust against failures and outages. \glspl{oss} will serve as a tool to guarantee the fulfillment of services by facilitating closed-loop control and management of network slices. 
At the same time, operators and infrastructure owners providing slicing services to verticals must generate diversified offers with slices dedicated to services that reflect verticals' needs. \glspl{bss} will be necessary to control such a diversified environment and to implement dynamic billing and pricing mechanisms for each slice. Examples of initiatives confirming this trend are Open APIs and Open Digital Architecture led by the industry association TM Forum~\cite{tmf_website}, \gls{onap}~\cite{onap_5g} and the open source project OpenSlice~\cite{openslice}.
Since these benefits affect both business and performance aspects of 5G networks, slicing has become pivotal to 5G systems. 
In this context, the open source community has led to the development of a variety of solutions to integrate slicing algorithms into the 5G ecosystem~\cite{BARAKABITZE2020106984}. 

Section~\ref{sec:frameworks} will survey the most relevant and well-established open source projects enabling the delivery and handling of network slicing technologies for network \gls{ran} and core.

\subsection{Multi-access Edge Computing}
\label{sec:mec}

5G systems will leverage advanced and high-performance signal processing and transmission techniques for the highest data rates and QoS possible. 
However, these technologies alone are not enough to meet the stringent throughput and latency requirements of many 5G applications. 
For instance, tactile applications as those for virtual and augmented reality, rely upon near real-time processing and interaction with the environment. 
To be effective these technologies require sub-millisecond transmission times~\cite{xiang2019reducing}. 
Furthermore, they involve a significant amount of computation (e.g., augmented reality processors hinge on GPU acceleration), which would result in excessive delay and poor user experience if performed in data centers from a distant cloud.
With network technologies of the past meeting the strict constraints of these practical applications was almost considered an utopian task. 
5G networks, on the other hand, leverage a simple, yet effective, approach to network architecture design based on softwarization principles to bring services and functionalities to the edge of the network~\cite{mao2017survey}.

In this context, \acrfull{mec} (the technology formerly known as Mobile Edge Computing) has been identified as the solution of the needed \textit{edgification} process. 
\gls{mec} introduces an innovative design shift where essential components of the architecture (both hardware and software) are moved closer to users. 
By building on edge computing, content caching, \gls{nfv} and \gls{sdn}, \gls{mec} provides an effective solution to the latency and throughput demands of 5G applications~\cite{giust2018multi}.
\gls{mec} (i) moves content and functionalities to the edge, meaning that data only sporadically needs to traverse the \gls{cn}, thus resulting in low latency and in an offloaded core, and (ii) enables localized service provisioning such as private cellular networking, \gls{iot} data collection and processing at the edge for health and environmental monitoring, and augmented reality for education, telesurgery and advanced industrial applications~\cite{kropp2019demonstration}. There have been several proposals on how to enable \gls{mec} in 5G networks. Solutions include those from \gls{etsi}, which defines the term \gls{mec} in~\cite{etsi2019mec}, and from the \gls{3gpp}, which has introduced open interfaces (in Releases~15 and~16) to integrate \gls{mec} apps with the softwarized 5G core~\cite{etsi20193gpp}. The interested readers are referred to~\cite{taleb2017multi,bruschi2019mobile} for analysis and reviews of different \gls{mec} architectures. Open source \gls{mec} frameworks and their enablers will be discussed in Section~\ref{sec:frameworks}.

\subsection{Intelligence in the Network}
\label{sec:intelligence}

Another key component of the 5G ecosystem is the application of machine learning and artificial intelligence-based technologies to network optimization~\cite{jiang2017machine}. 
The scale of 5G deployments makes traditional optimization and manual configuration of the network impossible. 
Therefore, automated, data-driven solutions are fundamental for self-organizing 5G networks. 
Additionally, the heterogeneity of use cases calls for a tight integration of the learning process to the communication stack, which is needed to swiftly adapt to quickly changing scenarios.

\begin{table*}[ht]
\centering
\caption{Open source \acrshort{ran} software.}
\label{tab:ran_software}
\small
\setlength{\tabcolsep}{1.75pt}
\footnotesize
\renewcommand{\arraystretch}{1.9}
\begin{tabular}[]{|c|c|c|c|c|c|c|c|}
\hline
\acrshort{ran} Software    & eNB    & gNB    & SDR UE	& \makecell{COTS UE Support}   & License   & \makecell{Main Contributor(s)}  & \makecell{Community Support} \\
\hline
\acrlong{oai}~\cite{oai_website}    & yes   & \makecell{under\\development}   & \makecell{yes\\(unstable)}  & yes   & \makecell{\acrshort{oai} Public\\License v1.1}   & \makecell{\acrshort{oai} Software Alliance, EURECOM} & mailing list \\
\hline
srsLTE~\cite{srslte_website}    & yes   & \makecell{under\\development}   & yes   & yes   & GNU AGPLv3    & \makecell{Software Radio Systems}  & mailing list \\
\hline
Radisys~\cite{radisys_website, oran_website}    & no    & \makecell{yes,\\(O-RAN)}    & no    & N/A   & \makecell{Apache v2.0,\\O-RAN Software\\License v1.0}   & Radisys   & no \\
\hline
\end{tabular}
\end{table*}

Learning techniques for 5G networks have been proposed for different applications. 
Use cases range from forecasting traffic demands to scale \gls{cn} resources~\cite{alawe2018improving}, to predicting \gls{harq} feedback~\cite{strodthoff2019enhanced} to reduce latency in \gls{urllc} flows, to beam adaptation in \gls{mmwave} vehicular networks~\cite{asadi2018fast}.
A summary of results from applying deep learning techniques can be found in~\cite{zhang2019deep, luong2019applications}. 

Notably, telecom operators have embraced the deployment of machine learning techniques for self-managed and self-optimized networks. 
The integration of machine learning in real deployments, however, faces several architectural and procedural challenges~\cite{polese2018machine}.
This is primarily because real-time network telemetry and data need to be collected and aggregated to allow the data intensive learning operations of training and inference. 
The previously discussed \gls{mec} paradigm has been proposed as an architectural enabler for applying machine learning to networking, with intelligent controllers deployed at the edge of the network and integrated to the \gls{ran}. 
A software-based framework that implements this paradigm is \gls{oran}~\cite{oranwp}, which envisions a \acrfull{ric} interfaced with \glspl{gnb} and \glspl{enb}, for monitoring, learning, and performing closed-loop actuation.
We discuss the \gls{oran} architecture in detail in \sec{sec:oran}.

\section{The \acrlong{ran}}
\label{sec:cloud_ran}

This section describes the open source libraries and frameworks for~4G and~5G cellular networks to deploy a software-defined \gls{ran}.
The most relevant of these open source software frameworks and their features are listed in Table~\ref{tab:ran_software}.

\subsection{\acrlong{oai}}
\label{sec:oairan}

The \gls{oairan}~\cite{kaltenberger2020openairinterface, nikaein2014openairinterface} provides software-based implementations of \gls{lte} base stations (\enbs), \ues and \gls{epc} (\acrshort{oaicn}; see \sec{sec:core_network}) compliant with \gls{lte} Release~8.6 (with an additional subset of features from \gls{lte} Release~10).

The \gls{oairan} source code is written in C to guarantee real-time performance, and is distributed under the OAI Public License~\cite{oairan_license}, a modified version of the Apache License v2.0 that allows patent-owning individuals and companies to contribute to the \gls{oai} source code while keeping their patent rights.
Both the \enb and \ue implementations are compatible with Intel~x86 architectures running the Ubuntu Linux operating system. (An experimental version for the CentOS~7 is under development.)
Several kernel- and BIOS-level modifications are required for these implementations to achieve real-time performance, including installing a low-latency kernel, and disabling power management and CPU frequency scaling functionalities.

\paragraph*{\textbf{\acrshort{enb} Implementation}}

At the physical layer, the \enb can operate in \gls{fdd} and \gls{tdd} configurations with $5$, $10$, and $20\:\mathrm{MHz}$ channel bandwidths, corresponding to $25$, $50$, and $100$ \glspl{prb}.
As for the transmission modes, it supports \gls{siso}, transmit diversity, closed-loop spatial multiplexing, \gls{mumimo}, and $2\times2$ \gls{mimo}. Channel quality reports are sent through standard-compliant \glspl{cqi} and \glspl{pmi}. Finally, \gls{oairan} also supports \gls{harq} at the \gls{mac} layer. 

In \gls{dl}, \gls{oairan} implements synchronization signals used by \ues to acquire symbol and frequency synchronization (\gls{pss} and \gls{sss}), and channels that carry information on the \gls{dl} configuration used by the \enb (\gls{pbch}) and on the \gls{dl} control channel (\gls{pcfich}).
The \gls{oairan} \enb also implements the \gls{pdcch}, which carries scheduling assignments of the \ues and \gls{dl} control information, and the \gls{pdsch}, which transports data intended for specific \ues.
Finally, ACKs/NACKs for the data received in uplink from the \ues are sent through the \gls{phich}, while broadcast and multicast services are provided through the \gls{pmch}.

In \gls{ul}, it supports the \gls{prach}, which is used by \ues to request an \gls{ul} allocation to the base station, as well as channels carrying reference signals from the \ue to the \enb (\gls{srs} and \gls{drs}).
Data from the \ues to the \enb is carried by the \gls{pusch}, while the \gls{pucch} is used to transmit \gls{ul} control information.
Modulations up to 64~\gls{qam} and 16~\gls{qam} are supported in \gls{dl} and \gls{ul}, respectively.

The \gls{eutran} stack of the \enb implements the \gls{mac}, \gls{rlc}, \gls{pdcp}, and \gls{rrc} layers
and provides interfaces to the core network with support for IPv4 and IPv6 connectivity (see \sec{sec:4g_5g_architecture} for a detailed description of these layers).
As for the \gls{mac} layer scheduling, \gls{oairan} implements a channel-aware proportional fairness algorithm commonly used in commercial cellular networks, as well as greedy and fair round-robin scheduling algorithms.

The \gls{enb} can be interfaced with both commercial and open source \glspl{epc} (\eg \acrshort{oaicn} and Open5GS; \sec{sec:core_network}), and with a number of \glspl{sdr}, including Ettus Research~\cite{ettus} B-series \glspl{usrp}, \eg \gls{usrp} B210, and X-series, \eg \gls{usrp} X310 (see \sec{sec:hardware} for a comprehensive description of compatible hardware platforms).
However, to the best of our knowledge, at the time of this writing the \enb application executed over \glspl{usrp} X-series appears to be less than fully stable.
Both \glspl{cots} smartphones and \glspl{sdr} can be used as \ues.
However, \gls{oai} privileged the development of the \enb application rather than the \ue one, which may result in connectivity issues between the two.

\gls{oairan} also includes a simulation environment implementing layer-2 and layer-3 functionalities only, without the need to interface with any external \gls{sdr} device. Being transparent to layer-1 procedures, the simulation environment provides a useful tool to evaluate the performance of algorithms and protocols at the upper-layers.  
Finally, \gls{nr}-compliant applications for base stations (\gnb), \ues and core network (\gls{5gc}) are currently being developed~\cite{kaltenberger2019openairinterface, kaltenberger2020openairinterface}.
At the time of this writing, a major \gls{nr} release has not been announced yet.

\paragraph*{\textbf{Sample Use Cases}}

\acrlong{oai} has witnessed recent widespread adoption by both academia and industry.
For instance, Kaltenberger et al.\ leveraged \gls{oai} to build a Cloud-\gls{ran} Massive \gls{mimo} testbed with \glspl{rru} built from commodity hardware~\cite{kaltenberger2017massive}.
Foukas et al.\ proposed Orion~\cite{foukas2017orion} and FlexRAN~\cite{foukas2016flexran}, two \gls{ran}-oriented centralized network virtualization solutions based on \gls{oai}.
Liu et al.\ implemented a learning-assisted network slicing solution for cyber-physical systems on \gls{oai}~\cite{liu2019learning}, while D'Alterio et al.\ leveraged \gls{oai} to prototype and experimentally evaluate the performance of a \gls{uav}-based \enb~\cite{dalterio2019quality}.

Fujitsu is integrating and testing \gls{oai} in commercial units of its proprietary infrastructure~\cite{fujitsu_oai}, while WindyCitySDR is leveraging \gls{oai} to create low-bandwidth mobile phone data networks~\cite{windycitysdr}.
InterDigital and SYRTEM are developing \gls{mmwave} software solutions and devices based on the \gls{oai} implementation~\cite{interdigital_oai, syrtem}.

\begin{figure}[ht]
    \centering
    \includegraphics[width=0.8\columnwidth]{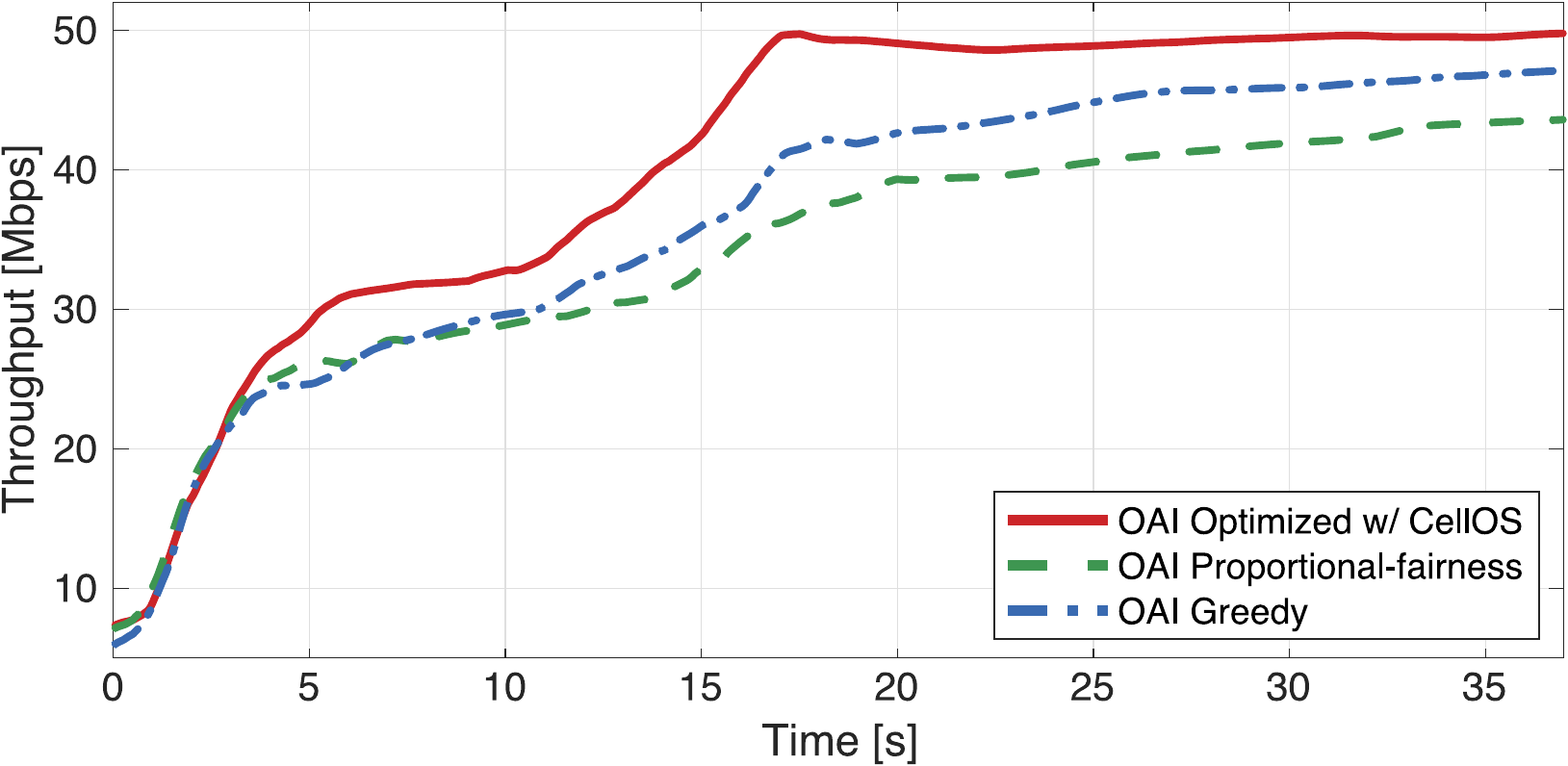}
    \caption{Software-defined optimization with \acrlong{oai} and CellOS~\cite{bonati2020cellos}.}
    \label{fig:oai_scheduling}
\end{figure}

The full potential of an open and softwarized approach to cellular networking is demonstrated by Bonati et al.~\cite{bonati2020cellos}.
Specifically, a softwarized automatic optimization framework with \gls{ric} functionalities, called CellOS, is instantiated on a network with \enbs featuring an enriched version of \gls{oai}.
In the experimental setting~3 \enbs serve a total of~9~\ues (\gls{cots} smartphones).
\fig{fig:oai_scheduling} compares the throughput achieved by the CellOS-driven automatic user scheduling optimization to that achieved by the \gls{oairan} proportional-fairness and greedy scheduling algorithms.

The CellOS optimized approach increases the network throughput significantly, and reduces the convergence time to stable high throughput with respect to the other schedulers. 
This simple yet effective experiment shows the importance of gaining access and reconfiguring via software network parameters and protocols (the scheduling algorithm, in this case). 
Without open and programmable software such as OAI, it would have been unfeasible to change scheduling policies baked into hardware components and improve network performance swiftly and automatically.

\subsection{srsLTE}
\label{sec:srslteran}

Similarly to \gls{oairan}, srsLTE~\cite{gomez2016srslte, puschmann2017implementing} provides software implementations of \gls{lte} \enb, \ue, and \gls{epc} (discussed in \sec{sec:core_network}) compliant with LTE Release~10 (with some features from higher versions, e.g., \gls{nr} Release~15).
The software suite is written in the C and C++ programming languages and it is distributed under the GNU AGPLv3 license~\cite{gnuapgl3}.
srsLTE is compatible with the Ubuntu and Fedora Linux distributions.
It does not require any kernel- or BIOS-level modifications to achieve real-time performance (unlike \gls{oai}; \sec{sec:oairan}). (Disabling CPU frequency scaling is recommended.)

\paragraph*{\textbf{\acrshort{enb} Implementation}}

At the physical layer, the \enb implementation supports \gls{fdd} configurations with channel bandwidths of $1.4$, $3$, $5$, $10$, $15$, and $20\:\mathrm{MHz}$, corresponding to configurations from~$6$ to~$100$ \glspl{prb}.
The available transmission modes are single antenna, transmit diversity, \gls{cdd}, and closed-loop spatial multiplexing.
The channels supported in \gls{dl} and \gls{ul} are the same of \gls{oai} (\sec{sec:oairan}), with modulations up to $256$~\gls{qam}.

Similar to \gls{oai}, the \gls{eutran} stack of srsLTE    \enb implements the \gls{mac}, \gls{rlc} (\gls{tm}, \gls{am}, and \gls{um} modes are supported), \gls{pdcp}, and \gls{rrc} layers (\sec{sec:4g_5g_architecture}).
The \enb interfaces with the \gls{cn} through the \gls{s1ap} and \glspl{gtp} interfaces, and supports IPv4 connectivity.
It can be used to serve both \gls{cots} and \gls{sdr} \ues, which can be implemented through srsLTE \ue application.

\paragraph*{\textbf{\acrshort{ue} Implementation}}

The \ue implementation features PHY, \gls{mac}, \gls{rlc}, \gls{pdcp}, and \gls{rrc} layers as the ones of the \enb.
Additionally, \srsue also includes a \gls{nas} layer that manages control plane communication between \ue and \gls{cn}, and a \gls{gw} layer. The latter supports IPv4 and IPv6 connectivity, and is used to create a virtual interface on the machine that runs the user application to tunnel IP packets from/to the RF front-end.

To authenticate users and \gls{cn}, the \ue application supports both soft and hard \gls{usim} cards.
These are meant to contain values to uniquely identify the \ue in the network, such as the \gls{imsi}, the authentication key (K), the operator code (OP), and the phone number.
The soft \gls{usim} can work with both the XOR~\cite{3gpp.33.105} and Milenage~\cite{3gpp.35.206} authentication algorithms, and the previously mentioned \ue authentication values are stored in a configuration file.
The hard \gls{usim}, instead, requires a physical SIM card that needs to be programmed with the user parameters discussed above through a smart card reader/programmer.
This SIM card, then, needs to be connected to the host computer that runs the \ue application, for instance, through the same smart card reader/programmer.

\smallskip
As RF front-end, both \enb and \ue applications are compatible with several of the boards that will be described in \sec{sec:hardware}, including \gls{usrp} B- and X-series (\ie \gls{usrp} B210, B205mini-i, and X310), as well as limeSDR~\cite{limesdr}, and bladeRF~\cite{bladerf}.
To analyze some of the \enb and \ue capabilities in a controlled environment, srsLTE provides utilities to simulate dynamics such as uncorrelated fading channels, propagation delays, and Radio-Link failures between \enbs and \ues.
%
Finally, at the time of this writing, srsLTE is working toward \gls{nr} compatibility.
An initial support of \gls{nr} at the \gls{mac}, \gls{rlc}, \gls{rrc}, and \gls{pdcp} layers is included in the latest releases of the code.

\paragraph*{\textbf{Sample Use Cases}}

Several recent works have been using srsLTE to investigate the security of \gls{lte} networks.
Bui and Widmer proposed OWL, an srsLTE-based framework to capture and decode control channel of \gls{lte} devices~\cite{bui2016owl}.
OWL is then leveraged by Meneghello et al.\ in~\cite{meneghello2020smartphone}, and Trinh et al.\ in~\cite{trinh2019classification} to fingerprint \gls{lte} devices through machine-learning approaches.
Kim et al.\ designed LTEFuzz, a tool for semi-automated testing of the security of \gls{lte} control plane procedures~\cite{kim2019touching}, while Rupprecht et al.\
carried out a security analysis of \gls{lte} layer 2~\cite{rupprecht2019breaking}.
Yang et al.\ proposed and evaluated SigOver, an injection attack that performs signal overshadowing of the \gls{lte} broadcast channel~\cite{yang2019hiding}.
Singla et al.\ designed an enhanced paging protocol for cellular networks robust to privacy and security attacks~\cite{singla2020protecting}. 
The National Institute of Standards and Technology (NIST) built OpenFirst on top of srsLTE, a platform for first responders to test and validate \gls{lte} technologies focused on public safety communications~\cite{nist_openfirst}, while Ferranti et al. experimentally evaluated the performance of a \gls{uav}-based \enb leveraging srsLTE~\cite{ferranti2020skycell}.
\begin{figure}[ht]
    \begin{center}
        \subcaptionbox{\label{fig:srslte_slice0}}{\includegraphics[width=0.7\columnwidth]{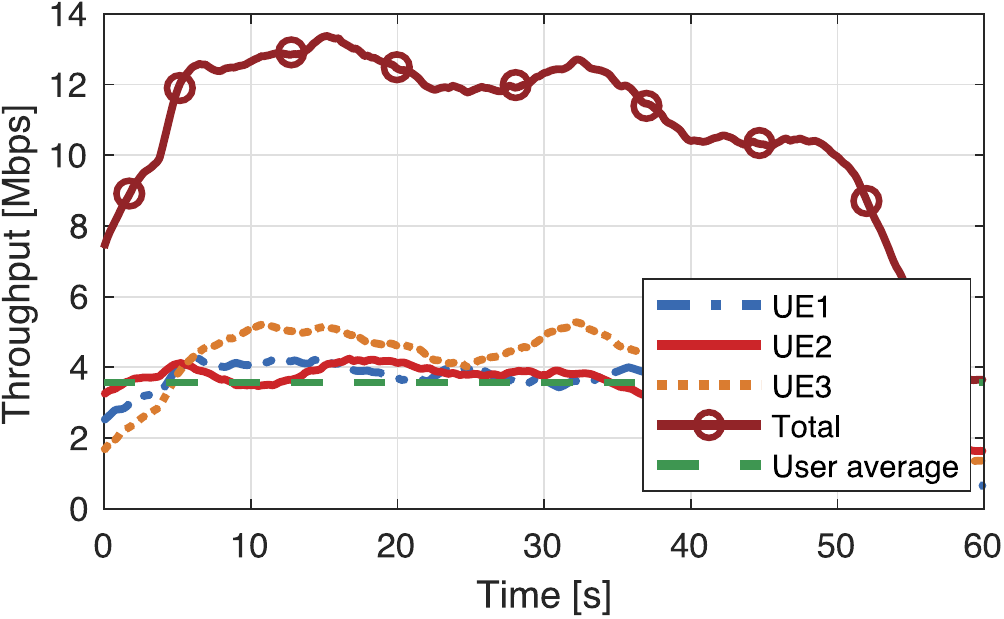}}\\\vspace{0.25cm}
        \subcaptionbox{\label{fig:srslte_slice1}}{\includegraphics[width=0.7\columnwidth]{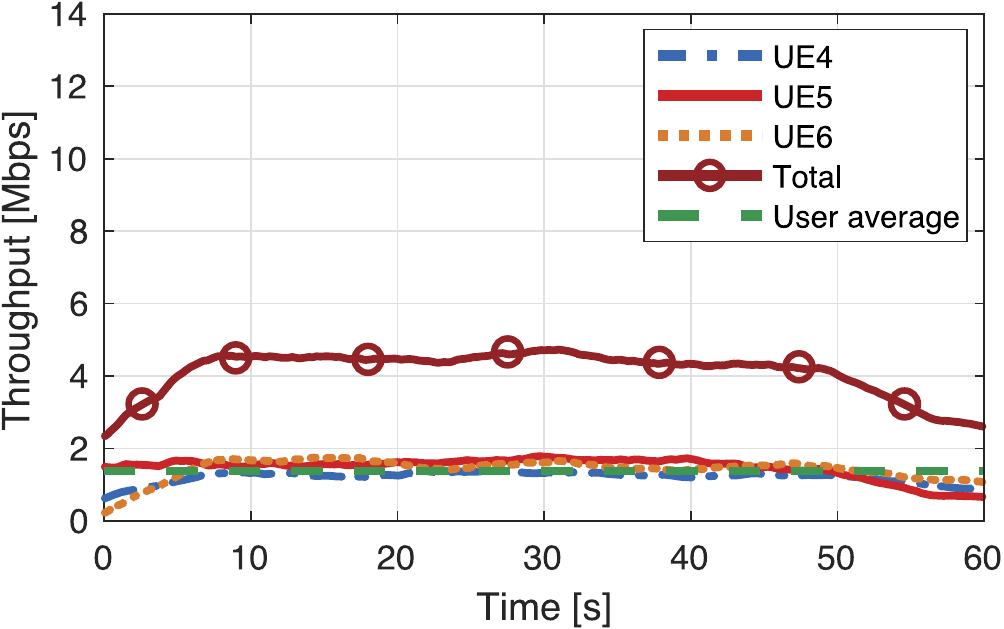}}
    \end{center}
    \caption{Softwarized per-slice \gls{qos} differentiation on srsLTE using Sl-EDGE~\cite{doro2020sledge} and CellOS~\cite{bonati2020cellos}: (a) Slice~1, premium service, and (b) slice~2, regular service.}
    \label{fig:srslte_slicing}
\end{figure}
Finally, D'Oro et al.\ proposed Sl-EDGE, an optimization-based \gls{mec} slicing framework to instantiate slice services on heterogeneous devices at the edge of the network~\cite{doro2020sledge}.

%
%
%
%
%
%
One use case of particular interest is that of RAN slicing. The authors of~\cite{bonati2020cellos} designed CellOS, a \gls{ric} which controls resources of the \enbs of the network by interfacing with different \gls{ran} software, such as \gls{oai} and srsLTE. In this case, the source code of srsLTE is extended to achieve differentiated service through 5G slicing technologies. \fig{fig:srslte_slicing}, shows experimental results in which 2~\enbs running srsLTE are serving 6~\gls{cots} \ues. First, the resource allocation for two slices, serving premium and regular users of the cellular network, is computed through Sl-EDGE~\cite{doro2020sledge}, then this is applied by CellOS~\cite{bonati2020cellos}. The network \glspl{enb} allocate 80\% of spectrum resources to slice~1 (premium service in \fig{fig:srslte_slice0}), and the remaining 20\% to slice~2 (regular service in \fig{fig:srslte_slice1}). As expected, the throughput of the premium \ues of the slice~1 outperforms that of the regular users of slice~2, with average gains of more than $2.5$x. 

\subsection{Radisys Open Source RAN Contributions}
\label{sec:radisys}

Radisys is a 4G/5G vendor that contributes to a number of open source software consortia, including O-RAN and several \gls{onf} initiatives~\cite{radisys_website}. 

As part of O-RAN (\sec{sec:oran}), Radisys provides an open source implementation of the \gls{3gpp} \gls{nr} stack for the \gls{gnb} \gls{du}~\cite{radisys_du}. At the time of this writing, this does not represent a complete solution that can be deployed to run real-world experiments (as with \gls{oai} and srsLTE), as it lacks integration with open source \gls{cu} and \gls{ru} implementations. However, this represents a key first step toward the availability of an open source 5G \gls{gnb} based on the \gls{cu}/\gls{du} split principle described in \sec{sec:4g_5g_architecture}.

The currently available open source code, licensed according to the Apache License v2.0, provides a complete implementation of the \gls{mac} and \gls{rlc} layers.
%
%
The Radisys release also provides a layer that manages the operations of the \gls{du} and interfaces it with the \gls{cu}, the \gls{ru} and external controllers, when available. The codebase is aligned with Release 15 of the \gls{3gpp} \gls{nr} specifications. The \gls{nr} \gls{mac} uses the \gls{fapi} to interact with a scheduler, adapted from an LTE implementation. The \gls{rlc} layer supports the \gls{tm}, \gls{um} and \gls{am} modes (see \sec{sec:4g_5g_architecture} for details on these modes of operation).
Additionally, Radisys has open sourced a full implementation for 4G \glspl{enb}, licensed with AGPLv3~\cite{radisys_4g}. However, this implementation concerns the firmware of a specific Qualcomm chipset FSM9955, thus representing a solution for 4G small cell hardware vendors rather than an alternative to srsLTE and \gls{oai}.

\section{Core Network}
\label{sec:core_network}

In this section, we describe the main open source solutions for the 4G and 5G core networks, \ie \gls{epc} and 5G Core, respectively.
A summary of the solutions discussed in this section is shown in Table~\ref{tab:core_software}.

\begin{table*}[ht]
\centering
\caption{Open source \acrshort{cn} software.}
\label{tab:core_software}
\small
\setlength{\tabcolsep}{1.75pt}
\footnotesize
\renewcommand{\arraystretch}{1.9}
\begin{tabular}[]{|c|c|c|c|c|c|}
\hline
\acrshort{cn} Software    & EPC   & 5G Core   & License   & \makecell{Main Contributor}  & \makecell{Community Support} \\
\hline
\acrlong{oai}~\cite{oai_website}    & yes   & \makecell{under development}   & Apache v2.0   & \makecell{\acrlong{oai} Software Alliance, EURECOM} & mailing list \\
\hline
srsLTE~\cite{srslte_website}    & yes   & no   & GNU AGPLv3    & \makecell{Software Radio Systems}    & mailing list \\
\hline
Open5GS~\cite{open5gs_website}  & yes   & \makecell{under development}    & GNU AGPLv3    & Open5GS   & \makecell{mailing list / forum} \\
\hline
OMEC~\cite{omec_website}    & yes   & compatible    & Apache v2.0   & \makecell{\acrshort{onf}, Intel, Deutsche Telekom, Sprint, AT\&T}   & mailing list \\
\hline
free5GC~\cite{free5gc_website}  & no    & yes   & Apache v2.0   & free5GC   & forum \\
\hline
\end{tabular}
\end{table*}

\subsection{\acrlong{epc}}
\label{sec:epc_softwares}

Implementations of the 4G \gls{epc}, discussed in details in \sec{sec:4g_5g_architecture}, typically include components for the \acrfull{mme}, the \acrfull{hss}, the \acrfull{sgw}, and the \acrfull{pgw}.

The \gls{mme} is responsible for control messages to establish connection with the \ues, paging and mobility procedures.
It includes the \gls{nas} signaling and security features, as well as tracking area list management, \gls{pgw}/\gls{sgw} selection, \ue authentication, and reachability procedures.
It also takes care of bearer management, \ie a tunnel between \ue and \gls{pgw} in the case of \gls{epc}, and between \ue and \gls{upf} in the case of \gls{5gc}~\cite{3gpp.23.401, 3gpp.23.501}.
Moreover, it supports protocols for control plane signaling between \gls{epc} and \gls{eutran}, reliable message-level transport service.
Tunneling protocols for \gls{udp} control messaging are also provided, as well as protocols for authentication, authorization and charging of \ues.
%

The \gls{hss} implements the user database, and stores information on the subscribers, \eg identity and key.
It is also responsible for user authentication.
It provides interfaces for user provisioning in the \gls{hss} database, as well as interfaces to connect to the \gls{mme}.

The \gls{sgw} and \gls{pgw} components carry packets through the \gls{gtp} for both user and control planes, \ie through the \gls{gtpu} and \gls{gtpc}, which use \gls{udp} as transport protocol.
Packet routing and forwarding, IP address allocation to \ues, and paging are also supported.
Open source implementations of the \gls{lte} \gls{epc} are provided by \acrlong{oai} (with \gls{oaicn}), srsLTE (with srsEPC), Open5GS, and \gls{omec}.

\begin{table}[ht]
\centering
\caption{Implemented \acrshort{epc} interfaces.}
\label{tab:epc_interfaces}
\setlength{\tabcolsep}{1.5pt}
\footnotesize
\renewcommand{\arraystretch}{1.5}
\begin{tabular}[]{|c|c|c|c|c|}
\hline
Interface    & \acrlong{oai}~\cite{oai_website} & srsLTE~\cite{srslte_website} & Open5GS~\cite{open5gs_website} & OMEC~\cite{omec_website} \\
\hline
\multicolumn{5}{|c|}{\acrshort{mme} / \acrshort{hss}} \\
\hline
S1-MME  & x & x & x & x \\
\hline
S6a     & x & x & x & x \\
\hline
S10     & x & - & - & - \\
\hline
S11     & x & x & x & x \\
\hline
\multicolumn{5}{|c|}{\acrshort{sgw} / \acrshort{pgw}} \\
\hline
S1-U    & x & x & x & x \\
\hline
S5 / S8 & - & x & x & x \\
\hline
S11     & x & x & x & x \\
\hline
SGi     & x & x & x & x \\
\hline
Gx      & - & - & x & x \\
\hline
\end{tabular}
\end{table}

A summary of the most relevant \gls{3gpp} interfaces implemented by each of these \gls{epc} softwares is shown in Table~\ref{tab:epc_interfaces}:
\begin{itemize}
    \item \textit{S1-MME:} It enables the flow of the S1-AP control application protocol between \gls{eutran} and \gls{mme}.
    \item S6a: It connects \gls{mme} and \gls{hss}, it is used for user authentication and authorization and to transfer user subscriptions.
    \item \textit{S10:} Control interface among different \glspl{mme}.
    \item \textit{S11:} Control plane interface between \gls{mme} and \gls{sgw} used to manage the \gls{eps}.
    \item \textit{S1-U:} It enables the flow of user plane data between \gls{eutran} and \gls{sgw}.
    \item \textit{S5/S8:} It provides user plane tunneling management, and control services between \gls{sgw} and \gls{pgw}.
    \item \textit{SGi:} It connects the \gls{pgw} to the Internet.
    \item \textit{Gx:} It allows the transfer of \gls{qos} policies and charging rules from the \gls{pcrf} to the \gls{pcef} in the \gls{pgw}.
\end{itemize}

\paragraph*{\textbf{\gls{oaicn}}} \gls{oaicn} is written in the C and C++ programming languages, and it is distributed under the Apache License v2.0~\cite{kaltenberger2020openairinterface}.
It is compatible with Intel~x86 architectures running the Ubuntu Linux operating system.
Kernel modifications similar to those discussed in \sec{sec:oairan} for \gls{oairan} are required to guarantee real-time capabilities.
Dynamic \gls{qos} with establishment, modification and removal of multiple dedicated bearers, and policy-based \gls{qos} update are also features implemented by the \gls{oaicn} \gls{mme}.
\gls{tft} operations, such as fault detection, filter rules, and IP-filters are also provided.
Finally, implicit (\eg service request failures) and explicit (\eg bearer resource and delete commands) congestion indicators are supported, along with multi-\gls{apn}, paging, and restoration procedures.

\paragraph*{\textbf{srsEPC}}
The \gls{epc} implementation included in the srsLTE software suite, namely, srsEPC, is written in C++ and distributed under the GNU AGPLv3 license~\cite{gomez2016srslte}.
It is compatible with the Ubuntu and Fedora Linux operating systems.
The \gls{hss} supports the configuration of \ue-related parameters in the form of a simple textual csv file.
\ue authentication is supported by XOR and Milenage authentication algorithms.
srsEPC enables per-user \gls{qci} and dynamic or static IP configurations.

%
%

\paragraph*{\textbf{Open5GS}}
This \gls{epc} is written in C and distributed under the AGPLv3 license~\cite{open5gs_website}.\footnote{Open5GS was previously known as NextEPC~\cite{nextepc_website}. The renaming happened in 2019.}
It is compatible with a variety of Linux distributions, such as Debian, Ubuntu, Fedora, and CentOS, as well as FreeBSD and macOS.
Differently from other \glspl{epc}, Open5GS supports the delivery of voice calls and text messages through the \gls{lte} network instead of relying on traditional circuit switching networks. This is achieved by leveraging \gls{volte} and SG-SMS solutions, respectively.
Moreover, Open5GS includes a \gls{pcrf} module, through which operators can specify network policies in real-time, including prioritizing a certain type of traffic. The implementation of \gls{5gc} functionalities is currently under development.

\paragraph*{\textbf{OMEC}}
This is a high performance open source implementation of \gls{lte} Release~13 \gls{epc} developed by the \gls{onf} together with telecom operators and industry partners, such as Intel, Deutsche Telekom, Sprint, and AT\&T~\cite{omec_website}.
\gls{omec} is built using a \gls{nfv} architecture to sustain scalability in large-scale scenarios such as those of 5G and \gls{iot} applications.
It is distributed under the Apache License v2.0, and offers connectivity, billing and charging features.
\gls{omec} can be used as a standalone \gls{epc}, or integrated in larger frameworks, such as \gls{comac} (see \sec{sec:other_frameworks}).

\paragraph*{\textbf{Sample Use Cases}}

Sevilla et al.\ developed CoLTE, a Community \gls{lte} project to bring cellular connectivity in rural areas that are not covered by traditional cellular service~\cite{sevilla2019experiences}.
CoLTE interfaces commercial \glspl{enb} (BaiCellsNova-233) with \gls{oaicn}, modified to include features such as user billing and accounting.
CoLTE is currently porting its implementation from \gls{oaicn} to Open5GS.
Core network functions for 5G \gls{nr}, instead, are being developed by bcom starting from the \gls{oaicn} implementation~\cite{bcom}.
Moreover, \gls{oaicn} is used as \gls{epc} inside the Magma framework (see \sec{sec:other_frameworks}).
%
%
Haavisto et al.\ use NextEPC (now Open5GS) to deploy a 5G open source \gls{ran}~\cite{haavisto2019opensource}; Lee et al.\ interface it with srsLTE to study security concerns of modern cellular networks~\cite{lee2019this}.

\subsection{5G Core}

An open source implementation of the 5G core is offered by free5GC and distributed under the Apache License v2.0~\cite{free5gc_website}.
It is written in the Go programming language, and it is compatible with machines running the Ubuntu Linux operating system.
This implementation, which was initially based on NextEPC (now Open5GS), supports the management of user access, mobility, and sessions (\gls{amf} and \gls{smf}), and the discovery of the services offered by other network functions (\gls{nrf}).
It also includes network functions to select which network slices to allocate to \ues (\gls{nssf}), to manage, store and retrieve user data (\gls{udm} and \gls{udr}), to perform \ues authentication within the network (\gls{ausf}).
Functions for the operation, administration and management of the core network (\gls{oam}), and to perform network orchestration, among others, are also included.

The \gls{3gpp} interfaces implemented by free5GC are:
\begin{itemize}
\item \textit{N1/N2:} Connect the \gls{amf} to the \gls{ue} and \gls{ran}, respectively. They are used for session and mobility management.
\item \textit{N3/N4/N6:} Connect the \gls{upf} to the \gls{ran}, \gls{smf}, and data network, respectively. They support user plane functions.
\item \textit{N8:} Connects the \gls{udm} and the \gls{amf}. It enables user authorization procedures.
\item \textit{N10/N11:} Connect the \gls{smf} to the \gls{udm} and \gls{amf}, respectively. They handle subscription and session management requests.
\item \textit{N12/N13:} Connect the \gls{ausf} to the \gls{amf} and \gls{udm}, respectively. They enable authentication services.
\end{itemize}

In addition to the \glspl{epc} that are evolving toward the 5G architecture (\sec{sec:epc_softwares}), an open source implementation of the \gls{5gc} is currently being developed by Hewlett Packard Enterprise (HPE)~\cite{hpe20205g}.

Software implementations of the data plane of the \gls{5gc} can also benefit from the acceleration introduced by domain-specific, platform-independent packet-processing languages such as P4~\cite{bosshart2014p4}, which is a protocol-independent programming language that instructs the networking hardware on how to process packets. Several P4-based \gls{upf} implementations have been proposed commercially and in the literature. Most of their software is not open source~\cite{RICARTSANCHEZ201880,ricart2019p4,kaloomUpf}.

\section{RAN and Core Frameworks}
\label{sec:frameworks}

This section describes several open source frameworks that operate both in the \gls{ran} and \gls{cn} domains. While the software described in Sections~\ref{sec:cloud_ran} and~\ref{sec:core_network} performs specific functions (e.g., \gls{enb}, \gls{ue}, or \gls{cn}), the frameworks that will be introduced in the following paragraphs are more general and broad in scope, and interact with individual components in the \gls{ran} and \gls{cn} for control, management, and coordination.

Table~\ref{tab:frameworks} compares the features, license and availability of the different frameworks and projects that will be presented throughout this section.

\begin{table*}[t]
\centering
\caption{Open frameworks and projects.}
\label{tab:frameworks}
\setlength{\tabcolsep}{1.5pt}
\footnotesize
\renewcommand{\arraystretch}{1.5}
\begin{tabular}[]{|c|c|c|c|c|c|}
\hline
Framework	& Main Focus	& Status	& License	& Main Members	& \makecell{Community\\Support} \\
\hline

\multicolumn{6}{|c|}{Mobile}\\
\hline
O-RAN~\cite{oran_website}   & \makecell{Virtualized, intelligent RAN}  & available & \makecell{Apache v2.0,\\O-RAN Software\\License v1.0}    & \makecell{O-RAN Alliance\\w/ telecom operators}    & no \\
\hline
COMAC~\cite{comac_website}   & \makecell{Agile service delivery at the edge}    & available &  Apache v2.0  & ONF   & mailing list \\
\hline
SD-RAN~\cite{sdran_website}  & \makecell{CU/DU control and user planes}    & \multicolumn{2}{c|}{under development}   & ONF   & N/A \\
\hline
Aether~\cite{aether_website}  & \makecell{5G/LTE, Edge-Cloud-as-a-Service (ECaaS)}   & \multicolumn{2}{c|}{under development}   & ONF   & N/A \\
\hline
Magma~\cite{magma_website}  & \makecell{\acrshort{cn} Orchestration}    & available & BSD   & Facebook  & \makecell{mailing list /\\forum} \\
\hline
OpenRAN~\cite{openran5gnr_website}  & \makecell{Programmable, disaggregated RAN w/ open interfaces}    & \multicolumn{2}{c|}{closed source} & \acrshort{tip} & no \\
\hline
\acrlong{rec}~\cite{akrainorec_website}  & \makecell{O-RAN RIC automated configuration /\\integration testing blueprint}  & available & Apache v2.0 & Akraino & no \\
\hline
Aerial~\cite{aerialsdk}  & \makecell{SDK for GPU-accelerated 5G \acrshort{vran}}    & early access & proprietary  & NVIDIA    & N/A \\
\hline

\multicolumn{6}{|c|}{Slicing}\\
\hline
5G-EmPOWER~\cite{5gempower_website}  & \makecell{Centralized controlled for heterogeneous RAN}   & available & Apache v2.0 & \makecell{FBK (in the framework\\of multiple EU projects)}    & no \\
\hline
FlexRAN~\cite{flexran_website}  & \makecell{Real-time controller for software-defined RAN}   & available & MIT License & \makecell{Mosaic5G Consortium}    & mailing list \\
\hline

\multicolumn{6}{|c|}{Edge}\\
\hline
CORD~\cite{cord_website}    & Data center for network edge  & available & Apache v2.0   & \makecell{ONF, AT\&T,\\Google, Telefonica}    & mailing list \\
\hline
LL-MEC~\cite{llmec_website}  & \makecell{Low-latency MEC and network slicing}    & available & Apache v2.0   & Mosaic5G Consortium  & \makecell{mailing list} \\
\hline
LightEdge~\cite{lightedge_website}  & \makecell{MEC services}    & available & Apache v2.0   & \makecell{FBK (in the framework\\of multiple EU projects)}  & \makecell{N/A}\\
\hline
\end{tabular}
\end{table*}

\subsection{O-RAN}
\label{sec:oran}

The \gls{oran} Alliance is an industry consortium that promotes the definition of an open standard for the \gls{vran}, with two goals~\cite{oranwp}. The first is the integration of machine learning and artificial intelligence techniques in the \gls{ran}, thanks to intelligent controllers deployed at the edge~\cite{polese2018machine}. The second is the definition of an agile and open architecture, enabled by well-defined interfaces between the different elements of the \gls{ran}. 
Since all \gls{oran} components must expose the same \glspl{api}, it is easy to substitute components with others offering alternative implementations of the same functionalities. This allows O-RAN-based~5G deployments to integrate elements from multiple vendors, thus opening the \gls{ran} market to third-party entities providing new functionalities and diversified services. Moreover, it makes it possible to adopt \gls{cots} hardware, in an effort to promote flexibility and reduce costs. Eventually, following the trend started with cloud-native infrastructures, the \gls{oran} Alliance also aims at promoting open source software as part of the consortium effort.

\paragraph*{\textbf{The O-RAN Architecture}}

\begin{figure*}[t]
	\centering
	\includegraphics[width=0.8\textwidth]{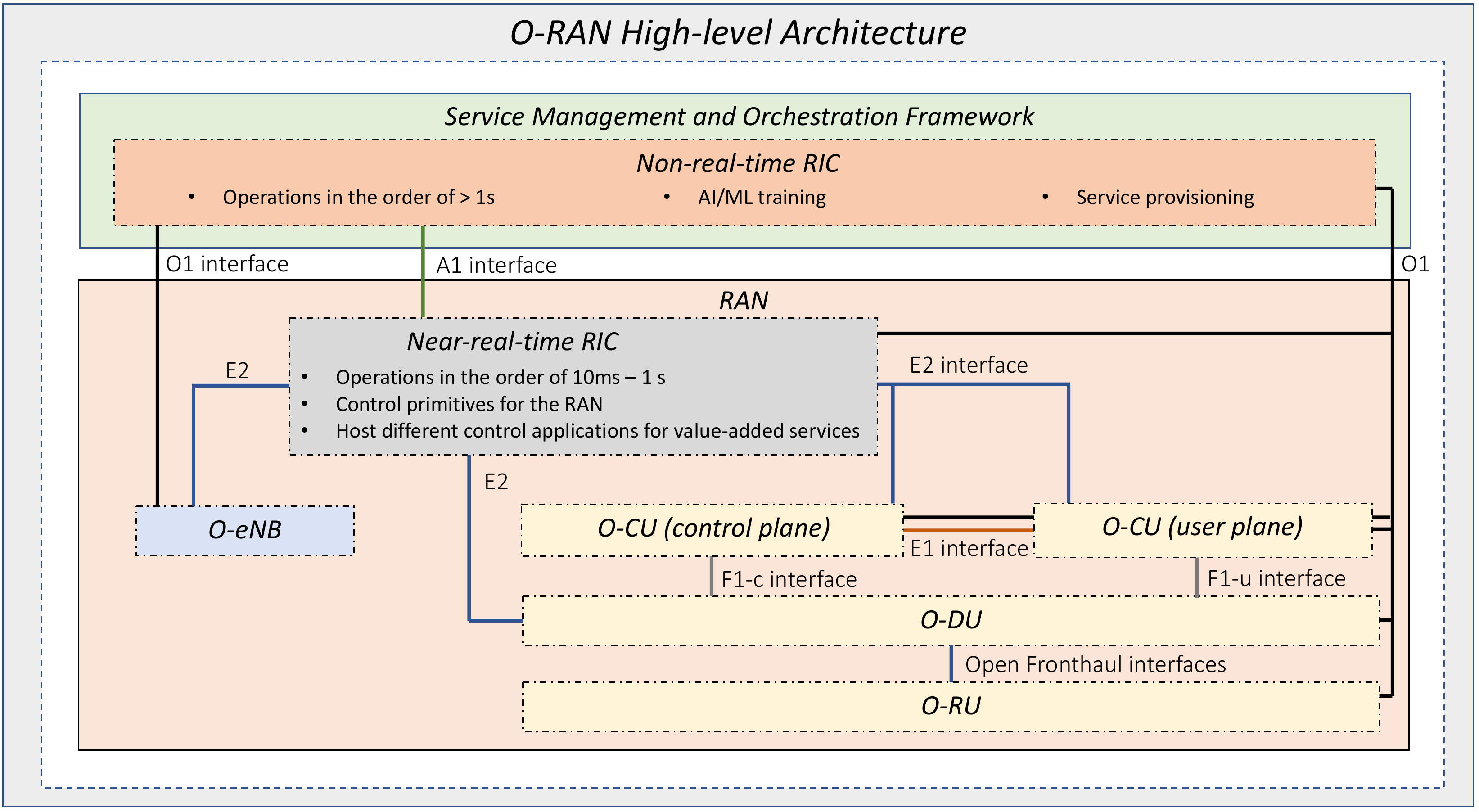}
	\caption{O-RAN high-level blocks and interfaces.}
	\label{fig:oran}
\end{figure*}

\fig{fig:oran} illustrates the high-level architecture and the interfaces of \gls{oran}~\cite{oranwp}.

With respect to \fig{fig:architecture}, \gls{oran} concerns the edge and the~4G and~5G \glspl{ran}. It is composed by a non-real-time and a near-real-time \acrfull{ric}, and by the \glspl{enb} and \glspl{gnb}. 
The service management and orchestration framework (top of \fig{fig:oran}) operates a non-real-time \gls{ric}, which performs control decisions with a granularity higher than one second. 
For example, it can provision the different functions of \gls{oran}, and train learning algorithms over data provided by the \gls{ran}, among others.
A near-real-time \gls{ric}, instead, performs a control loop with a much tighter timing requirement (with a decision interval as short as $10\:\mathrm{ms}$), relying on different start, stop, override, or control primitives in the \gls{ran}, e.g., for radio resource management. 
These \glspl{api} can be used by different applications installed on the near-real-time \gls{ric} (named xApps), which can be developed by third-party entities and pulled from a common marketplace. For example, through the near-real-time \gls{ric} and its xApps, an operator can control user mobility processes (e.g., handovers), allocate networking resources according to predicted paths for connected vehicles and \glspl{uav}, perform load balancing and traffic steering, and optimize scheduling policies~\cite{oranwpusecases}. The near-real-time \gls{ric} can also leverage machine learning algorithms trained in the non-real-time \gls{ric}. 
%
The remaining components of the O-RAN architecture concern the \gls{cu}/\gls{du}/\gls{ru} into which 5G \glspl{gnb} are split~\cite{3gpp.38.401}, and the 4G \glspl{enb}~\cite{3gpp.36.300} (bottom of \fig{fig:oran}). 
The \gls{cu} is further split into a control plane \gls{cu} and a user plane \gls{cu}. Among the different options investigated by the \gls{3gpp}, O-RAN has selected split 7-2x for the \gls{du}/\gls{ru} split~\cite{oran-fronthaul}, in which coding, modulation and mapping to resource elements are performed in the \gls{du}, and the inverse FFT, the cyclic prefix addition and digital to analog conversion are carried out in the \gls{ru}. Precoding can be done in either of the two.

\paragraph*{\textbf{O-RAN Interfaces}}

\gls{oran} is in the process of standardizing the interfaces between each of the components in \fig{fig:oran}. The two \glspl{ric} interact using the A1 interface, while the non-real-time \gls{ric} uses the O1 interface to interact with the \glspl{ru} and legacy 4G \glspl{enb}. The A1 interface~\cite{oran-a1} allows the non-real-time \gls{ric} to provide (i) policy-based guidance to the near-real-time \gls{ric}, in case it senses that its actions are not fulfilling the \gls{ran} performance goals; (ii) manage machine learning models, and (iii) provide enrichment information to the near-real-time \gls{ric}, for example from \gls{ran}-external sources, to further refine the \gls{ran} optimization. The O1 interface, instead, has operation and management functions, and strives at being compatible with existing standards to permit a seamless integration with existing management frameworks (\sec{sec:management})~\cite{oran-o1}. For example, it relies on the IETF Network Configuration Protocol (NETCONF)~\cite{RFC6241} and on several \gls{3gpp}-defined \glspl{api}. The non-real-time \gls{ric} uses O1 to (i) provision management, fault supervision, and performance assurance services; (ii) perform traces collection; (iii) start up, register, and update physical equipment, and (iv) manage communication surveillance services.

The near-real-time \gls{ric} exposes the E2 interface~\cite{oran-e2} to multiple elements, i.e., the \gls{cu}, the \gls{du} and the \gls{enb}. This interface only concerns control functionalities, related to the deployment of near-real-time \gls{ric} control actions to the nodes terminating the E2 interface, and to the management of the interaction of the \gls{ric} and these nodes~\cite{oran-arch-spec}. 

The E1 and F1 interfaces comply to the specifications from \gls{3gpp}. The~E1 interface runs between the control and user plane \glspl{cu}, and its main functions concern trace collection for specific \glspl{ue}, and bearer setup and management~\cite{3gpp.38.460}. The~F1 interface operates between the \glspl{cu} and \glspl{du}~\cite{3gpp.38.470}. It has two different versions, one for the control plane and one for the user plane. F1 transports signaling and data between \glspl{cu} and \glspl{du}, to carry out \gls{rrc} procedures and \gls{pdcp}-\gls{rlc} packet exchange. Finally, the interface toward the \gls{ru} is developed by the Open Fronthaul initiative inside \gls{oran}~\cite{oran-fronthaul}. This interface carries compressed IQ samples for the data plane, and control messages for beamforming, synchronization, and other physical layer procedures.

\begin{figure}[ht]
    \centering
    \includegraphics[width=\columnwidth]{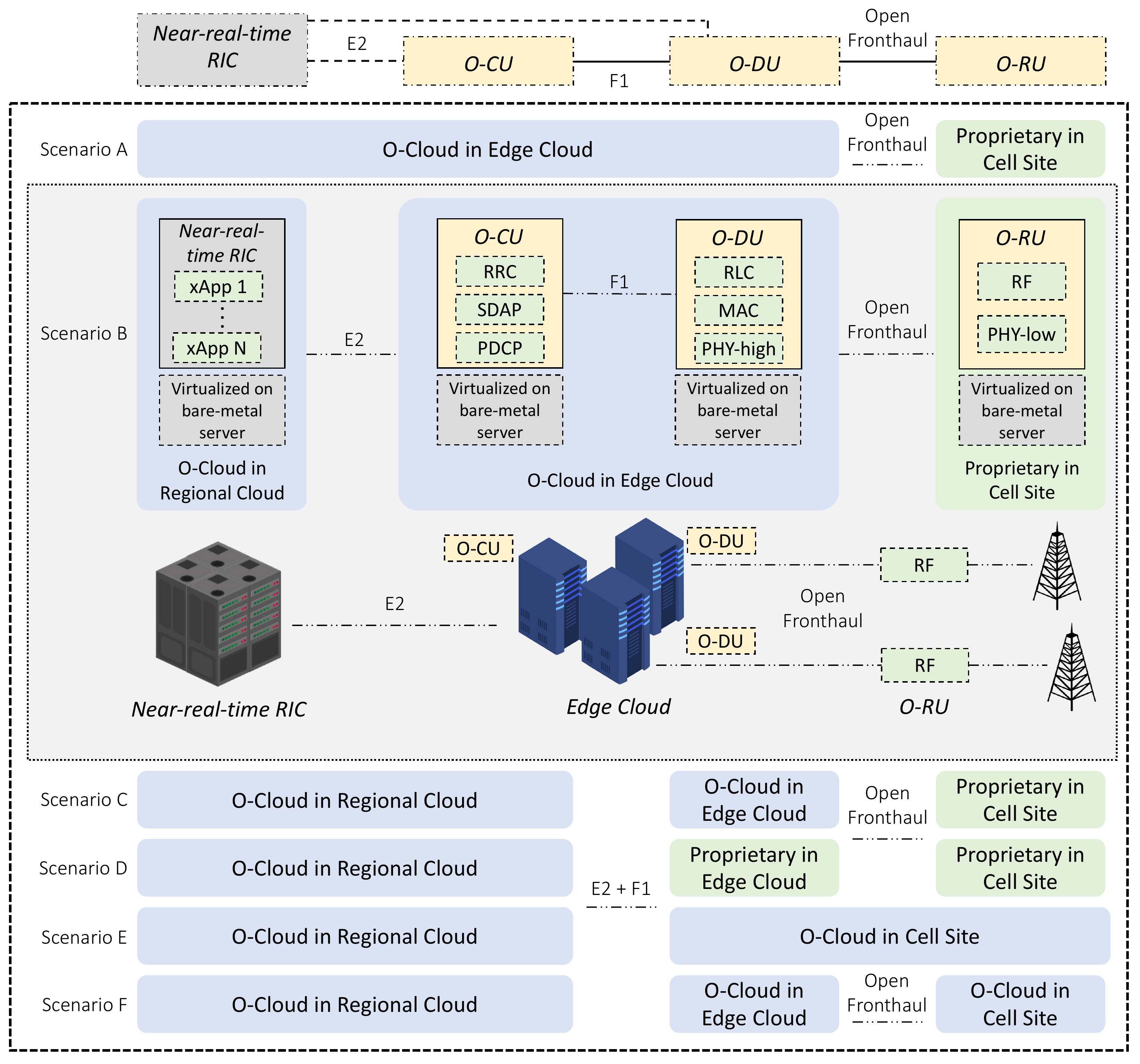}
    \caption{Logical (top) and physical (bottom) deployment options for O-RAN.}
    \label{fig:oran-deploy}
\end{figure}

\paragraph*{\textbf{O-RAN Deployment Options}}
O-RAN envisions different strategies for the deployment of its architecture in \textit{regional} and \textit{edge} cloud locations and at operator-owned cell sites~\cite{oranwpusecases}. Each facility could either run O-Cloud, i.e., a set of containers and virtual machines executing the O-RAN code with open interfaces, or be a proprietary site, which still uses the O-RAN open \glspl{api}, but could run closed-source code.
Both cases are illustrated in \fig{fig:oran-deploy}, which depicts the six different O-RAN deployment combinations indicated in~\cite{oranwpusecases}. 

In Scenario~A all the components except the \glspl{ru} are deployed at the edge of the network, co-located in the same data center that terminates the fronthaul fiber connectivity. 
Other alternatives foresee the \glspl{ric} and \glspl{cu} co-located at a regional cloud facility, with \glspl{du} and \glspl{ru} at the edge or on cell sites. 
%
The preferred deployment solution, however, is Scenario B, which deploys the \gls{ric} in the regional cloud, \glspl{cu} and \glspl{du} at the edge, and only the \glspl{ru} in the operator cell sites~\cite{oranwpusecases}.

\begin{table*}[t]
\centering
\caption{\acrshort{onf} frameworks interactions.}
\label{tab:framework_interactions}
\footnotesize
\renewcommand{\arraystretch}{1.3}
\begin{tabular}[]{|c|c|c|c|c|c|c|}
\hline
   & Aether~\cite{aether_website}    & COMAC~\cite{comac_website} & CORD~\cite{cord_website}  & SD-RAN~\cite{sdran_website}    & OMEC~\cite{omec_website} & ONOS~\cite{berde2014onos} \\
\hline
Aether~\cite{aether_website}  & - & x & x & x & x & x \\
\hline
COMAC~\cite{comac_website}   & x & - & x & x & x & x \\
\hline
CORD~\cite{cord_website}    & x & x & - & - & - & -\\
\hline
SD-RAN~\cite{sdran_website}  & x & x & - & - & - & x\\
\hline
OMEC~\cite{omec_website}    & x & x & - & - & - & -\\
\hline
ONOS~\cite{berde2014onos}    & x & x & - & x & - & -\\
\hline
\end{tabular}
\end{table*}

\paragraph*{\textbf{The O-RAN Software Community}}
Besides concerning standardization activities, the \gls{oran} Alliance has established a \textit{Software Community} in collaboration with the Linux Foundation for contributing open source~5G software that is compliant with the \gls{oran} specifications~\cite{oranwp}. The first two \gls{oran} releases date November~2019 (Amber release~\cite{amber_release}) and June~2020 (Bronze release~\cite{bronze_release}), and feature contributions from major vendors and operators, including AT\&T, Nokia, Ericsson, Radisys and Intel. These releases include Docker containers (which will be discussed in \sec{sec:virtualization_techniques}) and the source code for multiple \gls{oran} components:
\begin{itemize}
	\item The non-real-time \gls{ric}, with the A1 interface controller, and the possibility of managing machine learning and artificial models in the \gls{ran}.
 	\item The platform for the near-real-time \gls{ric}, with multiple applications, such as admission control, \gls{ue} manager, performance and measurement monitor, which talk to the \gls{du} through the E2 interface.
 	\item The \gls{du}, as previously discussed in \sec{sec:radisys}, and an initial version of the fronthaul library.
 	\item A framework for operation, administration, and maintenance, and the virtualization infrastructure.
 \end{itemize} 

Moreover, a simulator has been developed to test the functionalities of the different interfaces.

\paragraph*{\textbf{Expected Use Cases}}

The full realization of the \gls{oran} vision will revolutionize not only the modus operandi and business of telecom operators~\cite{oranwpusecases}, but also the world of those scientists, researchers and practitioners that will be able to run a modern, open source, full-fledged~5G control infrastructure in their lab and investigate, test and eventually deploy all sorts of algorithms (e.g., AI-inspired) for cellular networks at scale.
Researchers will be able to deploy and use the open source software provided by the \gls{oran} community to develop, test, and evaluate real-time \gls{ran} control applications. 
The \gls{oran} open source suite will enable 5G networking in a standardized environment, thus allowing reproducibility and easing future extensions. 
Moreover, deploying third-party xApps in the \gls{ric} (in collaboration with telecom operators) could enable experimentation running directly on \gls{oran}-compliant cellular networks. 
These tests could start in labs to be then they scaled up to larger deployments in the operator network, using the \glspl{ric} of both scenarios as a \emph{trait d'union}.

At the time of this writing, however, \gls{oran} is not at production-level. 
Therefore, future releases (\eg the Cherry release expected in December 2020) will attempt to complete integration of the different \gls{ran} components with the \glspl{ric}.
A parallel development effort is also being led by the SD-RAN project~\cite{sdran_website}, aiming at an open source, \gls{3gpp}-compliant \gls{ran} integrated with the \gls{oran} \gls{ric} and interfaces. 
According to~\cite{sdranwp}, the reference code will include the \gls{du} and \gls{cu}, interoperable with third-party \glspl{ru}, and a near-real-time \gls{ric} based on \gls{onos}~\cite{berde2014onos}.

\subsection{\acrlong{onf} Frameworks}
\label{sec:onf_frameworks}

The \acrfull{onf} is a consortium of several telecom operators that contribute open source code and frameworks used for the deployment of their networks. Specific examples include \gls{omec} (\sec{sec:epc_softwares}) and the aforementioned SD-RAN~\cite{sdranwp} and \gls{onos}~\cite{berde2014onos}. 
The \gls{onf} generally distinguishes between \textit{Component Projects}, which are frameworks and/or software that serve a specific purpose, and \textit{Exemplar Platforms}, which combine several Component Projects in a deployable, proof-of-concept reference design.
The different projects developed by the \gls{onf} are characterized by modular design, facilitating the integration of component projects, and providing the means to incorporate new open source projects. Additionally, some of them (e.g., \gls{onos} and Trellis) integrate P4 packet processing pipelines.
Table~\ref{tab:framework_interactions} summarizes the dependencies across different Component Projects and Exemplar Platforms.

\paragraph*{\textbf{Component Projects}}

The \gls{onf} currently overlooks the development of~10 open source component projects, concerning \gls{sdn}, transport networks, programmable networking hardware, and mobile networks~\cite{comp_projects}. 
In the last category, besides \gls{omec} and SD-RAN, notable efforts include \gls{cord}~\cite{peterson2016central}, which is an open source project (also part of the Linux Foundation portfolio) for deploying and managing edge cloud facilities for the \gls{mec} (\sec{sec:mec}).
The \gls{cord} framework is based on multiple software solutions that, together with reference hardware design, realize a reference \gls{mec} architecture based on \gls{sdn}, \gls{nfv} and cloud-native solutions.
\gls{cord} aims at (i) reducing deployment costs by using commodity hardware, and (ii) enabling innovative services, thanks to well-defined \glspl{api} for accessing edge computing facilities and multi-domain security. 
Moreover, \gls{cord} can be easily extended to address the heterogeneous requirements of different markets. 
In particular, two \gls{cord} architectures specific for mobile and residential services have been spawned off into two Exemplar Platforms (\gls{seba}~\cite{seba_website} and \gls{comac}~\cite{comac_website}).
\gls{cord} is one of the \gls{onf} projects with the largest number of contributions by the open source community. It includes detailed installation, operation and development guides~\cite{opencord_guide}, and a set of repositories with its source code~\cite{opencord_git}.

Another project related to software-defined mobile networks is \gls{onos}~\cite{berde2014onos}, an open source operating system for networking projects. 
While it has mostly been used for \gls{sdn} deployments in wired networks, \gls{onos} will provide a common substratum for SD-RAN and several Exemplar Platforms, such as Aether and \gls{comac}, described next.

\paragraph*{\textbf{Exemplar Platforms}}

An Exemplar Platform is given by extending a Component Project to implement a specific target or by combining and integrating multiple projects that can be deployed as a proof of concept. 
Among those currently available the following ones concern cellular and mobile networks~\cite{exemplar_platforms}.

\begin{itemize}
    \item \gls{comac}, which extends \gls{cord} into a platform that targets the integration of multiple access and \gls{cn} technologies, including 4G and 5G cellular networks, broadband, fiber and cable networks, and Wi-Fi deployments~\cite{comac_website}. The framework provides a common data plane in the core, which aggregates user data to and from different access technologies, and the possibility of managing users' subscriptions and identities with a single management platform. \gls{comac} is based on the \gls{seba} platform (a lightweight multi-access technology platform, which provides high-speed links from the edge of the network to the backbone of the infrastructure), and on multiple Component Projects, such as \gls{omec}, for the mobile core and edge, and \gls{cord} for the broadband subscriber management. Moreover, it will exploit \gls{oran} (with the SD-RAN implementation) for the control plane of the mobile cellular access. 
    The first \gls{comac} release~\cite{comac_release} provides instructions on how to configure the different software components to actually set up the overall platform (except for the SD-RAN portion that will be made available in future releases). 
    Additionally, a self-contained \gls{comac}-in-a-Box can be used to install the whole platform on a single server or virtual machine, to run end-to-end tests through an emulated data-plane (based on the \gls{oai} simulator, introduced in \sec{sec:oairan}) and the virtualized core and management environments~\cite{comac_box}.
    
    \item Aether, for streamlined deployment of private enterprise cellular networks \cite{aether_website}. 
    It combines three main elements, namely, a control and orchestration interface to the \gls{ran}, an edge cloud platform (the Aether edge), with support for cloud computing \glspl{api}, and a central cloud (the Aether core), for orchestration and management~\cite{aetherwp}. The Aether project will build and integrate several \gls{onf} efforts, including SD-RAN, \gls{onos}, \gls{cord} and OMEC. At the time of this writing, the source code and the deployment pipeline are not publicly available. When the code will be released, besides providing an opportunity for private 5G networks, Aether could be effectively used to deploy and manage integrated \gls{ran}-edge testbeds for 5G research and innovation.
\end{itemize}

\subsection{Other Frameworks and Projects}
\label{sec:other_frameworks}

Along with \gls{oran} and the \gls{onf} solutions, several open source communities (e.g., from 5G European projects) and companies have released frameworks and projects targeting connectivity, slicing and core-related functionalities. 
A few noteworthy examples are presented next.

\paragraph*{\textbf{5G-EmPOWER}}

5G-EmPOWER \cite{5gempower} is an operating system for heterogeneous RAN architectures. It consists of an open source and reprogrammable software platform abstracting the physical RAN infrastructure and providing high-level APIs to control RAN functionalities. The code of the platform is released under the Apache License v2.0~\cite{5gempower_git}. 

5G-EmPOWER embraces the \gls{sdn} philosophy to decouple control and data planes. This separation is obtained in practice via two main components, i.e., a centralized controller and a set of agents. The centralized controller (i) acts as an operating system with complete visibility of the physical infrastructure and its functionalities, and (ii) orchestrates the agents' actions via control directives sent through the OpenEmpower protocol~\cite{5gempower}. In turn, agents (i) run on each network element; (ii) abstract the underlying RAN-specific protocol implementations (e.g., LTE, Wi-Fi) to the controller, and (iii) modify the underlying protocol parameters according to the controller's directives. 

5G-EmPOWER currently supports several mobile \glspl{rat} such as LTE via srsLTE, Wi-Fi, and LoRa. 
The 5G \gls{nr} is not supported yet.
Integration of diverse \glspl{rat} is obtained through agents embedding specialized \textit{wrappers}, one for each \gls{rat}. While the general architecture of the agent is \gls{rat}-independent, the wrapper is \gls{rat}-specific. For instance, new \glspl{rat} (\eg 5G \gls{nr}) can be integrated by implementing new wrappers.

Despite the current lack of support for 5G \gls{nr}, 5G-EmPOWER already integrates relevant 5G-related technologies such as \gls{ran} slicing. Specifically, it provides software components that allow the instantiation of customized and isolated \gls{ran} slices on top of a shared physical infrastructure. Each \gls{ran} slice is created from a Slice Descriptor specifying \glspl{sla} and users belonging to each slice. A slice resource manager and a hypervisor are, then, in charge of admitting/revoking \gls{ran} slices and provisioning them with a certain amount of resources necessary to meet the corresponding \gls{sla}.

\paragraph*{\textbf{FlexRAN}} 

FlexRAN leverages abstraction and softwarization technologies to develop a \gls{rat}-independent \gls{ran} management platform~\cite{foukas2016flexran}. 
FlexRAN embraces \gls{sdn} principles to decouple control and data planes. The control plane is orchestrated by a real-time centralized controller, which controls a set of agents, one for each network element. FlexRAN implements a set of REST APIs in JSON format describing the northbound interface of FlexRAN. These APIs are used by the agents to interface with base stations, thus enabling control of the protocol stack and functionalities of the base stations (\ie \gls{mac}, \gls{rrc}, \gls{pdcp}).

FlexRAN directly interfaces with \gls{oai}. 
As such, it does not support 5G \gls{nr} communications yet. 
However, the northbound REST APIs can be used to specify and reconfigure slicing policies and requirements, providing support for 5G technologies such as \gls{ran} slicing. 
The FlexRAN code is available upon request and released as part of the Mosaic5G project under MIT license~\cite{mosaic5g_git}.

\paragraph*{\textbf{Magma}}

Magma is a framework developed by the Facebook Connectivity initiative for simplifying the deployment of cellular networks in rural markets~\cite{magma_website}. 
Notably, its goal is to avoid dependence on a specific access technology (\ie cellular or Wi-Fi) or on a generation of \gls{3gpp} core networks. 
Moreover, it avoids vendor lock-in for telecom operators, while offering advanced automation and federation capabilities. 
The latter is particularly relevant in rural and under-developed scenarios, as it allows the pooling of resources from multiple network operators. Magma is made up of three main components:
\begin{itemize}
   	\item The access gateway, which interfaces the access network to the \gls{cn}. The current Magma release supports an LTE \gls{epc}, and has been tested as termination point for the S1 interface of some commercial \gls{lte} base stations (see sections~\ref{sec:cloud_ran} and~\ref{sec:core_network} for more details).
   	\item A cloud-based orchestrator, which monitors the operations of the wireless network and securely applies configuration changes. It exposes an analytics interface providing control and traffic flow information.
   	\item A federation gateway, which is a proxy between the Magma core running in the access gateway and the network operator \gls{3gpp}-compliant core. This proxy exposes the \gls{3gpp} interfaces to the different \gls{cn} components, thus bridging the local mobile deployment with the operator backbone.
   \end{itemize}

\paragraph*{\textbf{LL-MEC}}
LL-MEC is an open source \gls{mec} framework for cellular systems compliant with \gls{3gpp} and \gls{etsi} specifications~\cite{nikaein2018ll}. This framework merges \gls{sdn}, edge computing and abstraction principles to provide an end-to-end platform where services requested by mobile users are executed on edge nodes of the network. LL-MEC consists of two main components: The Edge Packet Service controlling core network elements (e.g., routers and gateways) via OpenFlow APIs; and the Radio Network Information Service interfacing the data plane and physical \gls{ran} elements (e.g., eNBs) via the FlexRAN protocol~\cite{foukas2016flexran}.
Aside from \gls{mec} capabilities, LL-MEC supports network slicing for differentiated services applications with diverse latency and throughput requirements. 
The LL-MEC code is available upon request and released as part of the Mosaic5G project under Apache License v2.0~\cite{mosaic5g_git}.

\paragraph*{\textbf{LightEdge}}
LightEdge is a \gls{mec} platform for 4G and 5G applications compliant with \gls{etsi} \gls{mec} specifications~\cite{lightedge_mag}. LightEdge allows network operators to provide \gls{mec} services to mobile users through cloud-based applications. The framework provides a Service Registry summarizing services and applications registered to the \gls{mec} platform. LightEdge also includes modules and libraries for real-time information exchange across applications and services, and to perform traffic steering to and from the cellular network. LightEdge supports multiple eNBs and is compatible with several open source projects such as srsLTE, Open5GS, and srsEPC. The LightEdge code is available from the project repository under the Apache License v2.0~\cite{lightedge_git}.



\paragraph*{\textbf{OpenRAN}}
The \gls{tip} OpenRAN project aims at developing fully programmable \gls{ran} solutions that leverage general purpose hardware platforms and disaggregated software~\cite{openranwp, openran_website}. Albeit being closed source, OpenRAN implements open interfaces among the various elements of an NR-enabled cellular network, such as \gls{cu}, \gls{du}, and \gls{ru} (see \fig{fig:architecture}).
One of the projects spawned from OpenRAN is OpenRAN 5G NR, for defining a white-box gNB platform~\cite{openran5gnr, openran5gnr_website}. This platform allows equipment manufacturers to build flexible 5G \gls{ran} solutions for seamless multi-vendor support.

\paragraph*{\textbf{Akraino \acrshort{rec}}}
Akraino \gls{rec} is a blueprint to support and meet the requirements of the O-RAN \gls{ric} (\sec{sec:oran})~\cite{akrainorec_website, akrainorec_blueprint, akraino_repositories}. It is part of the Telco Appliance blueprint family~\cite{akrainorec_telco_appliance_blueprints}. Its features include automated configuration and integration testing to facilitate the management and orchestration of the virtualized \gls{ran}. The blueprint is made up of modular building blocks and provides an abstraction of the underlying hardware infrastructure, allowing O-RAN \gls{ric} to run on top of it, and to seamlessly interface with the provided APIs.

\paragraph*{\textbf{NVIDIA Aerial}}
NVIDIA Aerial is a set of \glspl{sdk} that allows to build \gls{gpu}-accelerated software-defined, cloud-native applications for the 5G \gls{vran}~\cite{aerialsdk}. At the time of this writing, Aerial provides two main \glspl{sdk}: cuBB and cuVNF.

\begin{itemize}
    \item The cuBB \gls{sdk} provides a highly-efficient~5G signal processing pipeline that runs \gls{phy}-layer operations directly on a \gls{gpu} to deliver high throughput.
    \item The cuVNF \gls{sdk} provides optimized input/output and packet placement in which 5G packets are directly sent to the GPU from compatible \glspl{nic}. Packets can be read and written directly from the \gls{nic} to the \gls{gpu} memory bypassing the \gls{cpu}, thereby reducing latency. This \gls{sdk} also allows developers to implement additional \glspl{vnf}, e.g., deep packet inspection, firewall and \gls{vran}.
\end{itemize}

\section{Open Virtualization and Management Frameworks for Networking}
\label{sec:management}

Besides \gls{ran} and \gls{cn} software, virtualization and management frameworks have an important role in the management and deployment of end-to-end, carrier-grade networks. 
Several communities and consortia have led the development of open source frameworks that have been deployed at scale by major telecom operators for the management of their physical and virtual infrastructure~\cite{yilma2019challenges,onap_website,jsan9010004}.

\begin{figure}[ht]
    \centering
    \includegraphics[width=0.9\columnwidth]{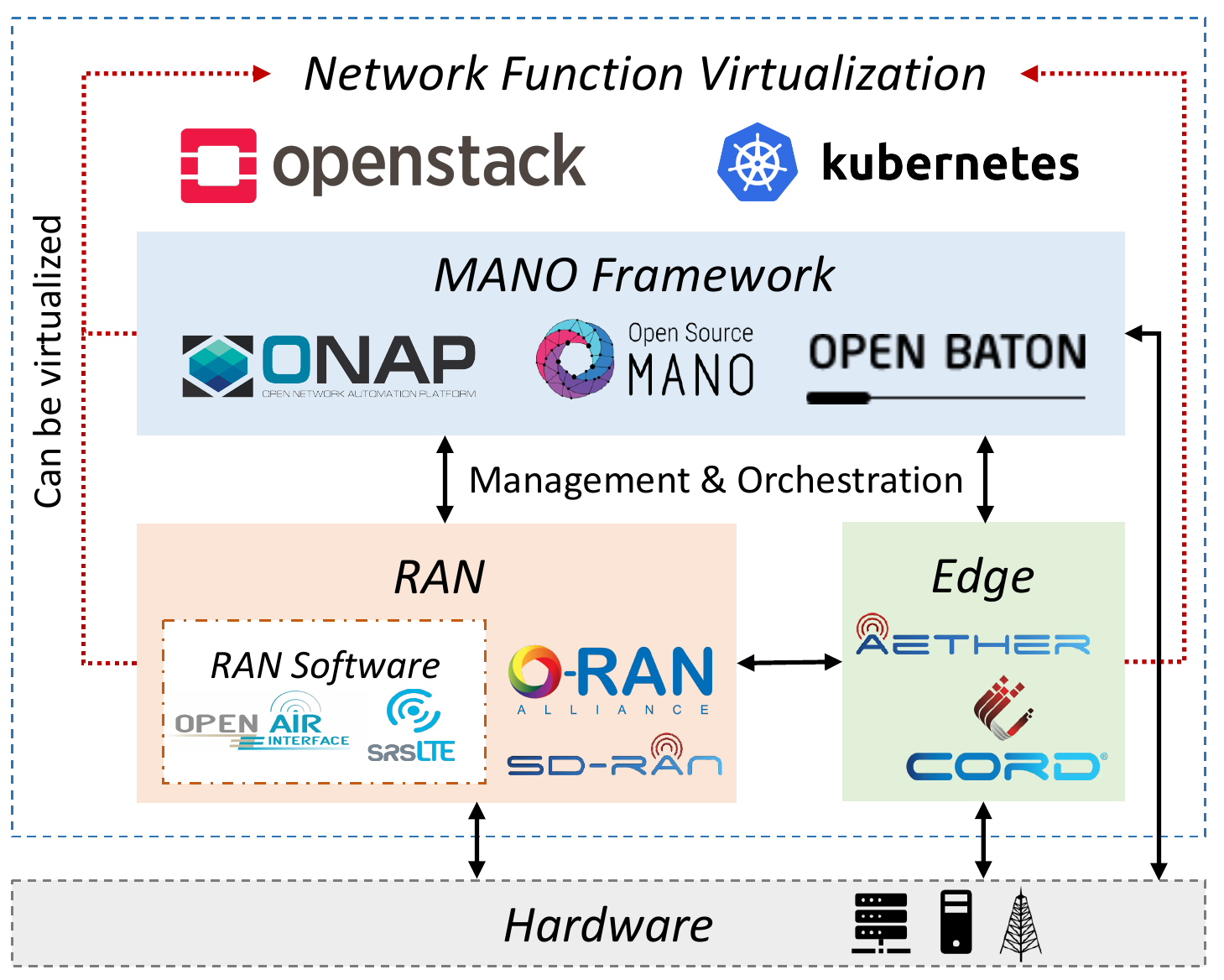}
    \caption{High-level relationship among \acrshort{mano}, \acrshort{ran}, and edge frameworks, and virtualization components.}
    \label{fig:mano_relationship}
\end{figure}

\gls{etsi} has defined a set of common features that an \gls{nfv} \gls{mano} framework should have, mainly for orchestrating network functions~\cite{ROTSOS2017203,mamushiane2019overview}.
\fig{fig:mano_relationship} depicts where these \gls{nfv} components fit in the 5G ecosystem. Table~\ref{tab:mano_frameworks} summarizes the main differences between the frameworks that will be described later in this section, i.e., \gls{onap}, \gls{osm}, and Open Baton.
\begin{table}[ht]
\centering
\caption{Comparison among different \gls{vnf} orchestrators.}
\label{tab:mano_frameworks}
\footnotesize
\setlength{\tabcolsep}{1.5pt}
\renewcommand{\arraystretch}{1.9}
\begin{tabular}{|c|c|c|c|}
        \hline
         & ONAP~\cite{onap_architecture} & OSM~\cite{osm_architecture} & Open Baton~\cite{carella2017prototyping} \\ \hline
         Community & \makecell{Linux Foundation\\w/ telecom operators} & \makecell{ETSI w/\\telecom operators} & \makecell{Fraunhofer\\FOKUS\\TU Berlin} \\ \hline
         \makecell{Compliance w/\\ETSI MANO} & in progress & yes & yes \\ \hline
         External \glspl{api} & \multicolumn{2}{c|}{\makecell{REST APIs (for external\\controllers, OSS/BSS, etc.)}} & Java SDK \\ \hline
         Network Services & \multicolumn{2}{c|}{VNFs and PNFs} & VNFs \\ \hline
         \multirow{2}{*}{Infrastructure} & \multicolumn{3}{c|}{Virtual Machines} \\
         \cline{2-4}
         & \makecell{Containers w/\\Kubernetes\\and Docker} & \makecell{Containers w/\\Kubernetes} & \makecell{Containers w/\\Docker} \\
         \hline
\end{tabular}
\end{table}
\gls{nfv} orchestrators are in charge of provisioning network services, i.e., combinations of physical and virtual network functions that can be chained together with a specific topology, managing their creation and life-cycle~\cite{gonzalez2018dependability}.
Notably, during the initialization of a network service, a basic configuration (0-day) is applied by default. Then, the \gls{mano} framework advertises the actual configuration for the function or service (1-day). Finally, updates (2-day configurations) can be deployed at a later stage. These operations are performed in concert by the different components of the \gls{nfv} orchestrator. Following the \gls{etsi} architecture~\cite{ROTSOS2017203,mamushiane2019overview}, an \gls{nfv} orchestrator is composed (i) of a subsystem that manages the virtualization infrastructure (e.g., \gls{vim} frameworks, such as OpenStack, Kubernetes, and Docker) and the connections to the physical hardware; (ii) of the actual \gls{mano} framework, and (iii) of the collection of \glspl{vnf} that it manages. These frameworks are equipped with southbound and northbound \glspl{api} to interact with other cellular infrastructure components, as shown in \fig{fig:mano_relationship}, such as edge frameworks for governing the \gls{ran} environment (e.g., Aether and \gls{cord}),
and \gls{ran} frameworks (see \sec{sec:frameworks} for more details). The latter include \gls{oran} and SD-RAN (described in Sections~\ref{sec:oran} and~\ref{sec:onf_frameworks}, respectively), which execute functions such as bringing intelligence to the network (\sec{sec:intelligence}) and interacting with open source software, such as \gls{oai} and srsLTE (\sec{sec:cloud_ran}).
These, in turn, focus on the \gls{cu}/\gls{du} split introduced by 5G \gls{nr}, and act as cellular base stations.

In the remainder of this section we discuss virtualization techniques and \glspl{vim} (Section~\ref{sec:virtualization_techniques}) and we describe popular \gls{mano} frameworks, such as \gls{onap}, \gls{osm}, and Open Baton (Sections~\ref{sec:onap}, \ref{sec:osm}, and~\ref{sec:baton}, respectively).

\subsection{Virtualization Techniques}
\label{sec:virtualization_techniques}

The \gls{nfv} paradigm decouples the services deployed in a network from the hardware infrastructure on which they run.
Applications are packaged into hardware-independent virtual machines, which can be instantiated on different physical machines.
This way, \gls{nfv} eliminates the need for hardware dedicated to each network function and enables scalability of network services.

\begin{figure*}[ht]
	\centering
	\includegraphics[width=0.8\textwidth]{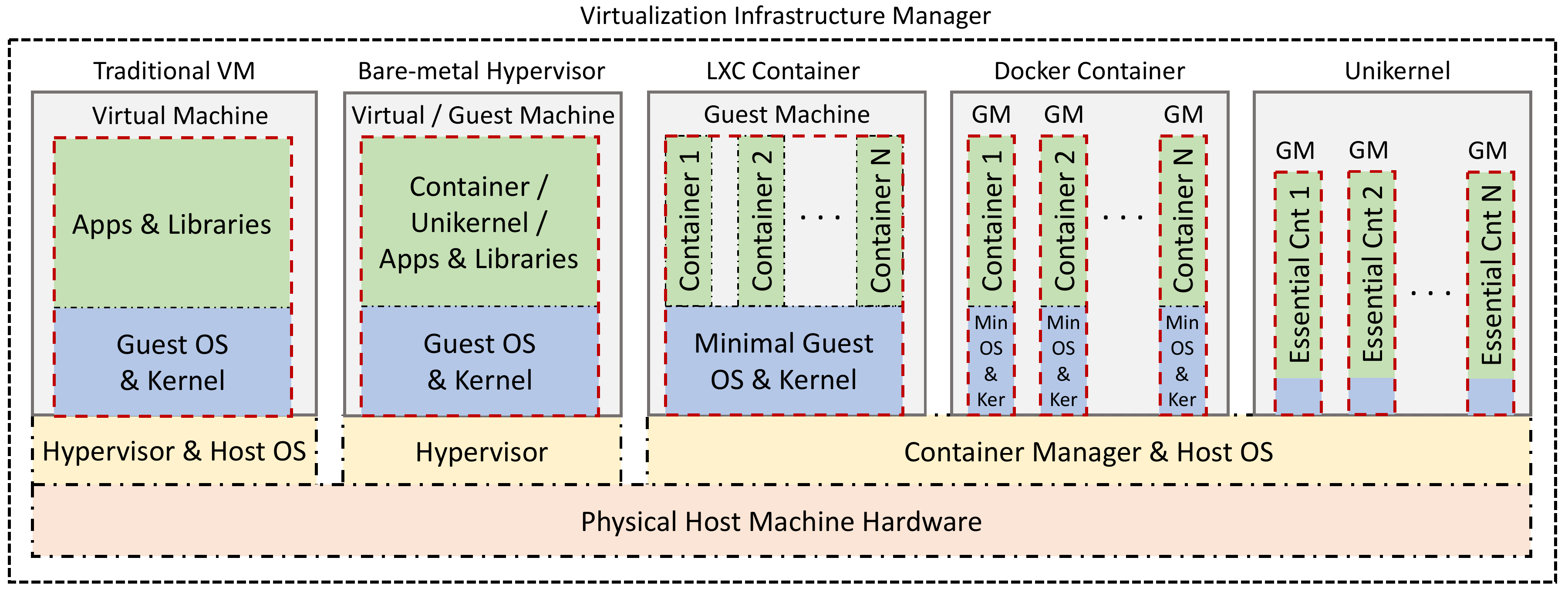}
	\caption{High-level \acrshort{nfv} architecture.}
	\label{fig:virtualization}
\end{figure*}

\gls{nfv}, whose high-level architecture is depicted in \fig{fig:virtualization}, provides many different ways to decouple applications and services, also known as \glspl{vnf}, from the general-purpose infrastructure on which they run, thus improving scalability and portability.
The most common approaches are: (i) Traditional \glspl{vm}; (ii) bare-metal hypervisors; (iii) containers, and (iv) unikernels.\footnote{Another virtualization solution that has recently been proposed is that of \textit{serverless computing}, or \textit{\gls{faas}}, in which the virtualization platform only spawns computing resources for specifying functions invoked by complex applications, without the need to manage a container or \gls{vm}. In this work, however, we focus primarily on the most common approaches in the \gls{nfv} domain.}
In the \gls{nfv} paradigm, a hypervisor can be used to create/run \glspl{vm}, containers, and unikernels, as well as to manage their resource allocation over the physical hardware. Finally, \glspl{vim} are leveraged to control the \gls{nfv} infrastructure at a higher level. 

\paragraph*{\textbf{Traditional Virtual Machines}}

A traditional \gls{vm} emulates a computer operating system through a guest operating system and kernel (\fig{fig:virtualization}).
To provide machine-level isolation, the \gls{vm} requires the virtualization of the hardware of the physical machine on which it runs (called ``host machine'').
This task is taken care of by the hypervisor, which is \textit{hosted} by the operating system of the physical machine and coordinates resource allocation between host and virtual machines.
In general, traditional \glspl{vm} are considered a resource-heavy approach because of the many hardware virtualization requirements (\eg virtual disk, CPU, network interfaces, etc.) that are needed to run the \glspl{vnf}.

\paragraph*{\textbf{Bare-metal Hypervisors}}

The approach of a bare-metal hypervisor \gls{vm} is similar to that of traditional \gls{vm}s, although the hypervisor \textit{directly runs} on the \textit{bare-metal hardware} of the host, without requiring a host operating system (\fig{fig:virtualization}).
Additionally, a bare-metal hypervisor can be used to run and manage a \textit{container or unikernel} (described next) instead of a full-fledged \gls{vm}. 

\paragraph*{\textbf{Containers}}

Containers are virtual environments that package a specific code and its dependencies to run applications and services in a virtualized way.
They are isolated from each other (through \textit{namespace isolation}) and share access to the operating system and kernel of the physical machine on which they run.
They only require a minimal guest operating system instead of the heavy and resource-wasteful hardware virtualization required by \glspl{vm} (\fig{fig:virtualization}).
Containers can be maintained both by a container manager interfaced with the operating system of the host machine, or by a hypervisor directly running on a bare-metal host machine.
The most widespread open source container virtualization systems are \gls{lxc}~\cite{lxc} and Docker~\cite{docker} ( \fig{fig:virtualization}):
\begin{itemize}

\item \textit{\acrshort{lxc}} was the first major implementation of the modern containers. It leverages control groups and namespace isolation to create virtual environments with separated networking and process space.

\item \textit{Docker} enables the creation of containers, and uses virtualization at the operating system level to deploy them on the physical machine. Differently from \gls{lxc}, on which it was initially based upon, Docker breaks applications, services, and dependencies into modular units and layers inside each container. Additionally, these layers can be shared among multiple containers, increasing the efficiency of Docker container images. 
Compared to \gls{lxc}, Docker containers lack some UNIX functionalities and subsystems.

\end{itemize}

\paragraph*{\textbf{Unikernels}}

Unikernels are minimal, lightweight, specialized images built with the sole purpose of running specific applications (\fig{fig:virtualization}).
They compile application services and dependencies into the executable virtual images, without including unnecessary components that would be, instead, included by a generic operating system.
This way, unikernels achieve better performance than traditional containers and virtual machines.
Since unikernels only include the software components that are needed to run the application of interest, they also improve the security of the system by exposing fewer functionalities that can be attacked by malicious entities.

Examples of unikernel systems are: (i) ClickOS~\cite{martins2013enabling}, IncludeOS~\cite{bratterud2015includeos} and OSv~\cite{kivity2014osv}, which focus on high-performance, low-latency, and secure applications; (ii) MirageOS~\cite{mirageos}, which includes several libraries that are then converted to kernel routines upon image compilation, and (iii) UniK~\cite{unik}, which deals with compilation and orchestration of unikernel images.

Unikernels applications for cellular networks include the following. 
Wu et al.\ that integrates Android system libraries into OSv to offload mobile computation for \gls{mcc} and \gls{mfc}~\cite{wu2018android}.
Valsamas et al.\ propose a content distribution platform for 5G networks that is based on unikernels such as ClickOS, OSv, and MirageOS~\cite{valsamas2018elastic}.
A performance comparison of the IncludeOS unikernel and Docker containers instantiated as \gls{vnf} for 5G applications is carried out in~\cite{filipe2019performance}. 







\paragraph*{\textbf{Hypervisors}}

A hypervisor is software that creates and runs virtual machines, the \textit{guest machines}, on a physical computer, called the \textit{host machine}.
Key tasks of an hypervisor include (i) providing isolation between virtual/guest machine and the host machine; (ii) managing allocation/reallocation of resources, such as CPU, memory and storage, to the guest machines, and (iii) scheduling of resources among host and guest machines.

There are two types of hypervisors: Type~1 and type~2 hypervisors.
The former are referred to as \textit{bare-metal hypervisors} and manage the guest operating system by running them directly on the host hardware, thus acting as an operating system for the host machine.
Examples of hypervisors of type~1 are Xen~\cite{barham2003xen} and VMware ESXi~\cite{esxi}.
Type~2, instead, are referred to as \textit{hosted hypervisors} and run on top of the host operating system as a software layer or application.
Examples of hypervisors of type~2 are Linux \gls{kvm}~\cite{kvm}, BSD bhyve~\cite{bhyve}, and Oracle VirtualBox~\cite{virtualbox}.

\begin{figure*}[t]
	\centering
	\includegraphics[width=0.8\textwidth]{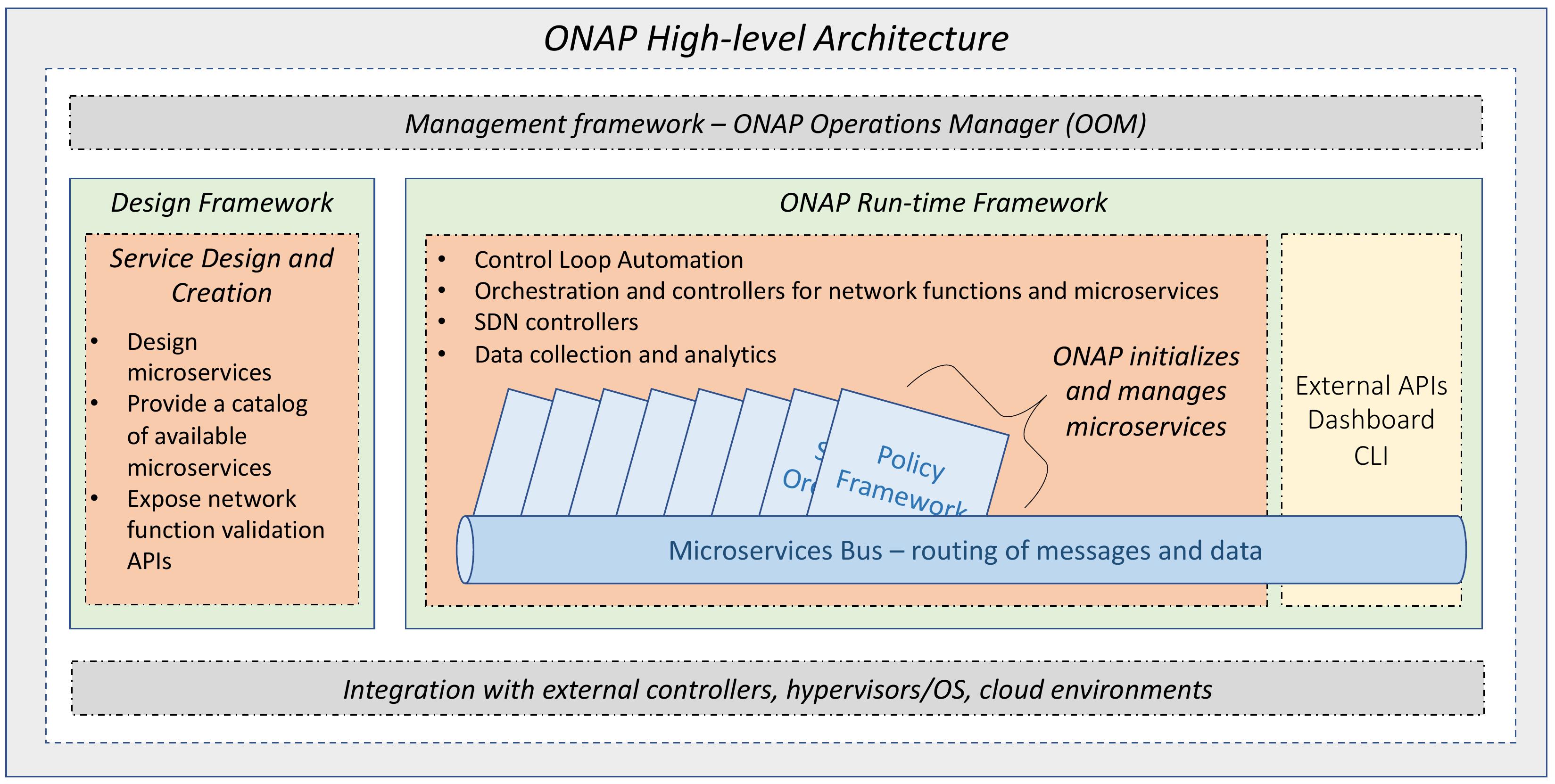}
	\caption{High-level architecture of the \gls{onap} framework.}
	\label{fig:onap}
\end{figure*}

\paragraph*{\textbf{\acrlong{vim}s}}

A \gls{vim} is in charge of control and management of the \gls{nfv} infrastructure and its resources, such as storage, computation, and networking resources, and coordinates the instantiation of virtual guest machines on the hardware of the physical host machines.
The \gls{vim} is part of \gls{mano} frameworks, such as those described in the remaining of this section.
Examples of \glspl{vim} are OpenStack~\cite{openstack} and Kubernetes~\cite{kubernetes}:
\begin{itemize}
    \item \textit{OpenStack} is a cloud computing software platform capable of controlling a plethora of heterogeneous resources, such as compute, storage and networking resources. Among its very many features, it can act as a \gls{vim}, managing the network infrastructure, virtual machines, containers, unikernels, \gls{vnf} services and applications.
    
    \item \textit{Kubernetes} provides automatic deployment, scaling, and management of virtual machines, containers, unikernels, and their applications through a set of primitives. Kubernetes abstracts and represents the status of the system through a series of objects. These are persistent entities that describe the \gls{vnf} or applications that are running on the Kubernetes-managed cluster, their available resources, and the policies on their expected behavior.
    In the past few years, a number of projects able to interact with Kubernetes have been launched to solve complex problems at the layers~2 and~3 of the protocol stack. These projects include Istio~\cite{istio} and \gls{nsm}~\cite{networkservicemesh}. Istio mesh services carry out traffic management, policy enforcement and telemetry collection tasks. \gls{nsm} interfaces with Kubernetes APIs to support advanced use cases and facilitates the adoption of new cloud native paradigms. Specifically, it allows network managers to seamlessly perform tasks such as adding radio services, requesting network interfaces, or bridging multiple layer~2 services.
\end{itemize}

\subsection{The \acrlong{onap}}
\label{sec:onap}

\gls{onap} is an \gls{nfv} framework developed as a project of the Linux Foundation, with AT\&T, China Mobile, Vodafone, China Telecom, Orange, Verizon and Deutsche Telekom as main mobile operator supporters. \gls{onap} is deployed in several commercial cellular networks, and vendors like Ericsson, Nokia, Huawei and ZTE, among others, provide \gls{onap} support and integration in their products~\cite{dublin_release, frankfurt_release}. Therefore, \gls{onap} represents one of the most advanced software-based solutions for commercial cellular networks, actively maintained and developed to meet production-level quality standards and satisfy new emerging requirements~\cite{kapadia2018onap}. 

\gls{onap} handles the design, creation, and life cycle management of a variety of network services. Network operators can use \gls{onap} to orchestrate the physical and virtual infrastructure deployed in their networks, in a vendor-agnostic way~\cite{onap_architecture}. In addition to common \gls{nfv} orchestrator functionalities (e.g., automated policy-driven management of the virtualization infrastructure and of the network services), \gls{onap} provides a design framework to model network applications and services as well as a framework for data analytics to monitor the services for healing and scaling.
Additionally, \gls{onap} provides a number of reference designs, \ie \textit{blueprints}.
These can be used to deploy the \gls{onap} architecture, depicted at a high-level in \fig{fig:onap}, in specific markets or for specialized use cases (\ie 5G networks or Voice over LTE deployments).
They have been tested in combination with their typical hardware configurations.

The main components of the \gls{onap} architecture~\cite{onap_architecture}, depicted in \fig{fig:onap}, are: (i) The Management Framework; (ii) the Design Framework, and (iii) the Run-time Framework.
The management framework, called \gls{oom}, orchestrates and monitors the lifecycle of the \gls{onap} components. The \gls{oom} leverages Kubernetes (Section~\ref{sec:virtualization_techniques}), and Consul~\cite{consulwp}, which enables service control, discovery, configuration, and segmentation.
Among its functionalities the most noteworthy are: (i) Component deployment, dependency manager, and configuration; (ii) real-time health monitoring; (iii) service clustering and scaling, and (iv) component upgrade, restart, and deletion. 

The design framework allows to create network services with a declarative modeling language, which makes it possible to specify requirements and functionalities of each service. It allows to model resources, services, products and their management and control functions, through a set of common specifications and policies. Additionally, it includes  service design and creation modules for the definition, simulation, and certification of systems assets, processes and policies. 
Finally, this module provides a database of existing services, and \glspl{api} for the validation of network functions. 

\gls{onap} run-time framework is made up of several software frameworks for most of its management and orchestration functionalities. In the run-time domain, a microservices bus allows communication and routing of messages and data among the different network functions initialized and managed by \gls{onap}. The run-time framework dispatches and terminates microservices, using an automated control loop, and collects data and analytics from the platform. The run-time component exposes \glspl{api}, a dashboard and a command-line tools with a unified interface to control the network infrastructure.
Finally, a southbound layer (gray box at the bottom of \fig{fig:onap}) can be used for the integration with external controllers, operating systems and cloud environments, while northbound \glspl{api} are offered to OSS/BSS, big data, and other relevant services.

\begin{figure*}[t]
	\centering
	\includegraphics[width=0.8\textwidth]{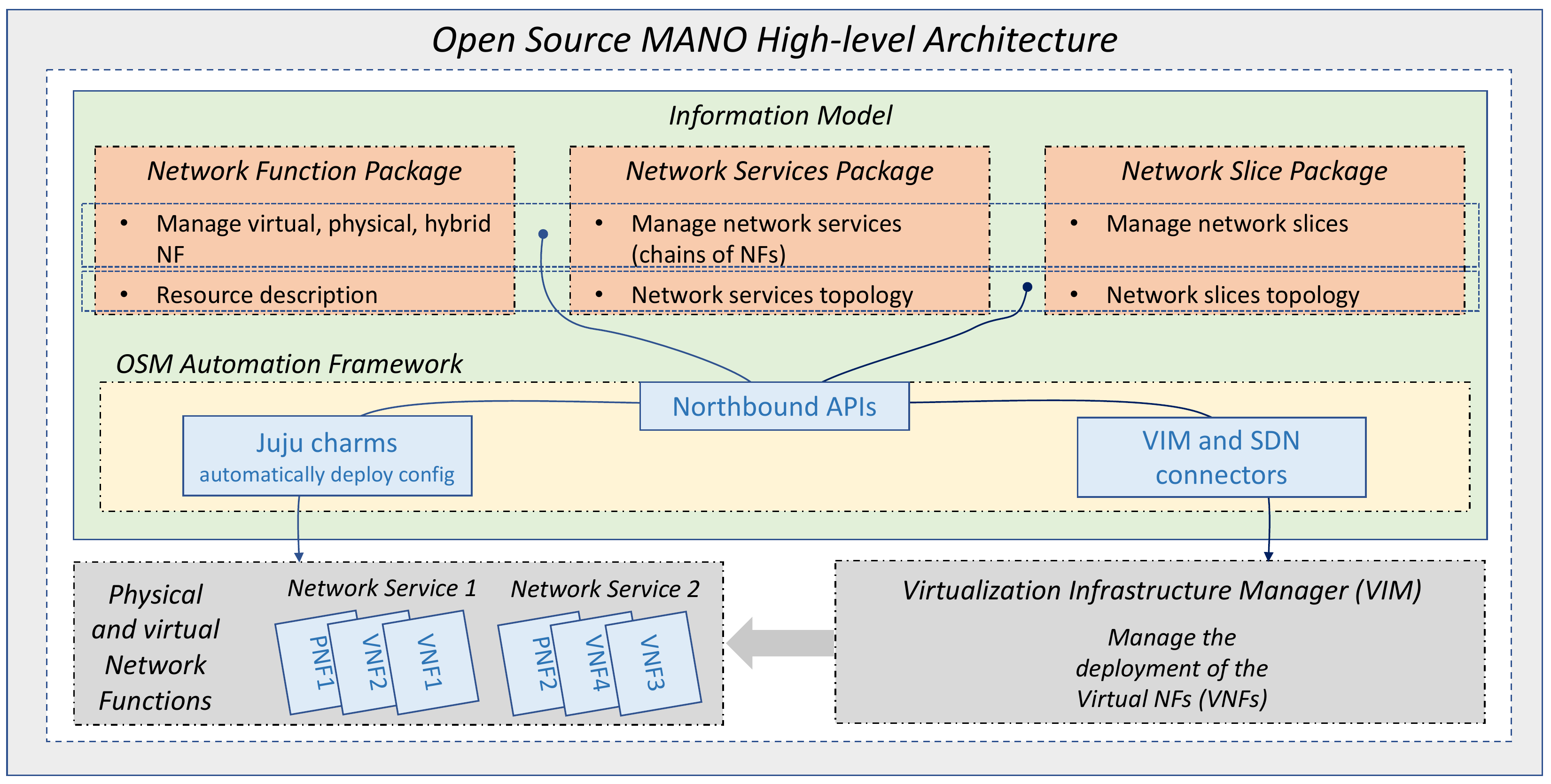}
	\caption{High-level architecture of the \gls{osm} framework.}
	\label{fig:osm}
\end{figure*}

\paragraph*{\textbf{Integration with 5G networks}}

Besides representing a general framework for the management and orchestration of mobile networks, \gls{onap} offers some key features that are relevant to 5G deployments. The main requirements that operators have identified are the need to support a hybrid infrastructure, with both physical and virtual appliances, edge automation, with the cloud geographically distributed in different edge locations, and real time analytics, which would enable closed-loop automation. 

The Dublin release (June 2019) has introduced a first iteration on the implementation of a 5G blueprint. The most noteworthy feature is the joint discovery of virtual and physical network functions. As previously mentioned (sections~\ref{sec:4g_5g_architecture} and~\ref{sec:cloud_ran}) the 5G \gls{ran} will not only deal with software components, but will also be in charge of real time signal and RF processing through physical hardware equipment. As such, \gls{onap} also includes discovery and integration procedures to gain awareness of both these \textit{virtual} and \textit{physical network functions}, and to properly manage them.
Additionally, the Dublin release includes preliminary support for network slicing.
%
Preliminary evaluation studies for the feasibility of network slicing in \gls{onap} are discussed in~\cite{slim2017towards,rodrigues20205g}. 
Finally, the \gls{onap} 5G blueprint currently supports a dynamic configuration and optimization of 5G network parameters. 
This relies on a data collection platform, which transfers in a matter of minutes relevant \glspl{kpi} from the network edge to central processing facilities, and analyzes these data to automatically apply optimizations, or scale resources as needed. 

The Frankfurt release (June 2020) builds upon the Dublin release by adding support for end-to-end network slicing and service orchestration. Additionally, it allows network designers to define control loops without having to wait for the next official \gls{onap} release. One of the most noteworthy features introduced by the Frankfurt release is the harmonization effort toward \gls{oran} compatibility through the O1 and A1 interfaces. This effectively aims at defining the specifications for managing the elements of the 5G \gls{ran}, such as \gls{cu}, \gls{du}, \gls{ru} (\sec{sec:oran} and \fig{fig:oran}).
Future releases will further integrate \gls{onap} and \gls{oran}, making it easier for telecom operators to deploy an integrated \gls{oran}/\gls{onap} solution~\cite{onap_oran}.

\begin{table*}[t]
\centering
\caption{Capabilities of SDRs and their integration with RAN software.}
\label{tab:hardware}
\setlength{\tabcolsep}{1.75pt}
\footnotesize
\renewcommand{\arraystretch}{1.9}
\begin{tabular}[]{|c|c|c|c|c|c|}
\hline
\gls{sdr}   &   \makecell{TX/RX Channels}   & Frequency Range   & \makecell{Instantaneous\\Bandwidth (up to)}   & \acrshort{ran} Software    & Target \\
\hline
bladeRF & 1 & $[300\:\mathrm{MHz}, 3.8\:\mathrm{GHz}]$  & $28\:\mathrm{MHz}$    & \acrshort{oai}, srsLTE    & \makecell{\acrshort{das} node, small cell} \\
\hline
bladeRF 2.0 micro   & 2 & $[47\:\mathrm{MHz}, 6\:\mathrm{GHz}]$ & $56\:\mathrm{MHz}$  & \acrshort{oai}, srsLTE  & \makecell{\acrshort{das} node, small cell} \\
\hline
Iris    & 2 & $[50\:\mathrm{MHz}, 3.8\:\mathrm{GHz}]$   & $56\:\mathrm{MHz}$    & \acrshort{oai}    & \makecell{\acrshort{das} node, small cell, cell tower} \\
\hline
LimeSDR & $4$~TX, $6$~RX    & $[100\:\mathrm{kHz}, 3.8\:\mathrm{GHz}]$  & $61.44\:\mathrm{MHz}$   & \acrshort{oai}, srsLTE  & \makecell{\acrshort{das} node,\\small cell} \\
\hline
\acrshort{usrp} B205mini-i  & 1 & $[70\:\mathrm{MHz}, 6\:\mathrm{GHz}]$ & $56\:\mathrm{MHz}$    & srsLTE    & \acrshort{das} node \\
\hline
\acrshort{usrp} B210    & 2 & $[70\:\mathrm{MHz}, 6\:\mathrm{GHz}]$ & $56\:\mathrm{MHz}$    & \acrshort{oai}, srsLTE    & \makecell{\acrshort{das} node, small cell} \\
\hline
\acrshort{usrp} N310    & 4 & $[10\:\mathrm{MHz}, 6\:\mathrm{GHz}]$ & $100\:\mathrm{MHz}$ & \acrshort{oai}  & \makecell{\acrshort{das} node, small cell, cell tower, rooftop} \\
\hline
\acrshort{usrp} X310    & \makecell{up to 2\\(daughterboards)}  & \makecell{[DC, $6\:\mathrm{GHz}$]\\(daughterboards)}  & \makecell{$160\:\mathrm{MHz}$\\(daughterboards)}  & \acrshort{oai}, srsLTE  & \makecell{\acrshort{das} node, small cell, cell tower} \\
\hline
\end{tabular}
\end{table*}

\subsection{\acrlong{osm}}
\label{sec:osm}
\acrfull{osm} is a \gls{mano} framework developed by a set of network providers, including Telefonica, BT and Telenor. The community also counts cloud and open source entities, such as Amazon and Canonical. Similar to \gls{onap}, the \gls{osm} framework is well developed and deployed in major cellular networks.

The goals and general architecture of \gls{osm} are introduced in~\cite{osm_architecture}, and shown at a glance in \fig{fig:nfv_high_level}.
%
Overall, the framework is an end-to-end network service orchestrator, tailored for deployment in mobile networks. 
%
\fig{fig:osm} describes the \gls{osm} architecture and its interactions with the network functions and \gls{vim} it manages, following the typical \gls{nfv} orchestrator structure described in the introduction to this section. The main logical components of \gls{osm} are: 
\begin{itemize}
	\item \textit{The Information Model:} It performs the modeling of network functions, services, and slices into templates called packages. This is enabled by a well-defined information model provided by the \gls{etsi} \gls{mano} framework~\cite{ROTSOS2017203}. Similarly to the design component of \gls{onap}, this allows telecom operators to analyze the requirements of the network and model the resources that need to be deployed for functions, services, and slices.
	\item \textit{The \gls{osm} Automation Framework:} It automates the life cycle of network services, from instantiation to scaling and, eventually, deletion. This is done by applying the information model to the actual deployed infrastructure, as shown in \fig{fig:osm}, through a northbound interface that the automation framework exposes to the different modeling components. The 0-day, 1-day and 2-day configurations of the actual services and functions are done through Juju Charms~\cite{jujucharm} (\ie tools to define, configure, and deploy services on cloud and bare-metal systems). 
\end{itemize}
Similarly to \gls{onap}, \gls{osm} has southbound and northbound \glspl{api} that can be exploited by other external services, such as other orchestrators and OSS/BSS, respectively~\cite{osm_api}.

\paragraph*{\textbf{Integration with 5G networks}}
\gls{osm} has published in December 2018 a 5G-ready release, which added the possibility of managing both virtual and physical network functions, network slicing, and a policy-based closed control loop, and extended the analytics and interface frameworks. Several 5G European projects have used and/or contributed to \gls{osm}. Metro-Haul focused on the design of an \gls{sdn}-based optical transport infrastructure for 5G networks, and developed an \gls{osm} component for the management of the infrastructure in a distributed wide area network~\cite{casellas2018metrohaul}. Similarly, 5G Tango has developed multiple components for \gls{osm}, including an emulator for the virtual infrastructure manager, and network slicing capabilities, while discussing and proposing possible extensions of the \gls{mano} concept for 5G into more advanced frameworks~\cite{soenen2019empowering,jsan9010004}. The 5GCity and 5G-MEDIA projects have used \gls{osm} as \gls{nfv} orchestrator for management frameworks of networks for smart cities and media distribution over a \gls{cdn}, respectively~\cite{colman2018city,rizou2018service}. Finally, 5G-TRANSFORMER has integrated \gls{osm} in its network slicing framework for the management of computing resources~\cite{oliva2018transformer}.

The author of~\cite{dreibholz2020flexible} investigates how to integrate \gls{osm} and \gls{oaicn}, to facilitate the deployment of fully-software-based solutions in testbeds and edge locations. Finally,~\cite{tranoris2018enabling} uses \gls{osm} to experiment with dynamic virtual network function placing in a 5G vehicular scenario.

\subsection{Open Baton}
\label{sec:baton}

Open Baton \cite{baton} is an open source project jointly developed by Fraunhofer FOKUS and TU Berlin aimed at providing a modular and reconfigurable framework for the orchestration of network services. The framework focuses on \gls{nfv} management and is fully-compliant with the \gls{etsi} NFV MANO specification. Its source code is available online under Apache License v2.0 \cite{opencbaton_git}. 

Open Baton provides a full-fledged ecosystem to instantiate and handle atomic \glspl{vnf}, as well as to compose them to create more complex network services. The framework has been designed to operate over a virtualized infrastructure. For this reason, Open Baton features drivers to directly interface with most \glspl{vim}, with specific support for OpenStack~\cite{openstack} (see \sec{sec:virtualization_techniques}).

Besides \gls{vnf} orchestration, Open Baton also provides support for multi-tenancy applications through network slicing and \gls{mec}~\cite{carella2017prototyping}. Specifically, Open Baton features a \gls{nse}, a Java-based external software component that interacts with Open Baton via dedicated \glspl{sdk}. The \gls{nse} allows network operators to specify \gls{qos} requirements for each network slice (e.g., minimum bandwidth for a target traffic class) in a clean and simple way via minimal JSON or YAML configuration files. Through Open Baton's \gls{vim} drivers, these configuration files are, then, dispatched to the \gls{vim}, which ultimately guarantees that each slice meets the set \gls{qos} requirements.

\begin{table*}[t]
\centering
\caption{5G Testbeds.}
\label{tab:testbeds}
\small
\setlength{\tabcolsep}{1.75pt}
\footnotesize
\renewcommand{\arraystretch}{1.9}
\begin{tabular}[]{|c|c|c|c|c|}
\hline
Testbed & Technology available & \makecell{5G Open Source Software} & Framework & Scenario \\
\hline
\acrshort{aerpaw}   & \makecell{5G and \acrshort{cr} for UASs}  & \multicolumn{2}{c|}{under development}   & City-scale outdoor \\
\hline
Arena   & \makecell{5G, \acrshort{cr}, massive \acrshort{mimo}}    & \gls{ran} \& Core   & N/A   & Large-scale office \\
\hline
Colosseum   & 5G, \acrshort{cr}  & \gls{ran} \& Core    & O-RAN \acrshort{ric}  & \makecell{Large-scale network emulator} \\
\hline
\acrshort{cornet}   & 5G, \acrshort{cr} & \gls{ran} \& Core & N/A   & Large-scale indoor \\
\hline
\acrshort{cosmos}   & \makecell{5G, mmWave, \acrshort{cr}, optical switching}  & \gls{ran} \& Core & \makecell{O-RAN\\components}  & \makecell{Indoor, city-scale outdoor} \\
\hline
Drexel Grid & 5G, \acrshort{cr} & \gls{ran} \& Core & N/A   & Large-scale indoor \\
\hline
\acrshort{fit} testbeds & \makecell{5G, \acrshort{cr}, \acrshort{iot}, \acrshort{nfv}} & \gls{ran} \& Core & \acrshort{osm}   & Large-scale indoor \\
\hline
IRIS    & \makecell{5G, \acrshort{cr}, Wi-Fi, WiMAX,\\cloud-\acrshort{ran}, \acrshort{nfv}, S-band} & \gls{ran} \& Core & N/A   & Indoor \\
\hline
NITOS   & \makecell{5G, \acrshort{cr}, Wi-Fi, WiMAX}   & \gls{ran} \& Core  & N/A & \makecell{Large-scale indoor and outdoor, office} \\
\hline
\powderrenew   & \makecell{5G, \acrshort{cr}, massive \acrshort{mimo},\\Network Orchestration}    & \gls{ran} \& Core & O-RAN \acrshort{ric}  & \makecell{Indoor, city-scale outdoor} \\
\hline
5TONIC  & \makecell{5G \acrshort{nfv}, network orchestration} & N/A   & \acrshort{osm} & Data center \\
\hline
\end{tabular}
\end{table*}

\section{\acrlong{sdr} Support for Open Source Radio Units}
\label{sec:hardware}

The open source software described throughout this article can be mostly executed on commodity hardware, except for the signal processing related to the physical layer, which generally runs on the \glspl{fpga} of the \glspl{sdr}. 
In this section, we discuss the main \gls{sdr} solutions compatible with the software suites described in \sec{sec:cloud_ran}. These platforms are pivotal in enabling researchers to deploy and experiment with end-to-end networks, even though they may not have access to carrier-grade hardware deployed by the major telecom operators.  

A summary of the capabilities of each \gls{sdr} is shown in Table~\ref{tab:hardware}. There we can find powerful \glspl{sdr} that can act as rooftop base stations, such as the \gls{usrp} N310, and cell towers, such as the \gls{usrp} X310 or arrays of Iris \glspl{sdr}. Smaller \gls{sdr} models, such as \glspl{usrp} B210, bladeRF/2.0 micro, and LimeSDR, instead, are powerful enough to operate as small cells, while, the ultra compact and lightweight \gls{usrp} B205mini-i can act as a \gls{das} node. A description of the capabilities of each of these \glspl{sdr} will be given in the following paragraphs.

\paragraph*{\textbf{\acrshort{usrp}}}

The \acrfull{usrp} are \gls{sdr} solutions produced by National Instruments/Ettus Research for designing, prototyping and testing wireless protocols and systems~\cite{ettus}:

\begin{itemize}

\item \textit{\acrshort{usrp} B210:} It is a full-duplex \gls{sdr} with two transmit receive channels.
It covers a frequency range from $70\:\mathrm{MHz}$ to $6\:\mathrm{GHz}$ with a real-time bandwidth of up to $56\:\mathrm{MHz}$.
It connects to the host computer through a USB~3.0 interface, and 
is compatible with  \gls{oai} and srsLTE discussed in \sec{sec:cloud_ran}.

\item \textit{\acrshort{usrp} B205mini-i:} It is a full-duplex \gls{sdr} with a frequency range from $70\:\mathrm{MHz}$ to $6\:\mathrm{GHz}$, and an instantaneous bandwidth of up to $56\:\mathrm{MHz}$.
Similar to \gls{usrp} B210, it connects to the host computer
through a USB~3.0 interface.
It is compatible with srsLTE.

\item \textit{\acrshort{usrp} X310:} It is an \gls{sdr} with two daughterboard slots, which enable up to two full-duplex transmit/receive chains.
The covered frequency range and instantaneous bandwidth vary according to the specific daughterboard model (from DC to $6\:\mathrm{GHz}$, and up to $160\:\mathrm{MHz}$).
This \gls{usrp} can connect to a host computer through a range of interfaces such as $1$~Gigabit Ethernet, dual $10$~Gigabit Ethernet, PCIe Express, and ExpressCard.
It is compatible with both \gls{oai} and srsLTE;

\item \textit{\acrshort{usrp} N310:} It is a full-duplex networked \gls{sdr} with four transmit/receive chains.
It covers the $[10\:\mathrm{MHz}, 6\:\mathrm{GHz}]$ frequency range with an instantaneous bandwidth of up to $100\:\mathrm{MHz}$.
It can be connected to a host computer
through $1$~Gigabit Ethernet, $10$~Gigabit, or Xilinx Aurora over two SFP+ ports, and is compatible with the \gls{oai} software.

\end{itemize}

\paragraph*{\textbf{bladeRF}}

The bladeRF denotes a series of full-duplex \gls{sdr} devices produced by Nuand~\cite{bladerf}.
They are available with different form factors and \gls{fpga} chips, and are compatible with both \gls{oai} and srsLTE. They and connect to the host computer through a USB~3.0 interface.

\begin{itemize}

\item \textit{bladeRF:} This \glspl{sdr} comes in two different configurations, i.e., blade\-RF x40 with a 40KLE Cyclone IV \gls{fpga}, and bladeRF x115 with a 115KLE Cyclone IV \gls{fpga}.
Regardless of the specific \gls{fpga} chip, the bladeRF \gls{sdr} has a single transmit/receive chain, which covers the $[300\:\mathrm{MHz}, 3.8\:\mathrm{GHz}]$ frequency range with up to $28\:\mathrm{MHz}$ of instantaneous bandwidth.

\item \textit{bladeRF 2.0 micro:} This model is equipped with two transmit/receive chains and supports $2 \times 2$ \gls{mimo} operations.
It is available with different \gls{fpga} chips, i.e., 49KLE Cyclone V \gls{fpga} chip (blade\-RF xA4), and 301KLE Cyclone V \gls{fpga} chip (bladeRF xA9).
It covers the $[47\:\mathrm{MHz}, 6\:\mathrm{GHz}]$ frequency range with an instantaneous bandwidth of $56\:\mathrm{MHz}$.

\end{itemize}

\paragraph*{\textbf{LimeSDR}}

This \gls{sdr} is produced by Lime Microsystems~\cite{limesdr}, has four transmit and six receive chains, and supports $2 \times 2$ \gls{mimo} operations.
It covers the $[100\:\mathrm{kHz}, 3.8\:\mathrm{GHz}]$ frequency range with an instantaneous bandwidth of up to $61.44\:\mathrm{MHz}$.
The LimeSDR is compatible with both \gls{oai} and srsLTE.
Other versions of this \gls{sdr} include the LimeSDR Mini, which has two channels instead of four, and the LimeSDR QPCIe, which enables $4 \times 4$ \gls{mimo} configurations instead of the $2 \times 2$ of the standard model. Support for these LimeSDR models has not been explicitly reported by either \gls{oai} or srsLTE.

\paragraph*{\textbf{Iris}}

This is a networked \gls{sdr} device with two transmit/receive chains, produced by Skylark Wireless~\cite{skylark}.
It works in the $[50\:\mathrm{MHz}, 3.8\:\mathrm{GHz}]$ frequency range and supports an instantaneous bandwidth of up to $56\:\mathrm{MHz}$.
This \gls{sdr} connects to the host computer through a $1$~Gigabit Ethernet interface, and is compatible with \gls{oai}.
It can be combined with additional hardware platforms provided by Skylark to boost its performance, while multiple Iris \glspl{sdr} can be grouped in arrays to enable massive \gls{mimo} operations (see Argos~\cite{argos}, and the \powderrenew \gls{pawr} testbed~\cite{powder, renew} described in \sec{sec:testbeds}).

\section{Testbeds} \label{sec:testbeds}

In this section we describe a number of testbeds that can be used to instantiate softwarized 5G networks by leveraging the open source utilities, frameworks and hardware components described in this article.
An overview of the capabilities of each testbed is given in Table~\ref{tab:testbeds}.

\paragraph*{\textbf{\acrlong{pawr}}}
The objective of the NSF-funded \acrfull{pawr} program is to enable experimental investigation of new wireless devices, communication techniques, networks, systems, and services in real wireless environments through several heterogeneous city-scale testbeds~\cite{pawr}.
The following are the \acrshort{pawr} platforms awarded so far and that are being built. 

\begin{itemize}

\item \textit{\powderrenew:} The combination of \acrlong{powder}~\cite{powder, breen2020powder} and \acrlong{renew}~\cite{renew, doostmohammady2020good} provides a testbed, namely \powderrenew, which covers an area of $6\:\mathrm{km^2}$ of the University of Utah campus in Salt Lake City, UT.
%
Its objective is to foster experimental research for a range of heterogeneous wireless technologies, including 5G, \gls{ran}, network orchestration, and massive \gls{mimo} technologies.
\powderrenew is equipped with cutting-edge compute, storage, and cloud resources, as well as state-of-the-art \glspl{sdr}.

Besides allowing users to install their own software suites, the testbed offers a series of ready-to-use ``profiles,'' which are instantiated on its bare-metal machines.
These include open source \gls{ran} software, \eg \gls{oai} and srsLTE, coupled with different \gls{epc} solutions, as well as components of the frameworks previously discussed, including the \gls{oran} \gls{ric}.
Finally, the \powderrenew platform has been used to demonstrate automated optimization of 5G networks in~\cite{bonati2020cellos}.

\item \textit{\acrshort{cosmos}:} The \gls{cosmos}~\cite{cosmos, raychaudhuri2020cosmos, kohli2020openaccess} is being deployed in the densely-populated neighborhood of West Harlem, New York City, NY. Upon completion, it will cover an area of $2.59\:\mathrm{km^2}$.
This testbed focuses on providing ultra-high-bandwidth and low-latency wireless communications, and it will have edge-computing capabilities.
Among others, \gls{cosmos} will allow researchers to experiment with mmWave and optical switching technologies.

\gls{cosmos} includes the \gls{orbit}~\cite{orbit}, a platform with a number of \glspl{usrp} X310, compatible with the open source \gls{ran} solutions described in \sec{sec:cloud_ran}.
The platform served as Open Test and Integration Center during the \gls{oran} plugfest, a proof of concept for demonstrating the potentials of multi-vendor interoperability of cellular networks~\cite{oran2019plugfest}.

\item \textit{\acrshort{aerpaw}:} The \gls{aerpaw}~\cite{aerpaw, marojevic2019experimental, sichitiu2020aerpaw} will be the first-ever wireless platform to allow large-scale \gls{uas} experimentation for 5G technologies and beyond.
\gls{aerpaw} will be deployed in the North Carolina Research Triangle and its features will include flying aerial base stations to provide cellular connectivity to ground users.

\end{itemize}

Additional \gls{pawr} platforms are scheduled to be selected in 2021, possibly providing testbeds for experimental research on rural broadband communication and networking.
As publicly funded testbeds, the \gls{pawr} platforms will be accessible to the wireless research community for experimental use. As such, each platform implements the fundamental requirements that will ensure reproducibility of experiments, interoperability with other platforms, programmability, open access to the research community, and sustainability.

\smallskip
\textit{\textbf{Colosseum}}
is the world's most powerful wireless network emulator~\cite{colosseum}.
It is housed at Northeastern University Innovation Campus in Burlington, MA.
It is composed of 21~server racks with 256~\gls{usrp} X310 \glspl{sdr}, half of which are controllable by experimenters, while the other half is allocated to Colosseum \gls{mchem}.
Through \gls{mchem} scenarios, researchers can seamlessly emulate entire virtual worlds with up to~$65,\!536$ concurrent wireless channels with realistic characteristics, such as path loss and fading.
This way, Colosseum assures the ultimate reproducibility of experiments in the sub-$6\:\mathrm{GHz}$ spectrum band.

At its core, Colosseum is a full-fledged data-center with state-of-the-art hardware, including more than $900\:\mathrm{TB}$ of networked storage, over $170$~high-performance servers, $320$~\glspl{fpga}, and full mesh high-speed connections.
Colosseum allows users to install and instantiate the open source \gls{ran} and core network solutions discussed in sections~\ref{sec:cloud_ran} and~\ref{sec:core_network}, as well as components of the frameworks detailed in \sec{sec:frameworks} (\eg the \gls{oran} \gls{ric}).
After having being used in the DARPA \gls{sc2}, Colosseum will be available to the research community shortly.


\smallskip
\textit{\textbf{Arena}}
is an indoor testbed composed of $24$~\gls{sdr} \glspl{usrp} ($16$~\glspl{usrp} N210 and $8$~\glspl{usrp} X310) stacked up in a radio rack, and controlled by $12$~Dell PowerEdge R340 high-performance machines in a server rack~\cite{bertizzolo2020arena}.
Servers and \glspl{sdr} are connected through dual $10$~Gigabit Ethernet connections to guarantee rapid and low-latency radio control and communication.
The \glspl{usrp} are connected to a grid of $8 \times 8$ antennas hung off the ceiling of a $208.1\:\mathrm{m^2}$ dynamic indoor office space through same-length cables that guarantee equal signal delays across the whole testbed.  
Moreover, Arena \glspl{sdr} are fully synchronized through $4$~Octoclock clock distributors to enable massive \gls{mimo} applications, among others.

The Arena testbed allows researchers to experimentally evaluate wireless protocols and solutions for indoor~5G deployments in an office-like environment.
For instance, Arena can be used to evaluate Wi-Fi and \gls{cr} solutions, to generate communication traces and data sets, and to evaluate the performance of standard compliant cellular networks through the \gls{oai} and srsLTE protocol stacks (\sec{sec:cloud_ran}).
Arena has been used to demonstrate future cellular networks capabilities~\cite{bertizzolo2019demo}, 5G \gls{ran} optimization~\cite{bonati2020cellos} \gls{ran} slicing~\cite{doro2020sledge, doro2020slicing}.

\smallskip
\textit{\textbf{5TONIC}}
includes data center and equipment for~5G virtual network experimentation~\cite{5tonic}.
It is composed of a \gls{nfv} infrastructure with high-performance servers and workstations running network orchestration and virtualization functions and a number of \gls{sdr} platforms and devices. 5TONIC allows users to run complex \gls{nfv} and orchestrations frameworks such as \gls{osm}.
5TONIC has been used for \gls{nfv} \gls{mano}~\cite{nogales2019design} and mmWave applications~\cite{roldan2018experiments}.

\smallskip
\textit{\textbf{\acrshort{fed4fire}}}
is a Horizon 2020~\cite{horizon2020} project to foster experimentally-driven research in the future Internet ecosystem~\cite{fed4fire}.
Among others, it includes a number of \emph{federated} testbeds for wireless, 5G, \gls{iot}, cloud, and big data applications.
Below, we describe the testbeds of the \acrshort{fed4fire} project targeting open source cellular networks research and compatible with the \acrshort{oai} and srsLTE \gls{ran} software tools (\sec{sec:cloud_ran}).

\begin{itemize}

\item \textit{NITOS} is a Future Internet Facility composed of an outdoor, an indoor, and an office testbed with both \glspl{sdr} and commercial nodes~\cite{nitos}.
The outdoor testbed comprises nodes with Wi-Fi, WiMAX, and \gls{lte} capabilities, while the indoor and the office testbeds are made up of of Icarus Wi-Fi nodes~\cite{icarus_node} deployed in an isolated environment.
NITOS has been used for \gls{mano}~\cite{karamichailidis2019enabling}, 5G distributed spectral awareness~\cite{chounos2019enabling}, and \gls{mec} applications~\cite{makris2019minimizing, passas2019pricing}, among others.


\item The \textit{IRIS} testbed focuses on Cloud-\gls{ran}, \gls{nfv}, and \gls{sdn} experimental research~\cite{iris_testbed}.
The testbed includes a number of ceiling-mounted \gls{sdr} devices supporting Wi-Fi, WiMAX, and 4G/5G technologies, as well as S-band transceivers.

\end{itemize}

\smallskip
\textit{\textbf{\gls{cornet}}}
is a testbed of $48$~\gls{sdr} nodes deployed in a four-story building in the Virginia Tech campus (Blacksburg, VA) that enables experiments on dynamic spectrum access and \gls{cr} research~\cite{cornet}.
\gls{cornet} allows users to perform 5G experimental research by leveraging open source software, such as \gls{oai} and srsLTE, or by emulating cellular signals through \gls{cots} equipment.
Among other applications, \gls{cornet} has been used to evaluate link adaptation in cellular systems~\cite{rao2019analysis}.

\smallskip
\textit{\textbf{\gls{fit}}}
is a French project for large-scale testbeds for wireless communications~\cite{fit}.
It includes: (i) \gls{fit} Wireless, which targets indoor Wi-Fi, \gls{cr}, and 5G applications (through \gls{oai}, for instance); (ii) \gls{fit} \gls{iot}-Lab, which concerns \gls{iot}-related experimentation, and (iii) \gls{fit} Cloud, which supports the other two by enabling \gls{sdn} and \gls{nfv} research through \gls{osm}, among other frameworks.
%

\smallskip
\textit{\textbf{Drexel Grid}}
is a testbed made up of $24$~\gls{sdr} devices ($20$~\glspl{usrp} N210 and 4~\glspl{usrp} X310) hung off the ceiling of a dedicated indoor room to evaluate diverse 5G and \gls{cr} wireless technologies~\cite{drexel_grid, dandekar2019grid}.
The \glspl{usrp} X310 of the testbed can be used with open source \gls{ran} software such as \gls{oai} and srsLTE (\sec{sec:cloud_ran}).
Additionally, this testbed includes a channel emulator with simulated nodes to evaluate wireless systems in a controlled and repeatable environment.
%

\section{Softwarized~5G: Limitations and Road Ahead} 
\label{sec:limitations}

Our work aimed at investigating how the ``softwarization of everything''---that is pervasive to current trends in computing, communication and networking---got its way into cellular networks, and in particular how it has not only revolutionized their fourth generation, but has also established a radically new way, both technical and commercial, to usher in the~5G era successfully.

Our overview has focused on the most recent advances in the open source and reprogrammable 5G ecosystem. 
We have listed and discussed a variety of heterogeneous, yet modular software and hardware components.
In particular, we have illustrated how their expected evolution is key to transitioning from the traditional black box approach of cellular network management to those white box principles that will bring both research and industry communities to swift innovation, shorter time-to-market and overall higher customer satisfaction.

By way of conclusion, we finally intend to point out that despite operators, vendors and scientists are paying considerable attention to the new software-defined technologies described in this article, these solutions are not ready for prime time on commercial 5G networks just yet.
Indeed, the road to celebrate this marriage needs overcoming a few show stoppers, which we describe below.

\begin{itemize}

\item {\bf Keep pace with the standards.} The cellular network community faces constant pressure to keep up with the specifications/technologies being introduced by new communication, networking and even programming standards. 
A notable example are the \gls{nr} and \gls{mmwave} communication technologies introduced as~5G enablers by~\gls{3gpp} and are currently being deployed in \textit{closed source} commercial networks. 
The \gls{ran} software libraries described in \sec{sec:cloud_ran} have not yet completed the development of the support for \gls{nr}, as such task requires a considerable effort in terms of coding and testing.
In this domain, open source network simulators have, so far, provided a more controlled development environment for the development and assessment of 5G solutions~\cite{mezzavilla2018end,patriciello2019e2e,pratschner2018versatile,oughton2019open}. The testing of real-world 5G software, especially for \glspl{mmwave}, is indeed extremely complex due to the lack of accessible open hardware for the software to run, which precludes testing important components, such as beam management. 
The platforms described in \sec{sec:hardware} are optimized for carriers below $6\:\mathrm{GHz}$, even though early prototypes of open \glspl{mmwave} boards are currently being developed~\cite{pi-radio, dhananjay2020calibrating, polese2019millimetera, haarla2020characterizing}. Similarly, most of the testbeds described in Section~\ref{sec:testbeds} focus on sub-$6\:\mathrm{GHz}$ deployments, with only a few (e.g., \gls{cosmos}) considering an extension to \glspl{mmwave}. The road ahead for the development of high-band 5G software-defined solutions, therefore, lies in a more concerted, joint software development effort, and in hardware platforms that can keep up with the requirements of the software community. Furthermore, the current lack of a mature code base for \gls{3gpp} NR \gls{ran} software libraries hinders the development of more advanced features even in the sub-$6\:\mathrm{GHz}$ spectrum, such as \gls{urllc} support (e.g., with mini slots~\cite{3gpp.38.321}) and multi-connectivity~\cite{38300}.

\item {\bf Latency and scalability issues.} 
The scalability of an \gls{sdr}-based system, both in terms of processing and computing requirements, depends on the number of signal processing operations, which are generally proportional to the available bandwidth~\cite{AKEELA2018106}. 
As the next generation wireless systems will deal with larger and larger bandwidths for higher data rates~\cite{38300}, the implementation of the radio stack and its software processing chains will have to be extremely efficient, robust, and count on powerful and reliable hardware.
Moreover, considering the tight latency and throughput requirements of many~5G use cases, the integration of software and hardware needs to be seamless, to deliver the best possible performance.
In this regard, although virtualized solutions add unprecedented flexibility to the network, they also come with new challenges. 
Specifically, these solutions rely upon resource sharing (which limits and regulates resource utilization among different processes~\cite{savi2019impact}), and virtualization requires additional interactions between the virtualized and bare-metal environments~\cite{xiang2019reducing}. Together, these aspects introduce additional latency which might not be tolerable for many 5G application and services. 

\item {\bf Limited contributions for \gls{ran} open source software.} The same large telecom operators and vendors driving the development of open source \gls{cn} and \gls{mano} frameworks are not showing the same level of attention to \gls{ran}-related projects. 
\gls{ran} efforts have indeed mostly come from academia or from smaller companies, with limited manpower and resources. 
As some sophisticated digital signal processing and implementations of the lower layers of the \gls{ran} stack are often source of intellectual property and product-bearing revenues for telecom businesses, major vendors and operators are not encouraged to release their solutions as open source. Recognizing this limitation, the \acrlong{oai} Software Alliance has licensed the \gls{oai} \gls{ran} implementation with a permissive license, which allows contributors to retain intellectual property claims (see \sec{sec:oairan}).
Additionally, the \gls{oran} Alliance is moving encouraging first steps toward an openly softwarized \gls{ran} (see \sec{sec:oran}), even though the current development efforts do not include also an open source software for the radio front-ends. 
Therefore, the wireless community should aim at increasing the support toward the development of complete and open \gls{ran} and radio software libraries, increasing the number of active contributors to the currently available open source \gls{ran} projects.

\item {\bf Lack of robust, deployable, and well-documented software.} As of now, most of the frameworks and libraries described in Sections~\ref{sec:cloud_ran}-\ref{sec:management} cannot be used in actual networks, as their open source component is either incomplete, requires additional integration and development for actual deployment, or lacks robustness. 
Moreover, to reach the quality of commercial solutions, the open source community should aim at delivering well-documented, easy-to-deploy, and robust software, specifying all dependencies and additional software components that guarantee the correct and efficient functioning of the system. For example, the container-driven development model as used in cloud-native computing could be adopted to simplify and expedite the software deployment process.

\item {\bf Need for secure open source software.} Heightened attention to software development following best practices for robustness and security is sorely required~\cite{cowan2003software}, to guarantee privacy, integrity, and security to the end users of softwarized networks. 
Openness already facilitates useful scrutiny of the code. Audits and reviews from the open source community can help prevent bugs and/or security holes, whose existence needs to be responsibly disclosed to the project maintainers~\cite{piantadosi2019fixing}.
Appropriate security, especially ``by design,'' however, is still lacking.
The exposure of \glspl{api} to third party vendors (e.g., for the \gls{ric} apps), for instance, could introduce new vulnerabilities in the network, in case the \glspl{api} are not properly securely designed, and contain weaknesses that can be exploited by attackers. 
It is clear that the security of the open source software that will be eventually deployed in 5G and beyond systems must be a key concern for the developers and telecom ecosystem. The wireless community, thus, should follow the best practices developed over the years by other open source communities (e.g., the Linux kernel), that constantly make it possible to tighten the security of open source products~\cite{wang2017how}.
    
\end{itemize}

All these road blocks are currently preventing, or considerably slowing down, the widespread and painless application of several of the softwarized solutions presented in this article. 
It is now the task of the wireless research and development community to transform these challenges into the opportunity to innovate further in the direction of truly realizing open, programmable, and virtualized cellular networks.

\footnotesize
\bibliographystyle{IEEEtran}
\bibliography{biblio}


\begin{appendices}
\section{Acronyms}
\label{sec:appendix_acronyms}
\renewcommand{\glossarysection}[2][]{}
\renewcommand{\arraystretch}{1.1}
\footnotesize
\setlength{\glsdescwidth}{0.75\columnwidth}
\printglossary[style=index,type=\acronymtype]
\normalsize
\end{appendices}

\begin{IEEEbiography}
[{\includegraphics[width=1in,height=1.25in,keepaspectratio]{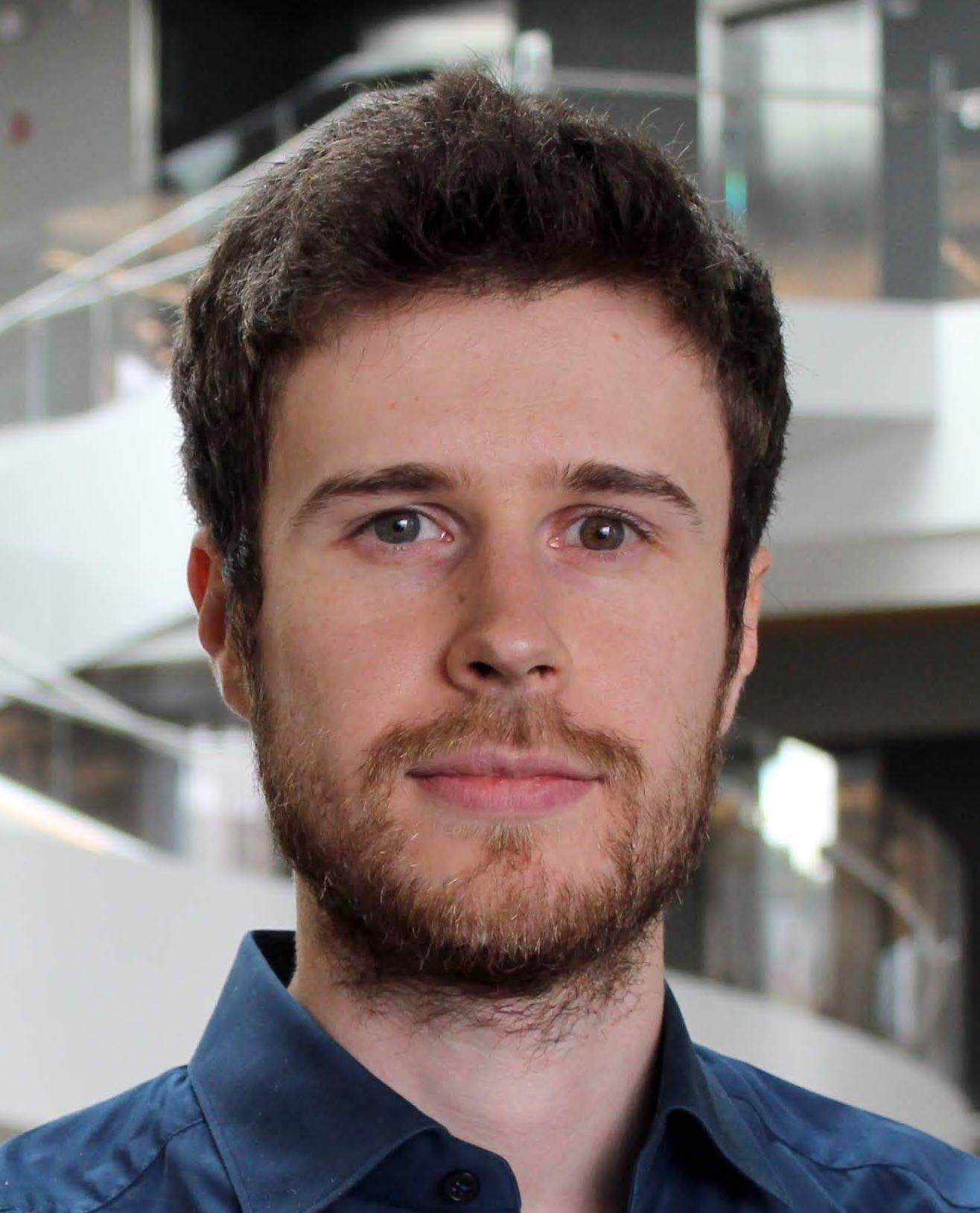}}]{Leonardo Bonati}
received his B.S. in Information Engineering and his M.S. in Telecommunication Engineering from University of Padova, Italy in 2014 and 2016, respectively. He is currently pursuing a Ph.D. degree in Computer Engineering at Northeastern University, MA, USA. His research interests focus on 5G and beyond cellular networks, network slicing, and software-defined networking for wireless networks.
\end{IEEEbiography}

\begin{IEEEbiography}
[{\includegraphics[width=1in,height=1.25in,keepaspectratio]{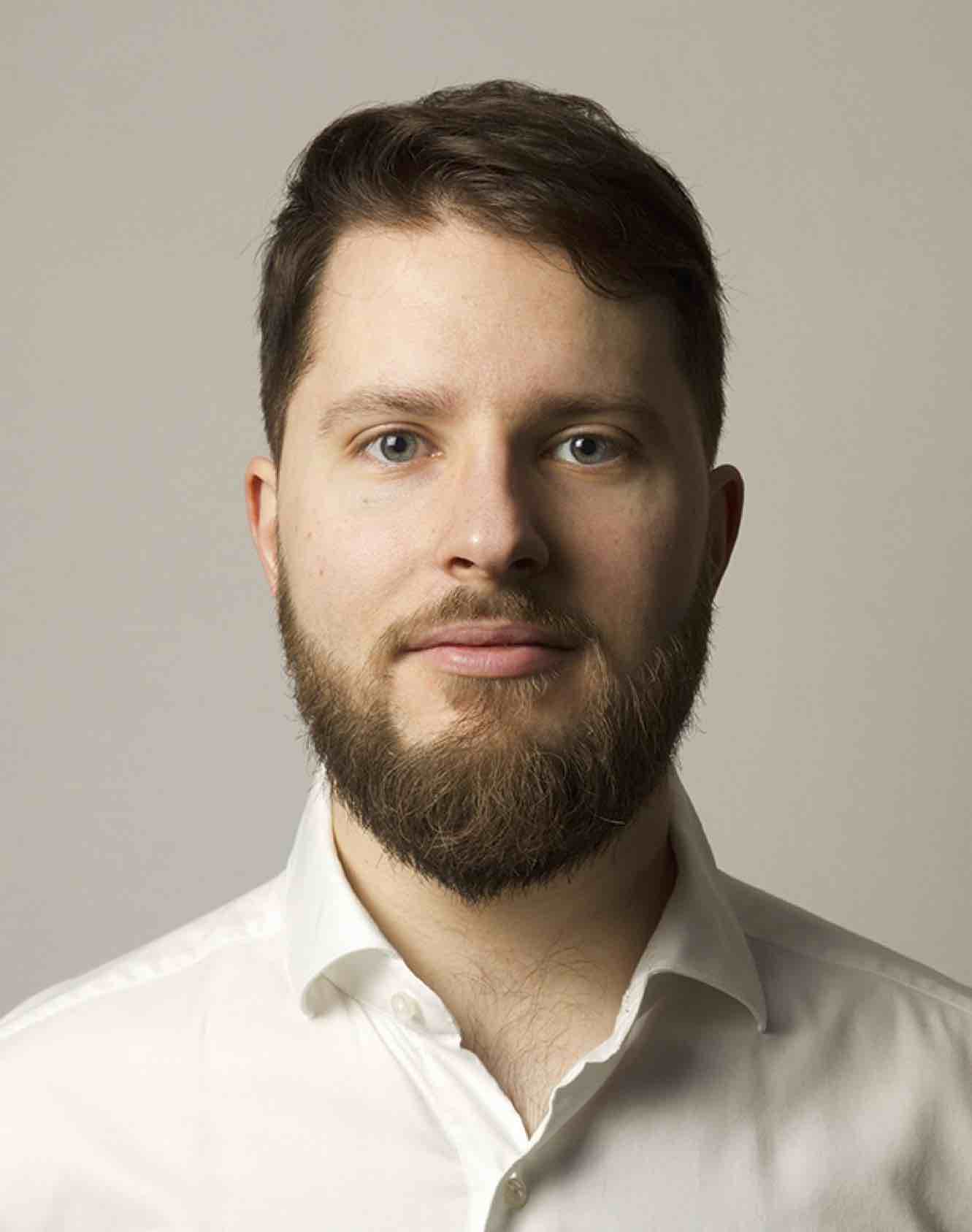}}]{Michele Polese} is an Associate Research Scientist at Northeastern University, Boston, since March 2020, working with Tommaso Melodia. He received his Ph.D. at the Department of Information Engineering of the University of Padova in 2020 under the supervision of with Michele Zorzi. He also was an adjunct professor and postdoctoral researcher in 2019/2020 at the University of Padova. During his Ph.D., he visited New York University (NYU), AT\&T Labs in Bedminster, NJ, and Northeastern University, Boston, MA. He collaborated with several academic and industrial research partners, including Intel, InterDigital, NYU, AT\&T Labs, University of Aalborg, King's College and NIST.
He was awarded with an Honorable Mention by the Human Inspired Technology Research Center (HIT) (2018), the Best Journal Paper Award of the IEEE ComSoc Technical Committee on Communications Systems Integration and Modeling (CSIM) 2019, and the Best Paper Award at WNS3 2019. His research interests are in the analysis and development of protocols and architectures for future generations of cellular networks (5G and beyond), in particular for millimeter-wave communication, and in the performance evaluation of complex networks. He is a Member of the IEEE.
\end{IEEEbiography}

\begin{IEEEbiography}
[{\includegraphics[width=1in,height=1.25in,keepaspectratio]{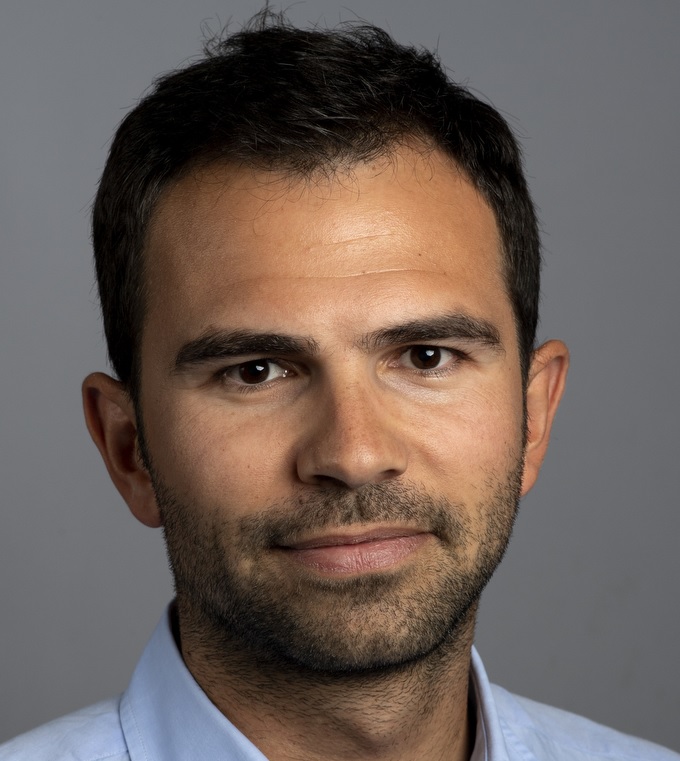}}]{Salvatore D'Oro}
received received his  Ph.D. degree from the University of Catania in 2015. He is currently a Research Assistant Professor at Northeastern University. He serves on the Technical Program Committee (TPC) of Elsevier Computer Communications journal and the IEEE Conference on Standards for Communications and Networking (CSCN) and European Wireless. He also served on the TPC of Med-Hoc-Net 2018 and several workshops in conjunction with IEEE INFOCOM and IEEE ICC. In 2015, 2016 and 2017 he organized the 1st, 2nd and 3rd Workshops on COmpetitive and COoperative Approaches for 5G networks (COCOA). 
Dr. D'Oro is also a reviewer for major IEEE and ACM journals and conferences. Dr. D'Oro's research interests include game-theory, optimization, learning and their applications to 5G networks. He is a Member of the IEEE.
\end{IEEEbiography}

\vfill

\begin{IEEEbiography}
[{\includegraphics[width=1in,height=1.25in,keepaspectratio]{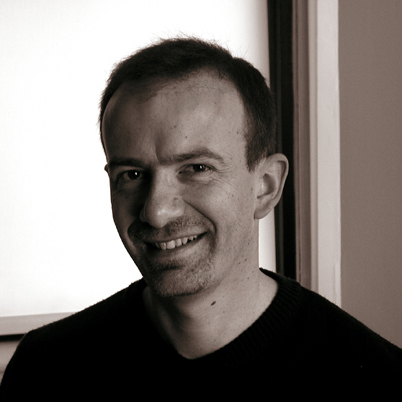}}]{Stefano Basagni}
is with the Institute for the Wireless Internet of Things and an associate professor at the ECE Department at Northeastern University, in Boston, MA. He holds a Ph.D.\ in electrical engineering from the University of Texas at Dallas (December 2001) and a Ph.D. in computer science from the University of Milano, Italy (May 1998). Dr. Basagni's current interests concern research and implementation aspects of mobile networks and wireless communications systems, wireless sensor networking for IoT (underwater and terrestrial), definition and performance evaluation of network protocols and theoretical and practical aspects of distributed algorithms. Dr. Basagni has published over nine dozen of highly cited, refereed technical papers and book chapters. His h-index is currently 45 (August 2020). He is also co-editor of three books. Dr. Basagni served as a guest editor of multiple international ACM/IEEE, Wiley and Elsevier journals. He has been the TPC co-chair of international conferences. He is a distinguished scientist of the ACM, a senior member of the IEEE, and a member of CUR (Council for Undergraduate Education).
\end{IEEEbiography}

\begin{IEEEbiography}
[{\includegraphics[width=1in,height=1.25in,keepaspectratio]{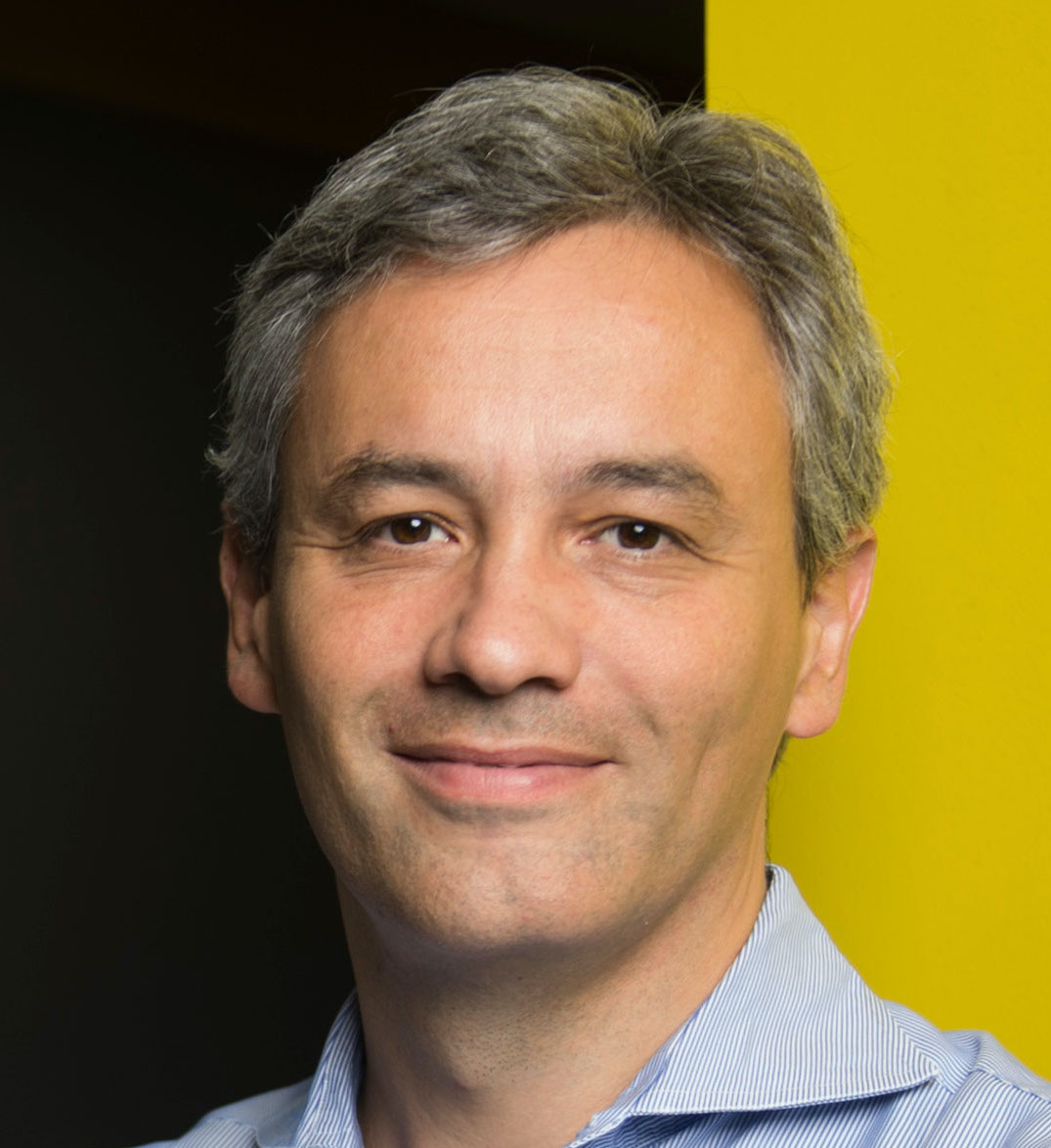}}]{Tommaso Melodia}
is the William Lincoln Smith Chair Professor with the Department of Electrical and Computer Engineering at Northeastern University in Boston. He is also the Founding Director of the Institute for the Wireless Internet of Things and the Director of Research for the PAWR Project Office. He received his Ph.D. in Electrical and Computer Engineering from the Georgia Institute of Technology in 2007. He is a recipient of the National Science Foundation CAREER award. Prof. Melodia has served as Associate Editor of IEEE Transactions on Wireless Communications, IEEE Transactions on Mobile Computing, Elsevier Computer Networks, among others. He has served as Technical Program Committee Chair for IEEE Infocom 2018, General Chair for IEEE SECON 2019, ACM Nanocom 2019, and ACM WUWnet 2014. Prof. Melodia is the Director of Research for the Platforms for Advanced Wireless Research (PAWR) Project Office, a \$100M public-private partnership to establish 4 city-scale platforms for wireless research to advance the US wireless ecosystem in years to come. Prof. Melodia's research on modeling, optimization, and experimental evaluation of Internet-of-Things and wireless networked systems has been funded by the National Science Foundation, the Air Force Research Laboratory the Office of Naval Research, DARPA, and the Army Research Laboratory. Prof. Melodia is a Fellow of the IEEE and a Senior Member of the ACM.
\end{IEEEbiography}

\vfill

\end{document}